\documentclass[a4paper,UKenglish,cleveref, autoref, thm-restate]{lipics-v2021}

\usepackage{hyperref}
\hypersetup{
	colorlinks=false,
	hidelinks,
	linkcolor=blue,
	filecolor=magenta,      
	urlcolor=cyan,
	pdfpagemode=FullScreen,
}

\usepackage{multicol}
\usepackage{paracol}
\usepackage{microtype}

\usepackage{pdfpages}

\usepackage{makecell}

\usepackage{placeins}
\usepackage{graphicx}
\usepackage{amssymb}
\usepackage{amsmath}
\usepackage{float}
\usepackage{braket}
\usepackage{caption}
\usepackage{enumitem}
\usepackage{verbatim}

\usepackage{tabto} 

\usepackage{booktabs}
\usepackage{pgfplots}
\usepackage{nicefrac}

\usepackage{tabulary}

\nolinenumbers

\usepackage{algpseudocode}
\algdef{SE}[DOWHILE]{Do}{doWhile}{\algorithmicdo}[1]{\algorithmicwhile\ #1}%

\usepackage[linesnumbered,titlenotnumbered,ruled]{algorithm2e}
\usepackage{tikz,pgffor}
\usetikzlibrary{arrows,backgrounds}
\usepackage{tkz-euclide}
\usetikzlibrary{arrows}
\usetikzlibrary{shapes}
\usetikzlibrary{calc}
\usetikzlibrary{automata}
\usetikzlibrary{positioning}
\usepackage{listings}
\usepackage{transparent}
\usepackage{calrsfs}

\usepackage{algpseudocodex}

\DeclareMathAlphabet{\pazocal}{OMS}{zplm}{m}{n}
\usepackage{amsthm}
\usepackage{thmtools}

\usepackage{thm-restate}

\newcommand{\nocontentsline}[3]{}
\newcommand{\tocless}[2]{\bgroup\let\addcontentsline=\nocontentsline#1{#2}\egroup}

\newcommand{\E}{\mathbb{E}}

\newcommand{\Exp}{\E}

\newcommand{\stateVass}{\ensuremath{p}}
\newcommand{\stateVassalt}{\ensuremath{q}}

\newcommand{\node}{\ensuremath{M}}

\newcommand{\transition}{\ensuremath{t}}

\newcommand{\prob}{\ensuremath{\mathbb{P}}}
\newcommand{\Prob}{\prob}

\newcommand{\length}{\mathit{len}}

\newcommand{\tin}{\mathit{In}}
\newcommand{\tout}{\mathit{Out}}

\newcommand{\simplecompset}{\ensuremath{Y}}
\newcommand{\multicompset}{\ensuremath{X}}
\newcommand{\states}{\mathit{St}}
\newcommand{\transitions}{\mathit{Trns}}
\newcommand{\counters}{\mathit{Count}}
\newcommand{\statesDemon}{\mathit{St_d}}

\newcommand{\statesProb}{\mathit{St_p}}

\newcommand{\countersset}{\mathit{Count}}

\newcommand{\Term}{\mathit{Term}}

\newcommand{\Czero}{\ensuremath{C_{zero}}}

\newcommand{\cMDstrats}[1]{cMD(#1)}
\newcommand{\stratscMD}[1]{\cMDstrats{#1}}
\newcommand{\cMD}{cMD}
\newcommand{\stratsMD}[1]{\cMDstrats{#1}}

\newcommand{\maxupdate}[1]{u_{#1}}

\newcommand{\strategyDemon}{\ensuremath{\sigma}}

\newcommand{\Q}{\ensuremath{\mathbb{Q}}}

\newcommand{\support}{\ensuremath{supp}}
\newcommand{\spanCone}{\ensuremath{span}}

\newcommand{\realeffect}{\mathrm{It}}

\newcommand{\mecs}{\mathit{mecs}}

\newcommand{\emptypointpair}{\ensuremath{\perp}}
\newcommand{\pointpair}{\ensuremath{\zeta}}

\newcommand{\Decompose}{Decomp}
\newcommand{\DecomposeComponent}{DecompPT}

\newcommand{\pathm}{\ensuremath{\pi}}

\newcommand{\MEC}{B}

\newcommand{\variablenotk}{\ensuremath{\ell}}

\newcommand{\vecmine}[1]{\vec{#1}}

\newcommand{\N}{\mathbb{N}}
\newcommand{\R}{\mathbb{R}}
\newcommand{\Z}{\mathbb{Z}}
\newcommand{\A}{\pazocal{A}}

\newcommand{\F}{\pazocal{F}}

\newcommand{\calC}{\pazocal{C}}
\newcommand{\calL}{\pazocal{L}}
\newcommand{\calT}{\pazocal{T}}

\newcommand{\calO}{\pazocal{O}}

\newcommand{\bigO}{\calO}

\newcommand{\size}[1]{|\!|#1|\!|}

\newcommand{\parent}{\ensuremath{Parent}}
\newcommand{\child}{\ensuremath{Children}}

\newcommand{\fsize}{{Size}}
\newcommand{\gresets}{{Resets}}

\newcommand{\bx}{\mathbf{x}}
\newcommand{\by}{\mathbf{y}}
\newcommand{\bz}{\mathbf{z}}
\newcommand{\bu}{\mathbf{u}}
\newcommand{\bv}{\mathbf{v}}
\newcommand{\bw}{\mathbf{w}}

\newcommand{\bincount}{\ensuremath{h}}

\newcommand{\ba}{\mathbf{a}}

\newcommand{\bp}{\mathbf{p}}

\newcommand{\bb}{\mathbf{b}}

\newcommand{\PTIMEfull}{\mathbf{PTIME}}

\makeatletter
\def\squarecorner#1{
	\pgf@x=\the\wd\pgfnodeparttextbox%
	\pgfmathsetlength\pgf@xc{\pgfkeysvalueof{/pgf/inner xsep}}%
	\advance\pgf@x by 2\pgf@xc%
	\pgfmathsetlength\pgf@xb{\pgfkeysvalueof{/pgf/minimum width}}%
	\ifdim\pgf@x<\pgf@xb%
	\pgf@x=\pgf@xb%
	\fi%
	\pgf@y=\ht\pgfnodeparttextbox%
	\advance\pgf@y by\dp\pgfnodeparttextbox%
	\pgfmathsetlength\pgf@yc{\pgfkeysvalueof{/pgf/inner ysep}}%
	\advance\pgf@y by 2\pgf@yc%
	\pgfmathsetlength\pgf@yb{\pgfkeysvalueof{/pgf/minimum height}}%
	\ifdim\pgf@y<\pgf@yb%
	\pgf@y=\pgf@yb%
	\fi%
	\ifdim\pgf@x<\pgf@y%
	\pgf@x=\pgf@y%
	\else
	\pgf@y=\pgf@x%
	\fi
	\pgf@x=#1.5\pgf@x%
	\advance\pgf@x by.5\wd\pgfnodeparttextbox%
	\pgfmathsetlength\pgf@xa{\pgfkeysvalueof{/pgf/outer xsep}}%
	\advance\pgf@x by#1\pgf@xa%
	\pgf@y=#1.5\pgf@y%
	\advance\pgf@y by-.5\dp\pgfnodeparttextbox%
	\advance\pgf@y by.5\ht\pgfnodeparttextbox%
	\pgfmathsetlength\pgf@ya{\pgfkeysvalueof{/pgf/outer ysep}}%
	\advance\pgf@y by#1\pgf@ya%
}
\makeatother

\pgfdeclareshape{square}{
	\savedanchor\northeast{\squarecorner{}}
	\savedanchor\southwest{\squarecorner{-}}
	
	\foreach \x in {east,west} \foreach \y in {north,mid,base,south} {
		\inheritanchor[from=rectangle]{\y\space\x}
	}
	\foreach \x in {east,west,north,mid,base,south,center,text} {
		\inheritanchor[from=rectangle]{\x}
	}
	\inheritanchorborder[from=rectangle]
	\inheritbackgroundpath[from=rectangle]
}

\tikzstyle{loop above}=[tran, to path={.. controls +(60:1) 
	and +(120:1) .. (\tikztotarget) \tikztonodes}]
\tikzstyle{loop below}=[tran, to path={.. controls +(240:1) 
	and +(300:1) .. (\tikztotarget) \tikztonodes}]
\tikzstyle{loop left}=[tran,  to path={.. controls +(150:1) 
	and +(210:1) .. (\tikztotarget) \tikztonodes}]
\tikzstyle{loop right}=[tran,  to path={.. controls +(330:1) 
	and +(30:1) .. (\tikztotarget) \tikztonodes}]
\tikzstyle{bigstoch}=[circle,draw,minimum size=8ex,inner sep=0pt,font=\Large, very thick,text centered, fill=blue!20, fill opacity=0.2, draw=black!80, text opacity=1]
\tikzstyle{stoch}=[circle,draw,minimum size=4ex,inner sep=0pt,font=\large, very thick,text centered, fill=blue!20, fill opacity=0.2, draw=black!80, text opacity=1]

\tikzstyle{nondet}=[square,,draw,minimum size=4ex,inner sep=0pt,font=\large, very thick,text centered, fill=blue!20, fill opacity=0.2, draw=black!80, text opacity=1]
\tikzstyle{stoch}=[square,,draw,minimum size=4ex,inner sep=0pt,font=\large, very thick,text centered, fill=blue!20, fill opacity=0.2, draw=black!80, text opacity=1]
\tikzstyle{random}=[circle,,draw,minimum size=4ex,inner sep=0pt,font=\large, very thick,text centered, fill=blue!20, fill opacity=0.2, draw=black!80, text opacity=1]
\tikzstyle{angelic}=[diamond,,draw,minimum size=4ex,inner sep=0pt,font=\large, very thick,text centered, fill=blue!20, fill opacity=0.2, draw=black!80, text opacity=1]
\tikzstyle{demonic}=[square,,draw,minimum size=4ex,inner sep=0pt,font=\large, very thick,text centered, fill=blue!20, fill opacity=0.2, draw=black!80, text opacity=1]
\tikzstyle{ran}=[circle,,draw,minimum size=4ex,inner sep=0pt,font=\large, very thick,text centered, fill=blue!20, fill opacity=0.2, draw=black!80, text opacity=1]

\tikzstyle{squar}=[regular polygon,regular polygon sides=4,draw,minimum size=4ex,inner sep=0pt,font=\large, very thick,text centered, fill=blue!20, fill opacity=0.2, draw=black!80, text opacity=1]

\tikzstyle{bigstoch2}=[circle,draw,minimum size=8ex,inner sep=0pt,font=\Large, very thick,text centered, fill=blue!20, fill opacity=0.2, draw=black!80, text opacity=1, double]
\tikzstyle{bigmin}=[diamond,draw,minimum size=8ex,inner sep=0pt,font=\Large, very thick,text centered, fill=blue!20, fill opacity=0.2, draw=black!80, text opacity=1]
\tikzstyle{max}=[rectangle,draw,minimum size=4ex,inner sep=0pt,font=\large, very thick,text centered, fill=blue!20, fill opacity=0.2, draw=black!80, text opacity=1]
\tikzstyle{bigtran}=[very thick,draw,-angle 60,font=\scriptsize, inner sep = 6pt]
\tikzstyle{tran}=[thick,draw,font=\scriptsize, inner sep = 6pt,-stealth]

\tikzstyle{label}=[shape=circle,draw,inner sep=0pt,minimum size=5mm]
\tikzstyle{line} = [draw, color=black, -latex]
\tikzstyle{max}=[thick,draw,color=black,minimum size=1.4em,inner sep=0em]
\tikzstyle{min}=[diamond,thick,draw,color=black,minimum size=1.4em,%
inner sep=0em]
\tikzstyle{mc}=[rounded corners,thick,draw,minimum size=1.4em,%
inner sep=.5ex]
\tikzstyle{tran}=[thick,draw,->,>=stealth,rounded corners]

\tikzstyle{tran right above}=[tran, to path={.. controls +(60:.8) and +(120:.8) .. (\tikztotarget) \tikztonodes}]
\tikzstyle{tran left above}=[tran, to path={.. controls +(120:.8) and +(60:.8) .. (\tikztotarget) \tikztonodes}]
\tikzstyle{tran left below}=[tran, to path={.. controls +(240:.8) and +(300:.8) .. (\tikztotarget) \tikztonodes}]
\tikzstyle{tran right below}=[tran, to path={.. controls +(300:.8) and +(240:.8) .. (\tikztotarget) \tikztonodes}]

\tikzstyle{loop right below}=[tran, to path={.. controls +(330:.8) and +(270:.8) .. (\tikztotarget) \tikztonodes}]
\tikzstyle{loop right above}=[tran, to path={.. controls +(30:.8) and +(90:.8) .. (\tikztotarget) \tikztonodes}]

\tikzstyle{bigstoch2}=[circle,draw,minimum size=8ex,inner sep=0pt,font=\Large, very thick,text centered, fill=blue!20, fill opacity=0.2, draw=black!80, text opacity=1, double]
\tikzstyle{bigmin}=[diamond,draw,minimum size=8ex,inner sep=0pt,font=\Large, very thick,text centered, fill=blue!20, fill opacity=0.2, draw=black!80, text opacity=1]
\tikzstyle{bigtran}=[very thick,draw,-angle 60,font=\scriptsize, inner sep = 6pt]

\title{Polynomial VASS MDPs} 

\author{Michal Ajdar{\'{o}}w }{ }{xajdarow@mail.muni.cz }{https://orcid.org/0000-0003-0694-0944}{}

\authorrunning{M. Ajdar{\'{o}}w} 

\Copyright{Michal Ajdar{\'{o}}w} 

\begin{document}
	
	\maketitle

	\begin{abstract}
			Markov decision process over vector addition system with states (VASS MDP) is a finite state model combining non-deterministic and probabilistic behavior, augmented with non-negative integer counters that can be incremented or decremented during each state transition. VASS MDPs can serve as abstractions of probabilistic programs, offering many decidable properties. In this paper, we develop techniques for analyzing the polynomiality of  asymptotic estimates for VASS MDPs.	That is, for every initial configuration of size \(n\), we consider the  number of transitions needed to reach a configuration with some counter negative.  We show that for any VASS MDP, either for all \(\epsilon>0 \) and all sufficiently large \(n\) the number of transitions is at least \(2^{n^{\frac{1}{2}-\epsilon}} \) with probability at least \(1-\epsilon \), or for any superpolynomial function \(f \), all \(\epsilon>0 \) and all sufficiently large \(n\) the number of transitions is upper bounded by \(f(n) \) with probability at least \(1-\epsilon \). Furthermore, if the VASS MDP is DAG-like then we can replace \(f(n)\) with \(n^{k+\epsilon} \), where \(k=2\cdot 3^{w\cdot d+|T|} \) and \(w,d\) and \(T \) are the number of MECs, counters and transitions of the VASS MDP, respectively. It is decidable in polynomial time which case holds.
	\end{abstract}
	
	
	\section{Introduction}
\label{sec-intro}
\newcommand{\term}{term}
\textbf{Probabilistic Programs \& VASS:} Probabilistic systems play an important role in various areas of computing such as machine learning \cite{Ghahramani:15-prob-ai-nature}, network protocol design~\cite{FosterKMRS:16-prob-netkat}, robotics~\cite{ThrunBurgardFox:2005-robotics}, privacy and security~\cite{BartheGGHS:16-diff-privacy-prob-couplings}, and many others. For this reason, verification of probabilistic systems receives considerable attention. As in the classical (non-probabilistic) setting, in probabilistic verification one typically constructs a suitable abstract model over-approximating the real behaviour of the system. In the past, the verification research was focused mostly on finite-state probabilistic models~\cite{BK:book}. However, the recent proliferation of general, Turing-complete \emph{probabilistic programming languages} necessitates the use of more complex models that can encompass multiple potentially unbounded numerical variables.

In the classical setting, one of the standard formalisms used for program abstraction are \emph{vector addition systems with states (VASS)}~\cite{KM69}. Intuitively, a VASS is a finite directed graph where every edge is labelled with a vector of integer counter updates of a fixed dimension~$d$. A \emph{configuration} $p\bv$ is specified by the current state $p$ and the current counter values $\bv$. The computation proceeds by moving along the edges in the graph and adding the labels to the counters vector. Since VASS themselves are not Turing-complete, they have many decidable properties, and they have been successfully used as program abstractions in termination and complexity analysis~\cite{SZV:amortized} as well as for reasoning about parallel programs~\cite{EN94,KM69} and parametrized systems~\cite{Bloem16,conf/icalp/AminofRZS15}. Applying such an abstraction to a probabilistic program yields a VASS MDP, which allows for a \emph{probabilistic choice} of a transition in some states. 

\noindent
\textbf{Termination Complexity:} One of the fundamental problems in program analysis is to evaluate a given program's runtime. In the classical setting, this problem emerges in various flavours, ranging from worst-case execution time-analysis~\cite{Wilhelmal:2008:wcet-survey,Cassez:11-wcet-survey} in real-time systems to obtaining bounds on the number of execution steps~\cite{GulwaniMehraChulimbi:09-speed}, analysing asymptotic~\cite{CFG17}, or amortized complexity~\cite{journals/toplas/0002AH12}. VASS-based abstractions were successfully used in the last scenario~\cite{SZV:amortized,SZV:difference-constraints}. In the probabilistic setting, several approaches to reason about the expected runtime of probabilistic programs were developed~\cite{KaminskiKMO:18-wp-expected-runtime-jacm,NgoCarbonneauxHoffmann:18-pp-resource-analysis}. The analysis is much more demanding than in the classical case.  In fact, reasoning about the termination behaviour of probabilistic programs is known to be harder than for deterministic programs (i.e. higher in the arithmetic hierarchy) \cite{KaminskiKatoenMateja:19-pp-hardness-act-inf}. 

In order to avoid multiple deficiencies of the expected runtimes  (e.g., non-compositionality \cite{KaminskiKMO:18-wp-expected-runtime-jacm}, see also \cite{Ajdarow2025}), a new measure of runtime for probabilistic systems,  called \emph{asymptotic estimates} has been recently proposed in \cite{AKCONCUR23}. Intuitively, it consists of a bound on all but \(\delta \) ratio of runs where \(\delta \) tends to \(0 \) as the size of the input tends to \(\infty \). These can be seen as a formalization of the folklore that  gambler's ruin \cite{gamblerruin} tends to terminate in quadratic time.   Asymptotic estimates are especially relevant with respect to almost-sure termination, i.e. whether a system terminates with probability \(1\), as a finite upper asymptotic estimate exists iff the given complexity is finite with probability \(1\) \cite{Ajdarow2025}. Almost-sure termination  has been studied for  probabilistic models such as Markov chains \cite{Baier2018}, probabilistic pushdown automata \cite{lmcs:2256}, probabilistic one-counter programs \cite{10.11452629599}, and \(k\)-d PCPs \cite{fdsfds}.


\noindent
\textbf{Our Setting:}
We study asymptotic estimates for VASS MDPs with demonic non-determinism. We consider a basic variant of VASS termination: the \emph{zero termination}, where the computation stops when some counter becomes negative. Two VASS complexity measures are commonly studied with respect to their asymptotic growth: \emph{termination complexity}, analogous to time complexity, and \emph{\(c\)-counter complexity}, analogous to space complexity.



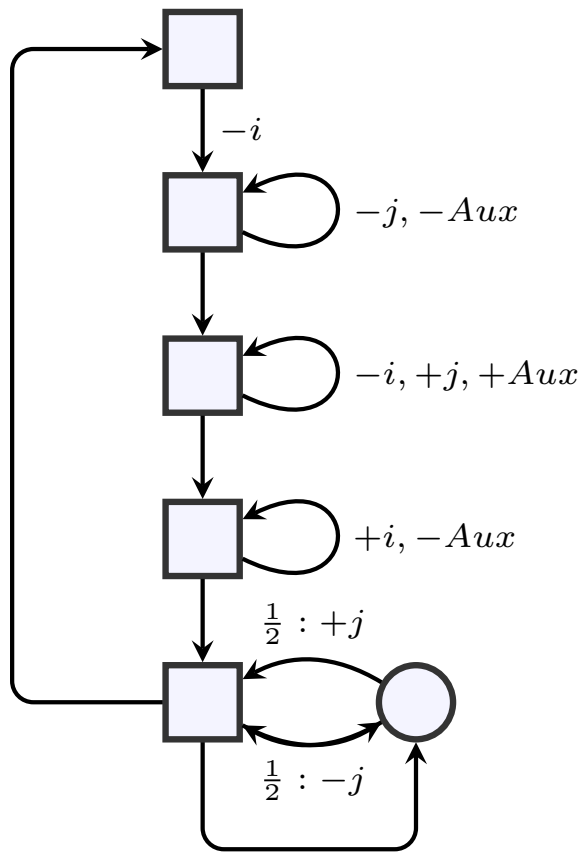
\begin{figure}[thb]
	\centering
	\lstset{basicstyle=\footnotesize}
	\begin{tabular}{l|l|l}
		\begin{minipage}{3.3cm}
			\begin{lstlisting}[escapeinside={(*}{*)}]
input i;
while(i>0) 
  i--;
  j:=i;
  while(j>0)
    if rand()<(*\(\frac{1}{2}\)*): 
      j--;
    else: 
      j++;
			\end{lstlisting}
			\vspace*{7em}
		\end{minipage}
		&
		\begin{minipage}{3.3cm}
			\begin{lstlisting}[escapeinside={(*}{*)}]
input i;
while(i>0) 
  i--;
  j:=0; Aux:=0;
  while(i>0)
    i--; 
    j++;
    Aux++;
  while(Aux>0)
    i++;
    Aux--;
  while(j>0)
    if rand()<(*\(\frac{1}{2}\)*): 
      j--;
    else: 
      j++;
			\end{lstlisting}
		\end{minipage}
		&
		\begin{minipage}{3cm}\hspace*{1em}
			\begin{tikzpicture}[scale=.9, every node/.style={scale=0.9}, x=1.4cm, y=1.2cm, font=\scriptsize]
				\foreach \x in {0,1,2,3,4}{ 
					\node [stoch] (t\x) at (0,-\x)  {};
				}
				
				\node [random] (r) at (1.12,-4) {};
				
				\node [] () at (2.45,-4) { };
				
				\foreach \x/\y in {1/2,2/3,3/4}{
					\draw [tran] (t\x) to node[right] {} (t\y);
				}
				\draw [tran] (t0) to node[right] {$-i$ } (t1);   
				\draw [tran,rounded corners] (t4) -- +(-1,0) --   +(-1,4) -- node[above] {} (t0);   
				\draw [tran,loop right] (t1) to node[right] {$-j,-Aux$   } (t1); 
				\draw [tran,loop right] (t2) to node[right] {$-i,+j,+Aux$ } (t2); 
				\draw [tran,loop right] (t3) to node[right] {$+i,-Aux$ } (t3); 
				
				\draw [tran,bend right] (t4) to node[below] {}  (r);

				%


				\draw [tran,bend left] (r) to node[below] {$\frac{1}{2}:-j$} (t4); 
				
				\draw [tran,bend right] (r) to node[above] {$\frac{1}{2}:+j$} (t4); 
				
				\draw [tran] (t4) to (0,-4.9) to (1.12,-4.9) to (r) ;
			\end{tikzpicture}
		\end{minipage}
	\end{tabular}
	\caption{A skeleton of a simple imperative program (left) and its VASS MDP model (right). }
	\label{figure-pvassexample}\label{fig-VASS-model}
\end{figure}
\begin{example}
	\label{exa-demonic}
	Consider the program skeleton of Fig.~\ref{fig-VASS-model}~(left). Since a VASS MDP cannot directly model the assignment \texttt{j:=i} and cannot test a counter for zero, the skeleton is first transformed into an equivalent program of Fig.~\ref{fig-VASS-model}~(middle), where the assignment \texttt{j:=i} is implemented using an auxiliary variable \texttt{Aux} and two \texttt{while} loops. Clearly, the execution of the transformed program is only longer than the execution of the original skeleton (for all inputs). For the transformed program, an over-approximating VASS MDP model is obtained by replacing conditionals with non-determinism, see Fig.~\ref{fig-VASS-model}~(right). Note that upper bounds on the complexity of this VASS MDP carry over to the original program.   This VASS MDP has a polynomial upper asymptotic estimate for its termination complexity.  Hence, the same upper bound is valid also for the original program skeleton. This also implies the original program terminates with probability \(1\).  		Using the results of this paper this  analysis can be performed fully automatically.
\end{example}

\noindent
\textbf{Our Contribution:} 
As our main result  (Theorem~\ref{theorem-main-polynomial-vs-exp}) we study the polynomiality of asymptotic estimates in VASS MDPs. Let \(\A \) be a VASS MDP and \(\F \) be either the termination or the \(c\)-counter complexity. We show  one of the following holds:
\begin{itemize}[topsep=3.5pt]
	\setlength{\itemsep}{-1pt}
	\item  every superpolynomial function \(f\) is an upper asymptotic estimate of \(\F \). Furthermore, if \(\A \) is DAG-like with \(w\) MECs,\footnote{Intuitively, a MEC (maximal end component) is a strongly connected sub-MDP in which nondeterminism can be resolved so that the component is never left, with probability 1.} then \(n^{2\cdot k} \) is an upper asymptotic estimate of \(\F \) where \(k=3^{w\cdot d+ |\transitions(\A)|} \). Moreover, if \(\F \) is \(c\)-counter complexity then also \(n^{k} \) is an upper asymptotic estimate of \(\F \).
	\item\(\F \) has a lower asymptotic estimate \(2^{\sqrt{n}} \). 
\end{itemize} 
It is decidable in polynomial time which case holds. We also show that for non DAG-like VASS MDPs the first case does \emph{not} necessarily imply  there exists a polynomial upper asymptotic estimate for termination/\(c\)-counter complexity (Lemma~\ref{lemma-non-dag-example}). 

Additionally, we also show an alternative, more intuitive definition of asymptotic estimates using asymptotic bounds of the \emph{failure-time} function (Theorem~\ref{observation-f-p-f}). This function is  widely studied in survival analysis \cite{survivalone} and has applications across multiple areas, such as engineering \cite{Kalbfleisch1980TheSAEngineering}, economics \cite{durationEconomics}, sociology \cite{allison1SOCIOLOGY},  medicine \cite{6c51df5f-27d3-3ed4-9311-1f6af36a828cMEDICINETRIALS}, and many others. This suggest analyzing asymptotic estimates might be of interest in these areas as well.

These results build on analogous results for VASS established in \cite{Leroux:Polynomial-termination-VASS,Zuleger:VASS-polynomial,AK:VASS-polynomial-termination}, techniques for VASS MDPs developed in \cite{AKCONCUR23}, and  combining them with novel techniques for probabilistic analysis. Mainly the deterministic/probabilistic decomposition for VASS MDPs (Section~\ref{sec-det-prob-decompositions}), where the deterministic part behaves almost like a  VASS (and VASS techniques work on it) and the probabilistic part is behaves nicely and can be fully captured using martingales.

\smallskip 

\noindent
\textbf{Related Work:} 
Traditional VASS decision problems such as reachability, liveness, or boundedness are computationally hard \cite{9719806,Lipton:PN-Reachability,MM:containment-Petri}, and other verification problems such as equivalence-checking \cite{Jancar:PN-bisimilarity-TCS} or model-checking \cite{Esparza:ModelChecking-AI} are even undecidable. In contrast to this, decision problems related to the asymptotic growth of VASS complexity measures have low complexity and for many important sub-classes even in \emph{polynomial time} \cite{BCKNVZ:VASS-linear-termination,Zuleger:VASS-polynomial,KLV:VASS-Grzegorczyk,Leroux:Polynomial-termination-VASS,AK:VASS-polynomial-termination}; see \cite{Kucera:Asymptotic-VASS-Analysis-SIGLOG} for a recent overview.

Regarding asymptotic analysis of VASS MDPs: Linearity of expected termination complexity for DAG-like VASS MDPs is decidable in polynomial time, and if it is not linear then it is in \(\Omega(n^2) \) \cite{BCKNV:probVASS-linear-termination}, and the same holds also for analyzing linearity of asymptotic estimates for  termination/\(c\)-counter complexity \cite{AKCONCUR23}.  This linear vs at least quadratic dichotomy does not hold for expected  \(c\)-counter complexity \cite{Ajdarow2025}. This shows that the expected \(c\)-counter complexity in VASS MDPs can take values that are not possible in VASS, since this dichotomy does hold for VASS \cite{Zuleger:VASS-polynomial,AK:VASS-polynomial-termination}. Additionally, asymptotic estimates of \(1\)-dimensional VASS MDPs are known to be fully classifiable in polynomial time \cite{AKCONCUR23}.

 
VASS analogy of our main result is shown in \cite{Leroux:Polynomial-termination-VASS}. Concretely, it shows that termination/\(c\)-counter complexity in VASS is either upper bounded by a polynomial or is in \(2^{\Omega(n)} \), with it being decidable which case holds in polynomial time.


	
\vspace{-0.2cm}
\section{Preliminaries}
\label{sec-prelim}

We use $\N$, \(\mathbb{N}_0\), \(\mathbb{Z}\), $\Q$,  and $\R$ to denote the sets of positive integers, non-negative integers, integers,  rational numbers, and real numbers, respectively.  Let $A$ be a finite index set. The vectors of $\R^A$ are denoted by bold letters such as $\bu,\bv,\bz,\ldots$. The component of $\bv$ of index $i\in A$ is denoted by $\bv(i)$. 
For every $n \in \R$, we use $\vecmine{n}$ to denote the constant vector where all
components are equal to~$n$.
The other
standard operations and relations on $\R$ such as
$+$, $\leq$, or $<$ are extended to $\R^A$ in the component-wise way. In particular,  $\bv \leq  \bu$ if $\bv(i) \leq \bu(i)$ for every index $i$. A function \(f\) is \emph{superpolynomial} or \emph{superpolylogarithmic}  if it holds \(f\in \Omega(n^k) \) or \(f\in \Omega(\log^k n) \) for all \(k\in \mathbb{N} \), respectively.




\subsection{VASS Markov Decision Processes}
\begin{definition}
	\label{def-VASS} 
	
	A \emph{Vector Addition System with States over Markov Decision Process} (VASS MDP) with counters \(\countersset(\A) \) is a tuple \(\A = \Big(\states(\A), \big(\statesDemon(\A),\statesProb(\A)\big),\allowdisplaybreaks\transitions(\A),P\Big)\) where 
\begin{itemize}[topsep=1.2pt,leftmargin=0.81cm,itemsep=-1.1pt]
	\item \(\countersset(\A) \) is a finite set of counters,
	\item $\states(\A) \neq \emptyset$ is a finite set of \emph{states} split into two disjoint subsets \(\statesDemon(\A)\) and \(\statesProb(\A)\) of \emph{non-deterministic} and \emph{probabilistic} states, respectively,
	\item $\transitions(\A) \subseteq \states(\A) \times \Z^{\countersset(\A)}\times \states(\A)$ is a finite set of \emph{transitions},
	\item $P$ is a function assigning to each $t\in \tout(p) $, where $p \in \statesProb(\A)$ and \(\tout(p)\subseteq\transitions(\A)\) is the set of all transitions of the form \((p,\bu,q) \), a positive rational probability so that  $\sum_{t \in \tout(p)} P(t) =1$. 
\end{itemize}
\end{definition}

The encoding size of $\A$ is denoted by $\size{\A}$, where the integers representing counter updates are written in binary and probability values are written as fractions of binary numbers. A  VASS MDP $\A$ is a \emph{VASS Markov chain} if  \(\statesDemon(\A)=\emptyset \), and a \emph{VASS} if \(\statesProb(\A)=\emptyset \). We say \(\A\) is \emph{\(d\)-dimensional} if \(|\countersset(\A)|=d \).  For every $p \in \states(\A)$, we use $\tin(p) \subseteq \transitions(\A)$ to denote the set of all transitions of the form $(q,\bu,p)$. The update vector $\bu$ of a transition $t = (p,\bu,q)$ is also denoted by $\bu_t$. The size of \(\A \) is 


A \emph{configuration} of $\A$ is a pair $p\bv$, where $p \in \states(\A)$ and $\bv \in \Z^{\countersset(\A)}$. If some component of $\bv$ is negative then $p\bv$ is \emph{terminal}. A \emph{computation} in $\A$ is a finite sequence of configurations  the form
$\alpha=p_0\bv_0, \ldots, p_k \bv_k$ where $(p_{i},\bv_{i+1}-\bv_i,p_{i+1}) \in \transitions(\A)$  for all $i<k$.  The \emph{length} of $\alpha$ is defined as $\length(\alpha) = k$. 
An \emph{infinite computation} in $\A$ is an infinite sequence $\pi = p_0\bv_0, p_1 \bv_1,p_2 \bv_2,\ldots$ such that every finite prefix of $\pi$ is a computation. Let $\Term(\pi)$ be the least $j$ such that $p_j\bv_j$ is terminal. If there is no such $j$, we put  $\Term(\pi) = \length(\pi)$.

 A \emph{strategy} of \(\A \) is a function $\sigma$ assigning to every computation $p_0\bv_0,\ldots,p_k \bv_k$ with $p_k \in \statesDemon(\A)$ a probability distribution over~$\tout(p_k)$. A strategy is \emph{counterless-memoryless (cM)} if it depends only on the last state $p_k$, and \emph{deterministic (D)} if it always returns a distribution assigning probability~\(1\) to a single transition; a strategy is \emph{cMD} if it is both. We denote by \(\stratsMD{\A} \) the set of all {\cMD}  strategies of \(\A\). Note that a {\cMD} strategy \(\sigma \) selects the same outgoing transition in every \(p\in \statesDemon(\A) \) every time \(p\) is visited. Hence we can  “apply” \(\sigma\) to \(\A \) by removing the other outgoing transitions, and declaring all states to be probabilistic. This way we obtain a VASS Markov chain which we denote by \(\A_\sigma \). 

 Every initial configuration $p\bv$ and every strategy $\sigma$ determine the probability space over computations initiated in $p\bv$ in the standard way. We use $\prob^\sigma_{p\bv}$ to denote the associated probability measure.
 For a measurable function $X$ over computations, we use $\Exp^\sigma_{p \bv}[X]$ to denote the expected value of~$X$. When  \(\sigma, p \) and/or \(\bv\) are clear from the context we omit them, i.e., we write only $\Exp[X]$, $\Exp_{p}[X],\dots$ etc.


We say that $q$ is \emph{reachable} from~$p$  if there exists a computation \(p\bv,\ldots, q \bv'\). We say  \(\A \) is \emph{strongly connected} if \(q\) is reachable from \(p\) for all \(p,q\in \states(\A) \).

\newcommand{\Cset}{S}
\newcommand{\Lset}{T}

An \emph{end component (EC)} of $\A$ is a pair $(\Cset,\Lset)$ where $\emptyset\neq \Cset \subseteq \states(\A)$ and  $\Lset \subseteq \transitions(\A)$ such that the following conditions are satisfied:
\begin{itemize}[topsep=1.7pt,itemsep=-1.6pt]
	\item  if $p \in \Cset \cap \statesDemon(\A)$, then $\tout(p)\cap \Lset\neq \emptyset$; 
	\item if $p \in \Cset \cap \statesProb(\A)$, then $\tout(p)\subseteq \Lset$; 
	\item if $(p,\bu,q) \in \Lset$, then $p,q \in \Cset$;
	\item for all $p,q \in \Cset$ we have that $q$ is reachable from $p$ and vice versa using only transitions from \(\Lset\). 
\end{itemize}

Note that if $(\Cset,\Lset)$ and $(\Cset',\Lset')$ are ECs with $\Cset \cap \Cset' \neq \emptyset$, then $(\Cset\cup \Cset', \Lset \cup \Lset')$ is also an EC. Hence, every $p\in \states(\A)$ either belongs to a unique \emph{maximal end component} (MEC), or does not belong to any EC. Also observe that each MEC can be seen as a strongly connected VASS MDP. (Intuitively, a MEC is a ``largest'' strongly connected sub‐VASS MDP where you can resolve nondeterminism so as never to leave.) 

We say a VASS MDP is \emph{DAG-like} if for each pair of distinct MECs $(\Cset,\Lset)$ and $(\Cset',\Lset')$ such that \(q\in\Cset \) is reachable from \(p\in \Cset' \) it holds that \(p \) is not reachable from \(q\) (i.e., the MEC decomposition is acyclic).

%

\subsection{Asymptotic Estimates for VASS MDPs}
\label{sec-comp-VASS-runs}
Let $\A $ be a VASS MDP, $c \in \countersset(\A)$, and $t \in \transitions(\A)$. For every computation  $\pi = p_0 \bv_0,p_1 \bv_1,\ldots$, we put
\begin{eqnarray*}
	\calL_\A(\pi)      & = & \Term(\pi)\\
	\calC_\A[c](\pi) & = & \sup \{\bv_i(c) \mid 0 \leq i < \Term(\pi)\}\\
	\calT_\A[t](\pi) & = & \big|\{i\in \mathbb{N}_0\mid 0\leq i< \Term(\pi) \textit{ and } (p_i,\bv_{i{+}1}{-}\bv_i,p_{i+1}) = t \}\big|
\end{eqnarray*}

We refer to the functions $\calL_\A$, $\calC_\A[c]$, and $\calT_\A[t]$ as \emph{termination}, \emph{$c$-counter}, and \emph{$t$-transition complexity}  respectively. These are the complexity measures for VASS runs used in previous works \cite{BCKNVZ:VASS-linear-termination,Zuleger:VASS-polynomial,KLV:VASS-Grzegorczyk,Leroux:Polynomial-termination-VASS,AK:VASS-polynomial-termination,AKCONCUR23} and can be seen as variants of the standard time/space complexities for Turing machines. When the underlying VASS MDP \(\A \) is clear from the context we omit it and write only $\calL$, $\calC[c]$, and $\calT[t]$ instead of $\calL_\A$, $\calC_\A[c]$, and $\calT_\A[t]$, respectively.




\begin{definition}
	\label{def-estimates}
	Let \(\A \) be a VASS MDP,  $f : \mathbb{R} \to \mathbb{R}$, and $\F$ be one of $\calL$, $\calC[c]$ or $\calT[t]$.
	
	We say that $f$ is a \emph{lower asymptotic estimate of $\F$} if for every $\varepsilon > 0$ there exists $p \in \states(\A)$ and a strategy $\strategyDemon$ such that
	\[
	\liminf_{n \to \infty} \ \prob^{\sigma}_{p \vec{n}}[\F \geq f(n^{1-\varepsilon}) ] \ = \ 1\,
	\]
	Similarly, we say that $f$ is an \emph{upper asymptotic estimate of $\F$} if for every $\varepsilon>0$, every $p \in \states(\A)$, and every strategy $\sigma$ it holds
	\[
	\limsup_{n \to \infty} 	\ \prob^{\sigma}_{p \vec{n}}[\F \geq f(n^{1+\varepsilon}) ] \ = \ 0
	\]	
	 Finally, we say $f$ is a \emph{tight asymptotic estimate of $\F$} if it is both a lower  asymptotic estimate and an upper asymptotic estimate of $\F$. 
	
	
\end{definition}


\textit{Due to space constraints we refer to Section~\ref{sec-asest-basic-prop} for more details/intuition as well as examples regarding asymptotic estimates, including the advantages of asymptotic estimates over analyzing the asymptotic of the expectations. See also \cite{Ajdarow2025}.}


The above definition is based on the one introduced in \cite{AKCONCUR23}.
An alternative, more intuitive, definition  of asymptotic estimates can be obtained using a natural notion widely studied in survival analysis \cite{survivalone}. Given a VASS MDP (or any other probabilistic system) one can ask the following natural question:
\begin{quote}
	\emph{How many steps must be executed for a given input size $n$ so that the probability of termination is at least \(p\)?}
\end{quote}
This question makes sense for every fixed $p<1$. Formally, consider the function \(f_p^\F:\mathbb{N}\rightarrow \mathbb{N}_{\infty} \) defined for each \(n\) as the smallest integer such that for every $q \in \states(\A)$, and every strategy $\sigma$ it holds \(\Prob_{q\vec{n}}^\sigma[\F\leq  f_p^\F(n)]\geq  p \) (or \(f_p^\F(n)=\infty \) if no such integer exists). We call \(f_p^\F \) \emph{failure-time}. 


In the next theorem we show that the asymptotic behavior of failure-time is very closely tied to asymptotic estimates.

 \begin{restatable}{theorem}{observationfpf}\label{observation-f-p-f}
	Let \(f:\mathbb{R}\rightarrow\mathbb{R} \) be such that \(\lim_{n\rightarrow\infty} \frac{f( n)}{f( n^{1+\epsilon})}=0\) for every \(\epsilon>0 \). Then:
	\begin{itemize}[topsep=1.5pt,itemsep=-1.0pt,leftmargin=0.3cm]
		\item \(f\) is a lower asymptotic estimate of \(\F \) iff for all \(\epsilon>0 \) and all \(p<1 \) it holds \(f_p^\F\in \Omega\big(f(n^{1-\epsilon})\big) \);
		\item \(f\) is an upper asymptotic estimate of \(\F \) iff for all \(\epsilon>0 \) and all \(p<1 \) it holds \(f_p^\F\in \calO\big(f(n^{1+\epsilon})\big) \).
	\end{itemize}
\end{restatable}
The proof of Theorem~\ref{observation-f-p-f} is mostly technical and can be found in Section~\ref{app-observation-f-p-f}. Note that the restriction \(\lim_{n\rightarrow\infty} \frac{f( n)}{f( n^{1+\epsilon})}=0\) is not particularly restrictive, in particular it holds for any  superpolylogarithmic function that is eventually either convex or concave (such as any polynomial or exponential function).

	
\section{Polynomiality of VASS MDPs}

\begin{theorem}\label{theorem-main-polynomial-vs-exp}
	Let \(\A \) be a VASS MDP. Then for each  \(\F\in \{\calL,\calC[c],\calT[t] \} \) one of the following holds:
	\begin{itemize}[topsep=4.0pt,itemsep=0.9pt]
		\item every superpolynomial function \(f\) is an upper asymptotic estimate of \(\F \). Furthermore, if \(\A \) is DAG-like with \(w\) MECs, then  \(n^{2\cdot k} \) is an upper asymptotic estimate of \(\F\) where \(k=3^{w\cdot |\countersset(\A)|+ |\transitions(\A)|} \). Additionally, if \(\F\) is \(\calC[c] \) then \(n^k \)  is also an upper asymptotic estimate of \(\F \).
		\item \(\F\) has a lower asymptotic estimate \(2^{\sqrt{n}} \). 
	\end{itemize}
	It is decidable in polynomial time which case holds.
\end{theorem}
Note that \(\calL=\sum_{t\in \transitions(\A)} \calT[t] \) so it suffices to consider only \(\calT[t] \) and \(\calC[c]  \). The main difficulty comes in proving Theorem~\ref{theorem-main-polynomial-vs-exp} for strongly connected VASS MDPs (Theorem~\ref{theorem-main-polynomial-vs-exp-sc}). This can then be extended onto DAG-like VASS MDPs  by analyzing each MEC separately. If there exists a MEC which is subexponential then the whole VASS MDP is subexponential, whereas if every MEC is polynomial then the whole VASS MDP is polynomial (Lemma~\ref{lemma-daglike-superpolynomial-upper-bound}), intuitively this is because asymptotic estimates are compositional \cite{Ajdarow2025}. This is the same approach for lifting analysis of strongly connected VASS onto VASS \cite{leroux:polynomialvsexponential,AK:VASS-polynomial-termination} or of strongly connected VASS MDPs onto DAG-like VASS MDPs \cite{BCKNV:probVASS-linear-termination,AKCONCUR23}. We  show this approach is also sufficient to analyze \emph{general} VASS MDPs (Section~\ref{sec-non-dag}). Once again, if a single subexponential MEC exists then the whole VASS is trivially subexponential, and if every individual MEC is polynomial then we show that every superpolynomial \(f\) is an upper asymptotic estimate of \(\F \) (Lemma~\ref{lemma-nondaglike-superpolynomial-pper-bound}). Intuitively, this is because to increase the degree of the bounding polynomial above any bound the computation has to cycle over MECs, but each such cycle has positive probability of ``falling lower'' in the MEC decomposition, and at the bottom no such cycling is possible.

This also gives us the \(\PTIMEfull \) upper bound of   Theorem~\ref{theorem-main-polynomial-vs-exp}, as there are at most linearly many MECs and each MEC can be seen as a strongly connected VASS MDP which can be analyzed in polynomial time (Theorem~\ref{theorem-main-polynomial-vs-exp-sc}).

One could also ask whether the case of every superpolynomial function being an upper asymptotic estimate of \(\F \)  implies there exists \emph{some} polynomial upper asymptotic estimate of \(\F\) even in non-DAG-like VASS MDPs. In general the answer is negative as in Lemma~\ref{lemma-nondag-example} we describe an example of a VASS MDP that has no polynomial upper asymptotic estimate for \(\calL \) but every superpolynomial function is an upper asymptotic estimate of \(\calL\).


\vspace{-0.3cm}
\section{Strongly Connected VASS MDPs}
\label{sec-something-some}

\begin{theorem}\label{theorem-main-polynomial-vs-exp-sc}
	Let \(\A \) be a strongly connected VASS MDP. For each \(c\in \countersset(\A)  \) (or \(t\in \transitions(\A)\)) one of these holds:
	\begin{itemize}[topsep=4.0pt,itemsep=0.9pt]
		\item \(\calC[c] \) (or \(\calT[t] \)) has an upper asymptotic estimate   \(n^k \) (or \(n^{2\cdot k} \))  where \(k=3^{|\countersset(\A)|+ |\transitions(\A)|} \);
		\item \(\calC[c] \) (or \(\calT[t] \)) has a lower asymptotic estimate \(2^{\sqrt{n}} \). 
	\end{itemize}
	It is decidable in polynomial time which case holds.
\end{theorem}


We describe a polynomial time algorithm which computes a set of counters \(C_k \) and transitions \(T_k \) (Section~\ref{sec-something}), and then we show that  for any \(c\in C_k \) and \(t\in T_k \) the first case holds (Lemma~\ref{lemma-polynomial-upper}), and for the rest the second case holds (Lemma~\ref{lemma-njnkjbn}). The polynomial time complexity of the algorithm is shown in Lemmas~\ref{theorem-poly-or-exp-k}~and~\ref{lemma-ci-ti-ai-ptime-computable}.  Before we can describe the algorithm we need to introduce the concepts  of \emph{multi-components} and \emph{simple components} (Section~\ref{sec-multicomponents}), as well as recall a dual linear constraint system from \cite{AKCONCUR23} (Section~\ref{sec-systems-intro}).


\subsection{Multi-components}
\label{sec-multicomponents}

	Let \(\A \) be a VASS MDP. A \emph{multi-component} of \(\A \) is a vector \(\bx\in \mathbb{R}^{\transitions(\A)} \) satisfying \(\bx\geq \vec{0} \) as well as both:\vspace{-0.3cm}
	
%
%

\setlength{\columnseprule}{0.4pt}
\begin{multicols}{2}
	for all $\stateVass\in \states(\A)$
	\begin{align}\label{eq-multicomponent-flow-cond}
		\sum_{t\in \tout(\stateVass)}\bx(t)&=\sum_{t \in \tin(\stateVass)} \bx(t)
	\end{align}
	
	\columnbreak  
	and for all $\stateVass\in \states_p(\A)$, $t\in \tout(\stateVass)$
	\begin{align}
		\bx(t) & =P(t) \cdot \sum_{t'\in \tout(\stateVass)} \bx(t') \label{eq-multicomponent-prob-split}
	\end{align}
\end{multicols}

\vspace{-0.05cm}
The \emph{effect} of \(\bx \) is \(\Delta(\bx)=\sum_{t \in \transitions(\A)} \bx(t)\cdot \bu_t \).  A multi-component \(\bx\) induces a VASS MDP \(\A_\bx \) created from \(\A \) by removing all transitions \(t\) with \(\bx(t)=0 \) (and any states which afterwards have no outgoing transitions). We say \(\bx \) is \emph{strongly connected} if \(\A_\bx \) is strongly connected. \( \bx\) \emph{contains} a state \(p\) if \(p\in\states(\A_\bx) \), and \(\bx\) \emph{contains} a transition \(t\) if \(\bx(t)>0 \) . We say \(\bx\) is \emph{centered} in a state \(\stateVass\) if \(\sum_{t\in \tout(\stateVass)}\bx(t)=1 \). If \(\bx\) is centered we use \(\stateVass_\bx\) to denote the state in which \(\bx \) is centered (if multiple such states exist we fix one of them as \(\stateVass_\bx \)). We say that \(\bx\) is a \emph{simple component} if \(\bx \) is centered in some state \(\stateVass\) and \(\A_\bx \) is a MEC of \(\A_\strategyDemon \) for some \(\strategyDemon\in \stratscMD{\A}\). 
For a simple component \(\by \) we use \(\strategyDemon_\by \) to denote a cMD strategy \(\strategyDemon\) such that \(\A_\by \) is a MEC of \(\A_\strategyDemon \).  We use \(\multicompset(\A) \) and \(\simplecompset(\A) \) to denote the sets of all multi-components and simple components of \(\A \), respectively.

\begin{remark}
\textit{Intuitively, multi-components can be seen as a probabilistic analogy of multi-cycles, and the vector \(\bx \) as its Parikh image. Simple components are then the corresponding analogy of simple cycles. We provide more intuition behind multi-components and simple components in Section~\ref{sec-multi-components-intuition}}.
\end{remark}

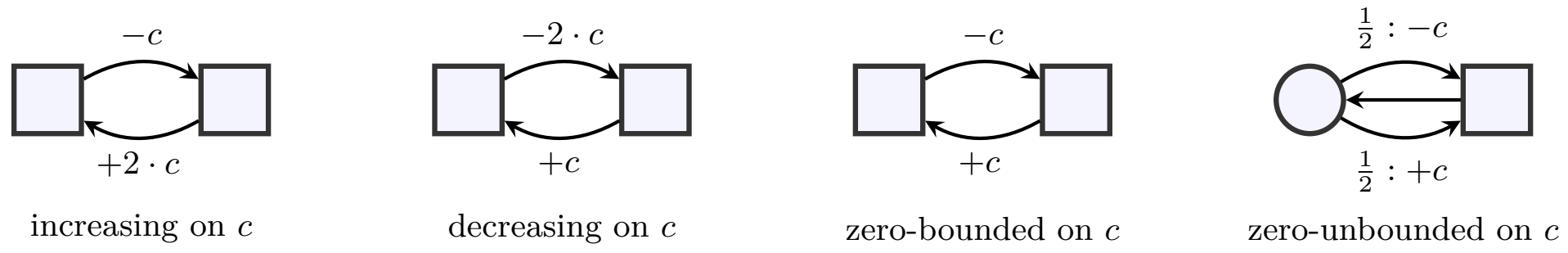
\begin{figure}[hbt]
	\centering
	\begin{tikzpicture}[scale=.9, every node/.style={scale=0.9}, x=2.1cm, y=2.1cm, font=\footnotesize]
		\node[demonic] (s1) at (0,0) {};
		\node[demonic] (s2) at (0.8,0) {};
		
		\node[] (name) at (0.4,-0.55) {increasing on \(c\)};
		
		
		\draw [tran, bend left] (s1) to node[above] {\parbox{5em}{\centering \( -c\)}} (s2);
		
		\draw [tran, bend left] (s2) to node[below] {\parbox{5em}{\centering \(+2\cdot c\)}} (s1);
		
		
		\begin{scope}[shift={(1.8,0)}]
			\node[demonic] (s1) at (0,0) {};
			\node[demonic] (s2) at (0.8,0) {};
			
			\node[] (name) at (0.4,-0.55) {decreasing on \(c\)};
			
			
			\draw [tran, bend left] (s1) to node[above] {\parbox{5em}{\centering \( -2\cdot c\)}} (s2);
			
			\draw [tran, bend left] (s2) to node[below] {\parbox{5em}{\centering \(+ c\)}} (s1);

		\end{scope}

		\begin{scope}[shift={(3.6,0)}]
			\node[demonic] (s1) at (0,0) {};
			\node[demonic] (s2) at (0.8,0) {};
			
			\node[] (name) at (0.4,-0.55) {zero-bounded on \(c\)};
			
			
			\draw [tran, bend left] (s1) to node[above] {\parbox{5em}{\centering \( -c\)}} (s2);
			
			\draw [tran, bend left] (s2) to node[below] {\parbox{5em}{\centering \(+ c\)}} (s1);
		\end{scope}
		
		\begin{scope}[shift={(5.4,0)}]
			\node[random] (s1) at (0,0) {};
			\node[demonic] (s2) at (0.8,0) {};
			
			\node[] (name) at (0.4,-0.55) {zero-unbounded on \(c\)};
			
			\draw [tran,] (s2) to (s1);
			
			\draw [tran, bend left] (s1) to node[above] {\parbox{5em}{\centering \(\frac{1}{2}: -c\)}} (s2);
			
			\draw [tran, bend right] (s1) to node[below] {\parbox{5em}{\centering \(\frac{1}{2}:+ c\)}} (s2);
		\end{scope}
	\end{tikzpicture}
	\caption{Examples of the different types of simple components. In the first three examples the simple component is \(\by=(1,1)\). In the last example there are two simple components \(\by_1=(\frac{1}{2},\frac{1}{2},1) \) and \(\by_2=(1,1,2) \) where the third transition is the one from the non-deterministic state. }
	\vspace{-0.5cm}
	\label{fig-simple-components-types-examples}
\end{figure}




We use \(\realeffect_\by \) to denote a random variable representing the effect of a computation in \(\A_\by \) started from \(p_\by \) until the first time \(p_\by\) is revisited. We call such computation a single \emph{iteration} of \(\by\). We use \(\support(\by)=\{\bv\mid \prob[\realeffect_\by=\bv]\neq 0 \} \) to denote the support of \(\realeffect_\by \). Given a counter \(c \) we say that a simple component \(\by \) of \(\A \)  is

\columnratio{0.391}
\begin{paracol}{2}
	\begin{itemize}[topsep=1.5pt,itemsep=-1.0pt,leftmargin=20pt]
		\item \emph{increasing} on \(c\) if $\Delta(\by)(c)>0$.
		\item \emph{decreasing} on \(c\) if $\Delta(\by)(c)<0$.
	\end{itemize}
	\switchcolumn
	\begin{itemize}[topsep=1.5pt,itemsep=-1.0pt]
		\item \emph{zero-bounded} on \(c\) if $\Delta(\by)(c)=0$ and $\prob(\realeffect_\by=0)=1$.
		\item \emph{zero-unbounded} on \(c\) if $\Delta(\by)(c)=0$ and $\prob(\realeffect_\by=0)<1$.
	\end{itemize}
\end{paracol}


\subsection{Constraint Systems \hyperref[fig-systems]{(I)} and \hyperref[fig-systems]{(II)}}
\label{sec-systems-intro}
An important tool for VASS asymptotic  analysis is the dual  constraint system from \cite{AKCONCUR23}. Due to space constraints the formal definitions are deferred to Section~\ref{sec-systems}~(Fig.~\ref{fig-systems}) and here we only provide the intuition. Note that we can find a maximal solution \(\bx,\ba,\bb \) of \hyperref[fig-systems]{(I)} and \hyperref[fig-systems]{(II)} in polynomial time.
 For clarity, let us start with simplified variants obtained for VASS. 

For VASS, a solution \(\bx\in \mathbb{Q}^{\transitions(\A)} \) of~\hyperref[fig-systems]{(I)} can be seen as a Parikh image \(\bx \) of a multi-cycle such that the effect $\Delta(\bx)$ of this multi-cycle  is non-negative on every counter. The objective of~\hyperref[fig-systems]{(I)} ensures  this multi-cycle contains as many transitions as possible, and \(\Delta(\bx) \) is positive on as many counters as possible. In VASS MDPs  \(\bx \) is now a multi-component. The objective of~\hyperref[fig-systems]{(I)} then maximizes the number of transitions contained in \(\bx \), and the number of counters \(\Delta(\bx) \) is positive on.

A solution \(\ba\in \mathbb{Q}^{\counters(\A)},\bb\in\mathbb{Q}^{\states(\A)} \) of~\hyperref[fig-systems]{(II)} for VASS can be seen as a ranking function \(\mathit{rank}:\states(\A)\times \mathbb{Q}^d \rightarrow \mathbb{Q} \) on configurations defined by $\mathit{rank}(\stateVass\bv)=\bb(\stateVass)+\sum_{c\in\countersset(\A)} \ba(c)\cdot \bv(c)$, such that the effect of every transition on \(\mathit{rank} \) is non-positive. The objective of~\hyperref[fig-systems]{(II)} ensures that as many transitions as possible decrease $\mathit{rank}$, and $\mathit{rank}$  depends on as many counters as possible (i.e., \(\ba(c)>0 \) for as many  \(c\in\countersset(\A)\) as possible). For VASS MDPs, this interpretation changes only for the outgoing transitions $t$ of probabilistic states \(\stateVass\in \statesProb(\A)\). Instead of considering the change of $\mathit{rank}$ caused by such $t\in \tout(\stateVass)$,  we now consider the expected change of $\mathit{rank}$ caused by executing a single step from $\stateVass$. The objective ensures that $\mathit{rank}$ depends on as many counters as possible, the value of $\mathit{rank}$ is decreased by as many outgoing transitions of non-deterministic states as possible, the expected change of $\mathit{rank}$ caused by performing a single step is strictly negative in as many probabilistic states as possible, and that as many probabilistic states as possible have at least one outgoing transition which strictly decreases \(rank\). 


Note that for a given ranking function $\mathit{rank}(\stateVass\bv)=\bb(\stateVass)+\sum_{c\in\countersset(\A)} \ba(c)\cdot \bv(c)$  we can for each transition \((p,\bu,q)\in\transitions(\A) \) declare \(\bu(\mathit{rank})=\bb(q)-\bb(p) +\sum_{c\in\countersset(\A)} \ba(c)\cdot \bu(c) \). Hence we can effectively treat \(\mathit{rank} \). We will also use \(r\) to denote \(\mathit{rank} \).

An important property is the duality of \hyperref[fig-systems]{(I)} and \hyperref[fig-systems]{(II)} which states for maximal solutions \(\bx,\ba,\bb \), that for each \(c\in \countersset(\A)\) either \(\Delta(\bx)(c)<0 \) or \(\ba(c)>0 \). Likewise for each \(t\in \transitions(\A) \) either \(\bx(t)>0 \) or any step which has  positive probability of taking \(t\) strictly decreases \(rank \) on average. See Section~\ref{sec-systems} for more details.

\subsection{Pseudo-Algorithm}
\label{sec-something}

Let \(\A_0=\A \) be a strongly connected VASS MDP. We define  \(T_0=\emptyset \) and \(C_0=\emptyset \).  For each \(i\geq1 \) we define \(\A_i,T_i,\) and \(C_i \) inductively as follows:

Let \(\bx_{i-1} \), \(\ba_{i-1},\bb_{i-1} \) be some maximal solution of  \hyperref[fig-systems]{(I)} and \hyperref[fig-systems]{(II)} for \(\A_{i-1} \), and \(r_{i-1} \) be the ranking function defined by \(\ba_{i-1},\bb_{i-1}\).
 Then we put 
\begin{itemize}[topsep=1.9pt,itemsep=-0.5pt]
	\item \(C_i=C_{i-1}\cup \{c\in \countersset(\A)\mid \Delta(\bx_{i-1})(c)=0 \} \)
	
	\item \(T_i=T_{i-1}\cup M_{i-1}  \) where \(M_{i-1}\) is the set of all transitions \(t\in \transitions(\A_{i-1})\) such that every simple component \(\by\) of \(\A_{i-1}\) with \(\by(t)>0\) contains a transition \(t'\) with \(\bu_{t'}(r_{i-1})< 0 \).
\end{itemize}

Let \(\A'_{i} \) be the VASS MDP \(\A \) restricted only to transitions  \(\transitions(\A_{i}')=\transitions(\A)\setminus T_i \) and let \(\MEC_1, \dots,\MEC_w \) be the MECs of \(\A'_{i} \). 
Then \(\A_i \) is the VASS MDP created from \(\A'_{i} \) by changing  the counter set to 
\[\countersset(\A_i)=\big(\countersset(\A)\setminus C_i\big)\cup \{(c,\MEC_j)\mid c\in C_i, 1\leq j\leq w \} \] where the effect of \(t\in \transitions(\A_i)=\transitions(\A)\setminus T_i \) on the counter \((c,\MEC) \) is defined as
 \[\bu_t\big((c,\MEC)\big)=\begin{cases}
	\bu_t(c) & \textit{ if } t \textit{ is a transition of } \MEC \\
	0 & \textit{ else}
\end{cases} \]

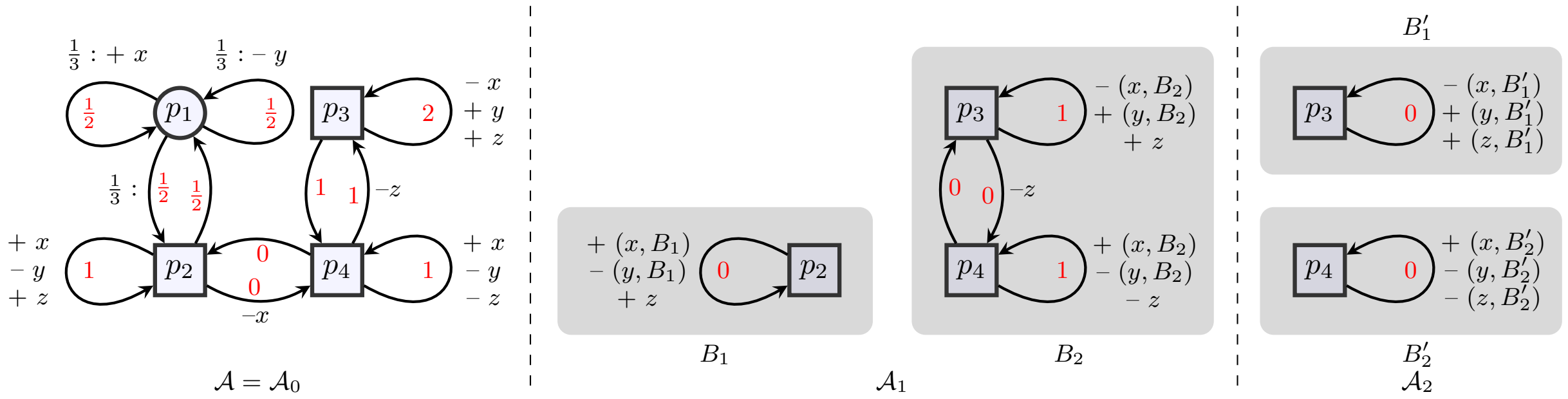
\begin{figure}[th]
	
	\makebox[\linewidth][c]{
		\begin{tikzpicture}[scale=.8, every node/.style={scale=0.85}, x=2.1cm, y=2.1cm, font=\footnotesize]
			
			\begin{scope}[shift={(0,0)}]
				
				\draw[dashed] (4.5cm,1.375cm) to (4.5cm,-3.1cm-1.42em);
				
				\node[] (X) at (1cm,-3.1cm-1em) {$\A=\A_0$};
				\node[random] (t1) at (0,0) {$\stateVass_1$};
				\node[stoch] (t2) [below of=t1,node distance=1.9cm] {$\stateVass_2$};
				\node[stoch] (t3) [right of=t1,node distance=1.9cm] {$\stateVass_3$};
				\node[stoch] (t4) [below of=t3,node distance=1.9cm] {$\stateVass_4$};

				\node[] () [left of=t2,node distance=1.1cm] {\color{red}\(1\)};
				
				\node[] () [left of=t1,node distance=1.1cm] {\color{red}\(\frac{1}{2}\)};
				\node[] () [right of=t1,node distance=1.1cm] {\color{red}\(\frac{1}{2}\)};
				
				\node[] () [right of=t3,node distance=1.1cm] {\color{red}\(2\)};
				
				\node[] () [right of=t4,node distance=1.1cm] {\color{red}\(1\)};
				
				\node[] () [below of=t1,node distance=0.9cm, xshift=-0.2cm] {\color{red}\(\frac{1}{2}\)};
					\node[] () [above of=t2,node distance=0.9cm, xshift=0.2cm] {\color{red}\(\frac{1}{2}\)};

				\node[] () [below of=t3,node distance=0.9cm, xshift=-0.2cm] {\color{red}\(1\)};
				\node[] () [above of=t4,node distance=0.9cm, xshift=0.2cm] {\color{red}\(1\)};
				
				\node[] () [right of=t2,node distance=0.9cm, yshift=-0.2cm] {\color{red}\(0\)};
				\node[] () [left of=t4,node distance=0.9cm, yshift=0.2cm] {\color{red}\(0\)};

				\node[] () at (-0.94cm,0.75cm) {{\parbox{5em}{\centering $\frac{1}{3}:$ + $x$}}};
				\draw [tran,loop left] (t1) to (t1);

				\node[] () at (0.9cm,0.75cm) {{\parbox{5em}{\centering $\frac{1}{3}:$ -- $y$}}};
				\draw [tran,loop right] (t1) to (t1);
				
				\draw [tran,loop left] (t2) to node[left] {{\parbox{2em}{\centering + $x$\\-- $y$\\+ $z$}}} (t2);

				\draw [tran,loop right] (t3) to node[right] {{\parbox{2em}{\centering -- $x$\\+ $y$\\+ $z$}}} (t3);
				
				\draw [tran,loop right] (t4) to node[right] {{\parbox{2em}{\centering + $x$\\-- $y$\\-- $z$}}} (t4);

				\draw [tran,bend right] (t1) to node[ left] {\parbox{1.48em}{\centering \(\frac{1}{3}:\)}} (t2);
				\draw [tran, bend right] (t2) to  (t1);
				
				\draw [tran, bend right] (t3) to  (t4);
				\draw [tran, bend right] (t4) to node[ right] {\parbox{1em}{\centering --\(z\)}} (t3);
				
				\draw [tran, bend right] (t2) to node[below] {\parbox{3em}{\centering --\(x\)}} (t4);
				
				\draw [tran, bend right] (t4) to node[above] {\parbox{3em}{\centering }} (t2);
			\end{scope}
			
			\begin{scope}[shift={(3.88,0)}]
				
				\fill [rounded corners,gray!30] (-3.3cm,-1.209cm) rectangle (0.75cm,-2.85cm);
				
				\node[] (XX) at (-1.275cm,-3.1cm) {$\MEC_1$};
				
				\fill [rounded corners,gray!30] (1.25cm,0.85cm) rectangle (5.13cm,-2.85cm);
				
				\node[] (XX) at (3.275cm,-3.1cm) {$\MEC_2$};

				\node[] (X) at (1cm,-3.1cm-1em) {$\A_1$};
				\node[] (t1) at (0,0) {};
				\node[stoch] (t2) [below of=t1,node distance=1.9cm] {$\stateVass_2$};
				\node[stoch] (t3) [right of=t1,node distance=1.9cm] {$\stateVass_3$};
				\node[stoch] (t4) [below of=t3,node distance=1.9cm] {$\stateVass_4$};

				\node[] () [left of=t2,node distance=1.1cm] {\color{red}\(0\)};
				
				\node[] () [right of=t3,node distance=1.1cm] {\color{red}\(1\)};
				
				\node[] () [right of=t4,node distance=1.1cm] {\color{red}\(1\)};

				\node[] () [below of=t3,node distance=0.9cm, xshift=-0.2cm] {\color{red}\(0\)};
				\node[] () [above of=t4,node distance=0.9cm, xshift=0.2cm] {\color{red}\(0\)};

				\draw [tran,loop left] (t2) to  node[left] {{\parbox{4em}{\centering + $(x, \MEC_1)$\\-- $(y, \MEC_1)$\\+ $z$}}}(t2);
				
				\draw [tran,loop right] (t3) to  node[right] {{\parbox{4em}{\centering -- $(x, \MEC_2)$\\+ $(y, \MEC_2)$\\+ $z$}}}(t3);
				
				\draw [tran,loop right] (t4) to node[right] {{\parbox{4em}{\centering + $(x, \MEC_2)$\\-- $(y, \MEC_2)$\\-- $z$}}} (t4);
				
				\draw [tran, bend left] (t3) to node[right] {\parbox{1em}{\centering --\(z\)}} (t4);
				\draw [tran, bend left] (t4) to (t3);

			\end{scope}
			
			\begin{scope}[shift={(6.01,0)}]

				\draw[dashed] (0.961cm,1.375cm) to (0.961cm,-3.1cm-1.42em);
				
				\fill [rounded corners,gray!30] (1.25cm,0.85cm) rectangle (5.13cm,-0.791cm);
				
				\node[] (XX) at (3.275cm,1.1cm) {$\MEC_1'$};
				
				\fill [rounded corners,gray!30] (1.25cm,-1.209cm) rectangle (5.13cm,-2.85cm);
				
				\node[] (XX) at (3.275cm,-3.1cm) {$\MEC_2'$};

				\node[] (X) at (3.275cm,-3.1cm-1em) {$\A_2$};
				\node[] (t1) at (0,0) {};
				\node[] (t2) [below of=t1,node distance=1.9cm] {$$};
				\node[stoch] (t3) [right of=t1,node distance=1.9cm] {$\stateVass_3$};
				\node[stoch] (t4) [below of=t3,node distance=1.9cm] {$\stateVass_4$};

				\node[] () [right of=t3,node distance=1.1cm] {\color{red}\(0\)};
				
				\node[] () [right of=t4,node distance=1.1cm] {\color{red}\(0\)};

				\draw [tran,loop right] (t3) to  node[right] {{\parbox{4em}{\centering -- $(x, \MEC_1')$\\+ $(y, \MEC_1')$\\+ $(z,\MEC_1')$}}}(t3);
				
				\draw [tran,loop right] (t4) to  node[right] {{\parbox{4em}{\centering + $(x, \MEC_2')$\\-- $(y, \MEC_2')$\\-- $(z,\MEC_2')$}}}(t4);
			\end{scope}
			
		\end{tikzpicture}
	}
	\caption{VASS MDPs \(\A_0,\A_1\), and  \(\A_2 \). The gray areas denote relevant MECs. The red numbers denote the value of a possible \(\bx(t) \), i.e., \(\bx_0\big((p_2,\vec{0},p_1)\big)=\frac{1}{2} \).}
	\label{fig:vass:example:algorithm:execution}
\end{figure}

\begin{example}
	Consider the example of the VASS MDPs \(\A_0,\A_1\), and  \(\A_2 \) from Fig.~\ref{fig:vass:example:algorithm:execution}.  Notice that in \(\A_0 \) the sum \(x+y \) is non-increasing in expectation. Hence it holds  \(\Delta(\bx_0)(x)= \Delta(\bx_0)(y)= 0 \). The self-loops over \(p_2\) and \(p_3\) both  increase \(z\) while canceling each other on \(x,y\); hence \(\Delta(\bx_0)(z)>0 \). This gives \(C_1=\{x,y \} \) as well as \(\bx_0\big((p_4,\vec{0},p_2)\big)=\bx_0\big((p_2,-x,p_4)\big)=0 \).   While the transitions \(\tout(p_1)\cup \tin(p_1) \) do satisfy \(\bx_0\big(\tout(p_1)\cup \tin(p_1)\big)>\vec{0} \) the self-loop \((p_1,-y,p_1) \) strictly decreases \(r_0=x+y \). Since each simple component \(\by\) containing \(p_1\) has \(\by\big(\tout(p_1)\cup \tin(p_1)\big)>\vec{0} \) it follows that \(\tout(p_1)\cup \tin(p_1)\subseteq T_1 \).

	Similarly in \(\A_1\), as no transition increases \((y,\MEC_1) \), the self loop over \(p_2\) is not included in \(\bx_1 \). The remaining transitions cannot increase \(z\) without simultaneously decreasing another counter; hence \(C_2=C_1\cup\{z \} \). The two self-loops over \(p_3 \) and \(p_4\) negate each other on all five counters, thus they are contained in \(\bx_1 \). Note that \(\A_3=\A_4=\dots \) are all empty.
\end{example}

We claim the above pseudo-algorithm is sufficient to analyze the polynomiality of strongly connected VASS MDPs in polynomial time (Lemma~\ref{lemma-ci-ti-ai-ptime-computable}). More precisely we have the following.

%
\begin{lemma}\label{theorem-poly-or-exp-k}
	Let \(\A \) be a strongly connected VASS MDP,  \(k \) be the smallest index such that \(\A_{k}=\A_{k+1} \), and \(0\leq \variablenotk \leq  k\).  For each \(c\in \countersset(\A)  \) (or \(t\in \transitions(\A)\)) we have the following:
	\begin{itemize}
		\item if \(c\in C_{\variablenotk} \) (or \(t\in T_\variablenotk \)) then \(\calC[c]\) (or \(\calT[t]\)) has an upper asymptotic estimate \(n^{3^{\variablenotk-1}} \) (or \(n^{2\cdot 3^{\variablenotk-1}} \));
		\item if \(c\notin C_k \) (or \(t\notin T_k \)) then \(\calC[c]\) (or \(\calT[t]\))  has a lower asymptotic estimate \(2^{\sqrt{n}} \). 
	\end{itemize}
	Furthermore, it holds \(k\leq|\countersset(\A)|+ |\transitions(\A)|\).
\end{lemma}

We prove the lower asymptotic estimates \(2^{\sqrt{n}} \) for the case when \(c\notin C_k \) (or \(t\notin T_k \)) in  Section~\ref{sec-exponential-lower} (Lemma~\ref{lemma-njnkjbn}). The  \(k\leq|\countersset(\A)|+ |\transitions(\A)|\) then follows from Lemma~\ref{lemma-k-converged} below.  We begin with the polynomial upper asymptotic estimates for the case when \(c\in C_{\variablenotk} \) (or \(t\in T_{\variablenotk}\)). 

\begin{restatable}{lemma}{lemmapolynomialupper}\label{lemma-polynomial-upper}
	For each \(\variablenotk\in \mathbb{N}_0 \) the following holds:
	\begin{itemize}
		\item If \(c\in C_{\variablenotk+1} \) (or \(t\in T_{\variablenotk+1} \))  then \(\calC[c]\) (or \(\calT[t] \)) has an upper asymptotic estimate \(n^{3^{\variablenotk}} \) (or \(n^{2\cdot 3^{\variablenotk}} \));
		%
	\end{itemize}
	
\end{restatable}

We prove this using induction over \(\variablenotk\). For each \(\variablenotk\) we show that every simple component containing a transition from \(T_i \) is either decreasing or zero-unbounded on \(r_{i-1} \). Hence all the transitions that can increase \(r_i\) on average are from \(T_{i-1} \) and we already have an upper bound for these from the induction step. This allows us to construct a supermartingale whose initial value is upper bounded by a polynomial and that either strictly decreases or behaves in ``zero-unbounded'' manner whenever a transition from \(T_i\) is taken. Simultaneously, if the supermartingale becomes negative then some counter must be negative as well. Thus we obtain the upper asymptotic estimates for \(\variablenotk\) from the results about \(1\)-dim VASS MDPs \cite{AKCONCUR23} by treating this supermartingale as a counter. A full proof of Lemma~\ref{lemma-polynomial-upper} can be found in Section~\ref{sec-lemma-polynomial-upper-proof}.\qed

\begin{lemma}\label{lemma-k-converged}
	Let \(h=|\countersset(\A)|+ |\transitions(\A)| \). It holds \(\A_{h}=\A_{h+1}=\dots \).
\end{lemma}
\begin{proof}
	Observe that  \(\A_i=\A_{i+1} \) implies  \(\A_i=\A_{i+1}=\A_{i+2}=\dots \) since each \(\A_j \) depends only on \(\A_{j-1} \). Furthermore \(\A_i\neq\A_{i+1} \) implies either \(C_i\subset C_{i+1} \) or \(T_i\subset T_{i+1} \). As \(C_{i}\subseteq \countersset(\A)\) and \(T_{i}\subseteq \transitions(\A) \) this can only hold for at most \(|\countersset(\A)|+|\transitions(\A)| \) many indices \(i\).
\end{proof}

Hence Theorem~\ref{theorem-poly-or-exp-k} implies that analyzing the  polynomiality of strongly connected VASS MDPs can be done polynomial time as per the following lemma.
\begin{restatable}{lemma}{lemmacitiaiptimecomputable}\label{lemma-ci-ti-ai-ptime-computable}
	Given \(C_i,T_i \) and \(\A_i \) we can compute \(C_{i+1},T_{i+1} \) and \(\A_{i+1} \) in time polynomial in \(||\A_0|| \). 
\end{restatable}
\(\bx_i,\ba_i \), and \(\bb_i \) are computable  in \(\PTIMEfull\).  Since \(r_i\) is non-increasing on average, \(M_i \) contains exactly those transitions that are not contained in some simple component that contains only transitions with \(\bu_t(r_i)=0 \). Hence it suffices to remove all transitions with \(\bu_t(r_i)\neq 0 \) and then keep iteratively removing any transition that does not belong to any MEC and put those removed into \(M_{i}\). A full proof of Lemma~\ref{lemma-ci-ti-ai-ptime-computable} can be found in Section~\ref{sec-lemma-ci-ti-ai-ptime-computable-proof}.\qed


%


	\newcommand{\eventloops}{E_{loops}}
\newcommand{\eventshortsucc}{E_{ShortSucc}}
\newcommand{\eventfinit}{E_{\fsize(3)=4}}
\newcommand{\eventfexp}{E_{\mathrm{Exp\fsize}}}
\newcommand{\eventdoublereset}{E_{ShortReset}}
\newcommand{\eventbigcounter}{E_{bigC}}
\section{Lower Asymptotic Estimate \(2^{\sqrt{n}}\)}
\label{sec-exponential-lower}
In this section we prove the following lemma regarding the lower asymptotic estimates for strongly connected VASS MDPs of Lemma~\ref{theorem-poly-or-exp-k}. 
\begin{lemma}\label{lemma-njnkjbn}
	If \(c\notin C_k \) (or \(t\notin T_k \)) then \(\calC[c]\) (or \(\calT[t] \)) has a lower asymptotic estimate \(2^{\sqrt{n}} \).
\end{lemma}
We prove this by constructing a strategy  \(\strategyDemon \), parameterized by \(0<\epsilon<\frac{1}{2} \), which for all  \(p\in\states(\A)\) satisfies \[\liminf_{n\rightarrow\infty}  \prob_{p\vec{n}}^\strategyDemon[\calC[c]\geq 2^{n^{\frac{1}{2}-\epsilon}} \textit{ and } \calT[t]\geq 2^{n^{\frac{1}{2}-\epsilon}} \textit{ for all } c\in \countersset(\A)\setminus C_k \textit{ and all } t\in\transitions(\A)\setminus T_k ] =1 \] 

\textbf{Section organization:} We describe the strategy \(\strategyDemon \) in Section~\ref{sec-strat-description}, and we prove it satisfies the claimed property in Section~\ref{sec-sigma-analysis} (Lemma~\ref{lemma-strategy-witnesses}). In order to be able to describe \(\strategyDemon \) we first introduce a novel technique for splitting any computation on VASS MDPs into deterministic and probabilistic parallel parts (Section~\ref{sec-det-prob-decompositions}), as well as a novel concept of iterating strongly connected multi-components (Section~\ref{sec-multi-comp-iterating}).

\textbf{Intuitively speaking} we show how to decompose any computation on a VASS MDP into two parallel parts, a deterministic part which behaves almost fully non-probabilistically (and whose effect is the expected effect of the computation), and a probabilistic part which can be fully captured using a constant number (that depends only on \(\A\)) of martingales. Then we use the same idea that was used in \cite{Zuleger:VASS-polynomial} to show a result analogous to Lemma~\ref{lemma-njnkjbn} in strongly connected VASS to iterate the deterministic part exponentially many times, and we only have to compensate for the probabilistic part of the computation. For counters \(c\notin C_k \) we show the increase in the deterministic part is fast enough to compensate for the potential decrease in the probabilistic part. This leaves having to compensate for the polynomial counters \(c\in C_k \), which can potentially die in as little as quadratic time in the probabilistic part (and remain unchanged in the deterministic part). These we compensate using ``resets'', i.e., whenever the probabilistic part is close to dying on any of these counters we schedule a computation which effectively ``resets'' these counters  back to their initial value in the probabilistic part. We show we can keep ``resetting'' the probabilistic part in this way for at least subexponentialy many steps with high enough probability, hence we can still reach subexponentiality in the deterministic part.

This is possible because, as we show, the cone generated by the support of any single iteration of a simple component in \(\A_k\) on the probabilistic part, restricted to counters from \(C_k \), is actually a vector space, and this vector space is a sub-vector space of \(\Delta\big(X(\A_k)\big)(C_k) \) (technically, here \(C_k\) are the ``local-copies'' of \(C_k\) from \(\A_k \) instead of the ones from \(\A \), i.e., the counters \((c,\MEC) \)).\footnote{In fact one can  show that any simple component of \(\A \) which does not have this property contains a transition \(t\in T_k\).} Hence whenever the probabilistic part changes too much on the counters from \(C_k \) we can schedule a computation which for these counters, in the deterministic part, exactly ``negates'' the accumulated effect in the probabilistic part. As each ``reset'' gives us a roughly at least \(n^2 \) steps before we need to ``reset'' again, and the ``reset'' itself is of length \(n\) this allows us to effectively compensate for the random effects in the probabilistic part on \(C_k \) for at least \(2^{n^{\frac{1}{2}-\epsilon}}\) steps with high enough probability without affecting the counters \(c\notin C_k \) too much. 

Note that while the technique from \cite{Zuleger:VASS-polynomial} allows to iterate all the way up to \(2^{n}\), the probabilistic part  limits us to be able to iterate only  up to \(2^{n^{\frac{1}{2}-\epsilon}}\) with high enough probability. This is because there is always an exponentially small probability of dying on the probabilistic part before we are able to schedule the next ``reset''.\footnote{In fact it can be shown that if the probabilistic part of every simple component of \(\A_k \) is zero-bounded on \(C_k \) then we can replace \(2^{n^{\frac{1}{2}-\epsilon}}\) with \(2^{n^{1-\epsilon}}\) in Lemma~\ref{lemma-njnkjbn}.} It remains open whether this is a fundamental property of VASS MDPs, or whether this is just a deficiency of our proof.

\subsection{Deterministic/Probabilistic Decomposition}\label{sec-det-prob-decompositions}
In this subsection we describe a novel technique for decomposing a computation in a VASS MDP into two concurrent parts, a \emph{deterministic} part which behaves almost non-probabilistically, and a  \emph{probabilistic} part. This decomposition is at the core of obtaining the subexponential lower asymptotic estimate of Lemma~\ref{lemma-njnkjbn}.

The starting point of this decomposition is the pointing decomposition recently developed in \cite{AKCONCUR23} which allows us to decompose a computation in a VASS MDP into an interweaving of a constant number (in \(n\)) of independent computations on VASS Markov chains. That is, as shown in \cite{AKCONCUR23}, we can view a strategy on \(\A \) as if, instead of choosing the next transition in non-deterministic states, it instead ``points'' (with some restrictions) at one of constantly many independent computations on VASS Markov chains, which then takes the next step from its own current state. The effect on the counters is then the effect of the transition taken in this VASS Markov chain.  Since these Markov chains quickly fall into MECs, and each of these MECs corresponds to a simple component of \(\A \), this effectively allows us to view each step of a computation as if pointing to some simple component \(\by \), except for up to a few steps which do not point at any simple component but which are asymptotically irrelevant (the expected number of such steps is in \(\bigO(1) \)). We formalize this concept of which simple component is being pointed at a given step in Section~\ref{sec-pointing}.



 Hence it suffices to do the deterministic/probabilistic decomposition only for iterations of individual simple components. Remember that \(\realeffect_\by \) denotes the effect of a single iteration of a simple component \(\by \). We split \(\realeffect_\by \) into two parts, a deterministic part whose effect is \(\E[\realeffect_\by] \) and a probabilistic part whose effect is \(\realeffect_\by-\E[\realeffect_\by] \).

Formally, given a simple component \(\by \) of \(\A \), we denote by \(\A_{co-\hat{\by}} \) a VASS Markov chain created from \(\A_\by \) by changing every transition \(t=(p,\bu,q) \) of \(\A_\by \) to \( t_{co-\hat{\by}}=(p,\vec{0},q)\) if \(q\neq p_\by \) and to \(t_{co-\hat{\by}}=(p,\E[\realeffect_\by],q) \) if \(q=p_\by \). Similarly, let \(\A_{\hat{\by}} \) be a VASS Markov chain created from \(\A_\by \) by changing every transition \(t=(p,\bu,q) \) of \(\A_\by \) with \(q=p_\by \) to \(t_{\hat{\by}}= (p,\bu-\E[\realeffect_\by],q)\) (and otherwise to \(t_{\hat{\by}}=t \)). \footnote{Note that the VASS MDPs \(\A_{\hat{\by}}\) and  \(\A_{co-\hat{\by}} \) can have non-integer transition effects on counters. All the notions for VASS MDPs extend onto VASS MDPs with real valued counters in the obvious way.}

We then define \(\hat{\by} \) and \(co-\hat{\by} \) to be the simple components of \(\A_{\hat{\by}} \) and \(\A_{co-\hat{\by}} \), respectively, that  for each \(t\in \A_\by\) satisfy \(\by(t)=\hat{\by}(t_{\hat{\by}})=co-\hat{\by}(t_{co-\hat{\by}}) \). To simplify the notation, in the rest of this paper we write \(\hat{\by}(t)\) and \(co-\hat{\by}(t) \) instead of \(\hat{\by}(t_{\hat{\by}})\) and \(co-\hat{\by}(t_{co-\hat{\by}})\), respectively.




We call \(co-\hat{\by} \) the \emph{deterministic part} and \( \hat{\by}\) the \emph{probabilistic part} of \(\by\). Notice that for each counter \(c\), \( \hat{\by}\) is either zero-bounded or zero-unbounded on \(c\), whereas \(co-\hat{\by} \) is never zero-unbounded on \(c\).

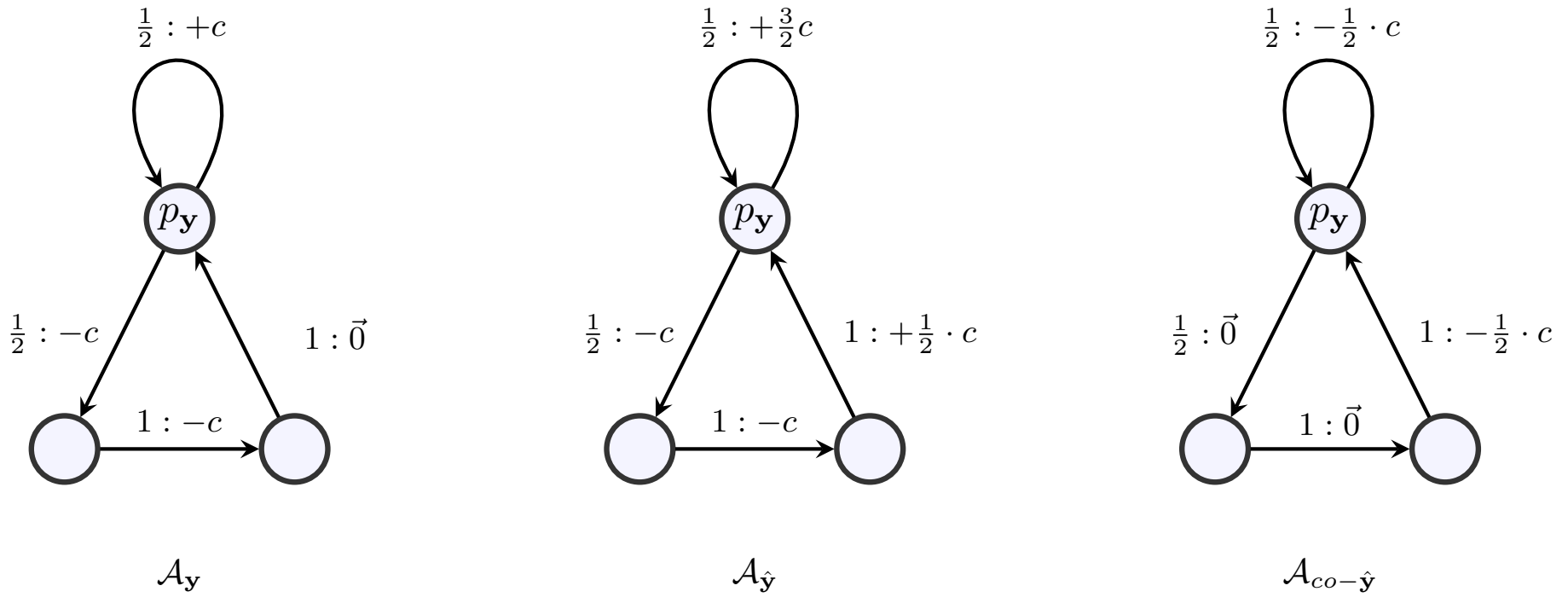
\begin{figure}[h]
	\centering
	\begin{tikzpicture}[scale=.85, every node/.style={scale=0.85}, x=2.1cm, y=2.1cm, font=\footnotesize]
		\node[random] (s1) at (0,0) {};
		\node[random] (s2) at (1,0) {};
		\node[random] (s3) at (0.5,1) {\(p_\by \)};
		
		\node[] (name) at (0.5,-0.55) {\(\A_\by\)};
		
		\draw [tran, loop above] (s3) to node[above] {\parbox{3.3em}{\centering \(\frac{1}{2}: + c\)}} (s3);
		
		\draw [tran] (s3) to node[above,left] {\parbox{3.3em}{\centering \(\frac{1}{2}: -c\)}} (s1);
		
		\draw [tran] (s1) to node[above] {\parbox{5em}{\centering \(1: -c\)}} (s2);
		
		\draw [tran] (s2) to node[above,right] {\parbox{5em}{\centering \(1:\vec{0}\)}} (s3);
		
		
		\begin{scope}[shift={(2.5,0)}]
			\node[random] (s1) at (0,0) {};
			\node[random] (s2) at (1,0) {};
			\node[random] (s3) at (0.5,1) {\(p_\by \)};
			
			\node[] (name) at (0.5,-0.55) {\(\A_{\hat\by}\)};
			
			\draw [tran, loop above] (s3) to node[above] {\parbox{5em}{\centering \(\frac{1}{2}: +\frac{3}{2} c\)}} (s3);
			
			\draw [tran] (s3) to node[above,left] {\parbox{3.3em}{\centering \(\frac{1}{2}: -c\)}} (s1);
			
			\draw [tran] (s1) to node[above] {\parbox{5em}{\centering \(1: -c\)}} (s2);
			
			\draw [tran] (s2) to node[above,right] {\parbox{5em}{\centering \(1:+\frac{1}{2}\cdot c\)}} (s3);
			
		\end{scope}

		\begin{scope}[shift={(5,0)}]
			\node[random] (s1) at (0,0) {};
			\node[random] (s2) at (1,0) {};
			\node[random] (s3) at (0.5,1) {\(p_\by \)};
			
			\node[] (name) at (0.5,-0.55) {\(\A_{co-\hat\by}\)};
			
			\draw [tran, loop above] (s3) to node[above] {\parbox{5em}{\centering \(\frac{1}{2}: -\frac{1}{2}\cdot c\)}} (s3);
			
			\draw [tran] (s3) to node[above,left] {\parbox{3.3em}{\centering \(\frac{1}{2}: \vec{0}\)}} (s1);
			
			\draw [tran] (s1) to node[above] {\parbox{5em}{\centering \(1: \vec{0}\)}} (s2);
			
			\draw [tran] (s2) to node[above,right] {\parbox{5em}{\centering \(1:-\frac{1}{2}\cdot c\)}} (s3);
		\end{scope}
	\end{tikzpicture}
	\caption{An example of  \(\hat{\by} \) and \(co-\hat{\by}\) for a given simple component \(\by \). Note that \(\Exp[\realeffect_\by](c)=-\frac{1}{2} \). Notice  that for each transition the sum of the effects in \(\A_{\hat{\by}} \) and \(\A_{co-\hat{\by}}\) is precisely the effect in \(\A_\by \). }
	\label{fig-simple-components-hat-examples}
\end{figure}

For a given computational step in \(\A \), let \(\by \) be the simple component this step pointed at and \(t\) the transition taken. Then we say the effect on the deterministic part of this step was the effect of \(t_{co-\hat{\by}}\) and the effect on the probabilistic part was the effect of \(t_{\hat{\by}} \). Steps which do not point to any simple component are treated as if both deterministic and probabilistic parts had effect \(\vec{0} \). The following lemma bounds the probability of the accumulated effect of the probabilistic part becoming too large too quickly.

\newcommand{\xconstant}{x}
\newcommand{\yconstant}{y}

\begin{restatable}{lemma}{lemmarandompartprobbound}
	\label{lemma-random-part-prob-bound}
	Let \(\hat{V}_i \) denote the accumulated effect on the counters in the probabilistic  part  after \(i\) steps of a computation. Let \(\xconstant,\yconstant\in \R \) and  \(\tau \) be a stopping time such that \(|\hat{V}_\tau(c)|\geq \xconstant \)  for any \(c\in \countersset(\A) \) for the first time. Then for each \(p\in\states(\A)\) and each strategy \(\strategyDemon \) it holds 
	\[\prob_p^\strategyDemon[\tau<\yconstant]
	\leq
	f\cdot \exp\big( \frac{-(\xconstant-g)^2
	}{
		h\cdot \yconstant
	}\big)
	\] for some constants  \(f,g\), and \(h\) that depend only on \(\A \).
\end{restatable}

We use properties of \hyperref[fig-systems]{(II)} to construct martingales that copy each counter up to at most a constant deviation \(g\), which then allows us to obtain the probability bound from Azuma inequality. A full proof of Lemma~\ref{lemma-random-part-prob-bound} can be found in Section~\ref{sec-lemma-random-part-prob-bound-proof}.\qed

\subsection{Iterating Multi-Components}
\label{sec-multi-comp-iterating}
In this section, we describe a novel procedure to iterate a strongly connected multi-component \(\bx\) from state \(p\) contained in \(\bx \). Towards this we  make use of the following technical lemma.
\begin{restatable}{lemma}{lemmadecomposemulticomponentsintocomponents}
	\label{lemma-decompose-multicomponents-into-components}
	
	Let \(\bx\in \mathbb{Q}^{\countersset(\A)} \) be a multi-component of \(\A \). Then \(\bx\) can be decomposed into a conical sum of simple components, that is \(\bx=\sum_{\by} a_\by \cdot\by\) where \(\by \) ranges over all the simple components of \(\A \) and \(0\leq a_\by\in \Q \) for all \(\by \). Furthermore, the decomposition can be done in such a way that \(a_\by >0 \) iff \(\by \) is a simple component of \(\A_\bx \).
\end{restatable} 

 We construct the decomposition by repeatedly subtracting multiples of simple components of \(\A_\bx \) from \(\bx \). A full proof of Lemma~\ref{lemma-decompose-multicomponents-into-components} can be found in Appendix~\ref{sec-lemma-decompose-multicomponents-into-components-proof}.\qed 

 Intuitively, a single iteration of \(\bx \) consists of iterating each simple component \(\by \) exactly \(a_\by \) times, while utilizing the interconnected structure of the simple component intersections obtained from \(\bx \) being strongly connected to move between the individual simple components.

Formally, let \(\by_1,\dots,\by_\ell \) be all the simple components of \(\A_\bx \) centered in \(p_1,\dots,p_\ell \). For simplicity we assume that the decomposition into simple components \(\bx=\sum_{i=1}^{\ell}a_i\cdot \by_i  \) satisfies  \(a_i\in \mathbb{N} \) for all \(i\leq \ell\) (and we will make sure this holds every time we use this procedure in this paper).

Let  \(G=(V,E) \) be a graph where \(V=\{\by_1,\dots,\by_{\ell} \} \) and \(\{\by_i,\by_j\}\in E \) iff \(\states(\A_{\by_i})\cap\states(\A_{\by_j})\neq \emptyset \). Let \(G_T \) be any spanning graph of \(G\). Wlog. we can declare \(\by_1 \) the root of \(G_T \) and assume that \(p\) is a state of \(\A_{\by_1}\). For each \(\by_i \) let \(\parent(\by_i) \) denote the parent of \(\by_i \) in \(G_T \) and \(\child(\by_i)=\{p_j \mid \parent(\by_j)=\by_i \}\). As per Lemma~\ref{lemma-component-recentering} we can wlog.  assume that \(\by_2,\dots,\by_\ell \) are centered in states \(p_2,\dots,p_\ell \) such that each \(p_i \) is also a state of \(\A_{\parent(\by_i)} \).

A single iteration of \(\bx \) consists of iterating each simple component \(\by_i \) exactly \(a_i\) times using the tree-like structure of \(G_T \) to move from one simple component to another. It starts by iterating \(\by_1 \) exactly \(a_1\) times, however this iterating may be temporarily halted in favor of iterating other simple components as follows. Let \(\by \) be the current simple component that is being iterated, whenever a state \( p_i\in \child(\by)\) is visited for the first time during all the iteratings of \(\by\),  the iterating of \(\by \) is temporarily halted and we instead iterate \(\by_i \) exactly \(a_i \) times, after which we resume the iterating of \(\by \) for the remaining number of times.

We say that the iteration of \(\bx \) succeeded if at the end we iterated each of the \(\by_1,\dots,\by_\ell \). Note that the only way for it to fail is if the iterating of some \(\by_i \) never visits some \(p\in\child(\by_i) \) even once during all \(a_i \) iterations, which happens with probability at most \(b^{a_i} \) for some constant \(b<1 \).

\subsection{Witness Strategy}
\label{sec-strat-description}

Let \(\bx \) and \(\ba,\bb \) be some maximal solutions of \hyperref[fig-systems]{(I)} and  \hyperref[fig-systems]{(II)} for \(\A_k \). Let
\begin{equation*}
	\Czero=\{c\in \countersset(\A)\mid \Delta(\bx)(c)=0 \} 
	\hfill \hspace{1.5cm}
	C_+=\{c\in \countersset(\A)\mid \Delta(\bx)(c)>0 \}\hspace{0.98cm}
\end{equation*}

Note that \(\Czero=C_k \) and \(C_+=\countersset(\A)\setminus C_k \).


We use \(\spanCone(\by) \) to denote the cone generated by  \(\support(\by) \). The core of our lower bound comes from the following lemma, which allows us to ``reset'' the probabilistic part of the computation in \(\A_k \) whenever it threatens to kill the run.


\begin{restatable}{lemma}{lemmavecspaces}\label{lemma-vecspaces}
	Let \(\MEC \) be an MEC of \(\A_k \) and let \(\by \) be a simple component of \(\MEC \). Then it holds that \(\spanCone(\hat{\by})(\Czero)\) is a sub-vector space of the vector space  \(\Delta\big(\multicompset(\MEC)\big)(\Czero)\).	
\end{restatable}
These being vector spaces follows from them having expected effect \(\vec{0}\) on \(\Czero\).  And if there were to exist \(\bu \) that is orthogonal to \(\Delta\big(\multicompset(\MEC)\big)(\Czero) \) but not to \(\spanCone(\hat{\by})(\Czero) \) then we can project effect vectors onto \(\bu \), thus obtaining a new counter \(c_\bu\). Since \(\hat{\by}\) would be zero-unbounded on \(c_\bu \) we can show that there exists a transition \(t\) of \(\MEC\) with \(t\in M_k\subseteq T_k \) thus contradicting Lemma~\ref{lemma-k-converged}. A full proof of Lemma~\ref{lemma-deviations-resetable} can be found in Section~\ref{sec-lemma-vecspaces-proof}.\qed

In order to simplify the notation we introduce the notion of bins. The idea is to virtually ``split'' the counter vector \(\bv \) into multiple separate counter vectors \(\bv_1,\dots,\bv_\bincount \) such that  \(\bv=\sum_{i=1}^{\bincount}\bv_i \). We call the individual vectors \(\bv_1,\dots,\bv_\bincount \) \emph{bins}. We reference the individual bins  with names, concretely we use: main-bin, looping-bin, \(\bx_1 \)-bin, \(\dots\), \(\bx_w \)-bin, and resetting-bin, where \(w \) is the number of MECs of \(\A_k \). A key point is that as long as every single bin is non-negative then the actual counter vector is also non-negative, while we have each computational step affect potentially different bin. Hence it suffices to show that all bins are non-negative to show non-termination.

 Let \(\bx=\sum_{\by\in \simplecompset(\A_k)} a_\by\cdot \by \) be a decomposition of \(\bx \) into simple components where \(a_\by>0 \) for each  \(\by\in \simplecompset(\A_k)\) (this decomposition exists from Lemma~\ref{lemma-decompose-multicomponents-into-components}). Since all the numbers are rational, we can wlog.  assume   \(a_\by\in \mathbb{N} \) for all \(\by \)  and using  Lemma~\ref{1dim-lemma-component-realeffekt-is-effect} also that\footnote{As otherwise we can simply take a rational multiple of \(\bx \) as our \(\bx\) (see Section~\ref{sec-systems}).}  \begin{equation}\label{eq-multi-comp-it-big}
	\sum_{\by\in \simplecompset(\A_k)} a_\by\cdot \E[\realeffect_\by](C_+)\geq \vec{3}
\end{equation}

Let \(\MEC_1,\dots,\MEC_w \) be all the MECs of \(\A_k \), and let \(\bx_1,\dots,\bx_w \) be the restrictions of \(\bx \) onto \(\MEC_1,\dots,\MEC_w \), respectively (i.e., \(\bx_i(t)=\bx_i(t) \) if \(t\) is in \(\MEC_i \) and \(\bx_i(t)=0 \) otherwise).  Let \(p_1,\dots,p_w \) be some states of \(\MEC_1,\dots,\MEC_w \), respectively.

\newcommand{\nepsilon}{\epsilon}

\newcommand{\epsilontwo}{\nepsilon[small]}

\newcommand{\epsilonthree}{\nepsilon[bound]}

\newcommand{\nepthree}{\nepsilon[small]}
\newcommand{\epsilonX}{\nepthree}
\newcommand{\epsilonANYSMALLDIFFERENT}{\nepthree}
\newcommand{\epsilonANYSMALL}{\nepthree}
\newcommand{\epsilonbigZ}{\nepthree}
\newcommand{\epsilonFIXINGDIFFERENTSMALL}{\nepthree}

\newcommand{\epsilonone}{\nepsilon[\bx]}

\newcommand{\epsilontenvar}{\nepsilon[loops]}
\newcommand{\epsilonfour}{\nepsilon[loops]}

\newcommand{\epsilonsingleiteration}{\epsilon[\realeffect]}

\newcommand{\espilonresets}{\nepsilon[resets]}

\newcommand{\epsev}{\nepsilon[1]}
\newcommand{\epsilonsix}{\epsev}
\newcommand{\epsilonbigO}{\epsev}
\newcommand{\epsilonbigY}{\epsev}

\newcommand{\epsilonkuwait}{\nepsilon[2]}

\newcommand{\epsilondoublebigOO}{\ensuremath{1.5}}

\newcommand{\epsilonSOMETHINGNEW}{\ensuremath{1.5}}
\newcommand{\epsilonNEWFIXSECOND}{\ensuremath{1.5}}

\newcommand{\epsilonTECHNICALHERE}{\ensuremath{1.5}}

\newcommand{\theepsilons}{\(\epsilontwo\),  \( \epsilonthree\), \( \epsilonone\), \( \epsilonsingleiteration\), \( \epsilontenvar \),  \( \espilonresets\), \(\epsev\), \(\epsilonkuwait\)   }

\textbf{Description of \(\strategyDemon\):} We are now ready to describe the strategy \(\sigma \), parameterized by \(0<\epsilon<\frac{1}{2} \), which for all  \(p\in\states(\A)\) satisfies \[\liminf_{n\rightarrow\infty}  \prob_{p\vec{n}}^\strategyDemon[\calC[c]\geq 2^{n^{\frac{1}{2}-\epsilon}} \textit{ and } \calT[t]\geq 2^{n^{\frac{1}{2}-\epsilon}} \textit{ for all } c\in C_+ \textit{ and all } t\in\transitions(\A)\setminus T_k ] =1 \] 
 Due to the complexity of the strategy we give only high level description.

  Let  \theepsilons be constants all satisfying \(0<\epsilon[*]<1 \) and whose values are given in Table~\ref{Table-eps-equations} in Section~\ref{sec-eps-assign}.  Note that these constants depend only on \(\epsilon \). (The exact values are not that important as these are purely technical constants to simplify the notation.)

\(\strategyDemon\) will ``remember'' (these can always be computed from the history) functions \(\fsize,\gresets:\mathbb{N}\rightarrow \mathbb{N}_0 \)  initialized as \(\gresets(i)=0 \) and \(\fsize(i)=1 \) for all \(i\). Let \(n\) be the size of the smallest counter in the initial configuration. Each bin is initialized to \(\lfloor\frac{\vec{n}}{\bincount}\rfloor \) where \(\bincount \) is the number of bins (and if these do not sum up to the counters vector then the remainder is put into another bin that is never used anywhere).



\(\strategyDemon \) begins by ``looping'' over all states of \(\A \) by choosing uniformly at random at each step. We say the  \(i\)-th loop finishes when each state of \(\A \) has been visited this way at least once since the \((i-1)\)-st loop finished (and the \(0\)-th loop is considered to have finished before the computation started). The effect of each step taken during such looping is put into the looping-bin. When the \(i\)-th loop finishes \(\strategyDemon \) sets \begin{equation}\label{eq-fsize-def}
	\fsize(i+1)= \big\lfloor\frac{2\cdot \fsize(i)\cdot \lfloor n^{1-\epsilontwo} \rfloor - \gresets(i)\cdot n^{1-\epsilonthree+\epsilonX}}{\lfloor n^{1-\epsilontwo} \rfloor}\big\rfloor \end{equation}
Additionally, when a state \(p_j\in \{p_1,\dots,p_w \} \) is visited for the first time during the \(i\)-th loop \(\strategyDemon \) calls the following sub-procedure, performing it in full before resuming the looping:

\begin{itemize}
	\item Repeat the following \(\fsize(i) \) times: Let \(\lfloor\frac{\vec{n}}{\bincount}\rfloor + \bv \) be the current counters vector in the \(\bx_j \)-bin, then:
	\begin{itemize}
		\item if \(|\bv(\Czero)|< n^{1-\epsilonthree} \)  iterate the multi-component \(\bx_j \cdot \lfloor n^{1-\epsilonone} \rfloor \);
		\item if \(|\bv(\Czero)|\geq  n^{1-\epsilonthree} \), first set \(\gresets(i)=\gresets(i)+1\), then \(\strategyDemon \) moves the vector \(\bv(\Czero) \) from the \(\bx_j \)-bin to the resetting-bin (thus ``resetting'' it to \(\lfloor\frac{\vec{n}}{\bincount}\rfloor \) on the counters from \(\Czero\)). Afterwards, \(\strategyDemon \) iterates the multi-component \(\bx_j \cdot \lfloor n^{1-\epsilonone} \rfloor+\bz \) where \(\bz \) is a multi-component of \(\A_{\bx_j} \) that can be decomposed as \(\bz=\sum_{\by\in \simplecompset(\A_{\bx_j})}  a_\by'\cdot \by\) and it holds \(\sum_{\by\in \simplecompset(\A_{\bx_j})} a_\by'\cdot\E[\realeffect_\by](\Czero)=- \bv(\Czero)\) that minimizes \(\sum_{\by\in \simplecompset(\A_{\bx_j})}a_\by' \) (its existence follows from  Lemma~\ref{lemma-deviations-resetable}). Finally, \(\strategyDemon \) moves the vector \(-\bv(\Czero) \) from the main-bin to the resetting-bin thus resetting it back to \(\lfloor\frac{\vec{n}}{\bincount}\rfloor\).

	\end{itemize}
	In both cases, during the iterating of the multi-component, the effect on the  deterministic part of the computation is put into the main-bin whereas the effect \(\hat\bu \) on the probabilistic part is split such that \(\hat\bu(\Czero) \) is added to the \(\bx_j \)-bin and \(\hat\bu(C_+) \) to the main-bin.  
\end{itemize}	

\begin{note}
		Technically, it can hold \(a_\by'\notin \N_0 \). As we can only iterate \(\bx_j \cdot \lfloor n^{1-\epsilonone} \rfloor+\bz\) when \(a_\by'\in \N_0 \), we actually replace \(\bz \) with \(\bz'= \sum_{\by\in \simplecompset(\A_{\bx_j})}  \lceil  a_\by'  \rceil\cdot \by\), while 		pretending we only iterated the \(a_\by'-\lfloor a_\by'\rfloor \) fraction of the last iteration of  each \(\by\) (this way in effect we obtain \(\bz \) instead of \(\bz' \)) and the effect of the extra ``precomputed'' fraction  is put into a separate precompute-bin. Additionally, \(\strategyDemon \) also remembers for each simple component how large fraction \(b_\by \)		of an iteration it still has ``precomputed'', and  the next time \(\strategyDemon \) is computing another \(\bz' \) it uses \(a_\by'-\min(a_\by',b_\by) \) instead of  \(a_\by' \)  	 while subtracting \(\min(a_\by',b_\by) \) from \(b_\by \), while taking the effect of the ``missing'' fraction of iteration from the precompute-bin. It is easy to 	see that this way at any point each simple component \(\by\) will be precomputed by at most a single iteration, and hence the precompute-bin will never change by more then the combined effect of at most a single iteration of all the simple components. For our purposes the precompute-bin is sufficiently large for this from our assumption of \(\eventshortsucc \) (defined in Section~\ref{sec-sigma-analysis} below). 
\end{note}

\subsection{Analysis of \(\sigma \) }
\label{sec-sigma-analysis}

\newcommand{\eventsmallprobsingle}{E_{smallProb}}

\newcommand{\eventresets}{E_{resets}}

\newcommand{\eventbzbound}{E_{{Small}\bz}}

In this section we prove the claim we made the  strategy \(\sigma\), as defined above, namely that it  witnesses the lower asymptotic estimates \(2^{\sqrt{n}} \) from Lemma~\ref{lemma-njnkjbn}.

\begin{lemma}	\label{lemma-strategy-witnesses}
	For all \(0<\epsilon<\frac{1}{2} \) and all \(p\in \states(\A) \) it holds \[\liminf_{n\rightarrow\infty}  \prob_{p\vec{n}}^\strategyDemon[\calC[c]\geq 2^{n^{\frac{1}{2}-\epsilon}} \textit{ and } \calT[t]\geq 2^{n^{\frac{1}{2}-\epsilon}} \textit{ for all } c\in \countersset(\A)\setminus C_k \textit{ and all } t\in\transitions(\A)\setminus T_k ] =1 \] 
	\end{lemma}
\begin{proof}
	To simplify the notation we explicitly assume \(n\) is sufficiently large. E.g., we say that \(100\cdot n<n^{1+\epsilon} \) without explicitly saying the ``for all sufficiently large \(n\)''.
	
We proceed to define several events whose intersection directly implies the statement from Lemma~\ref{lemma-strategy-witnesses}. Then we show  \(\strategyDemon \) satisfies each event with very high probability, and thus also their intersection. The events are:

\begin{itemize}
	
	\setlength{\itemsep}{1.95pt}
	\item \(\eventloops\) is the event that the looping-bin does not become negative on any counter before the \(\lfloor n^{\epsilonfour}\rfloor \)-th loop finishes. (Lemma~\ref{lemma-event-looping})
	\item \(\eventshortsucc \) is the event that, within the first \(n^{\epsilonfour} \) loops, no single iterating of a multi-component fails and each single iteration of any simple component takes at most \(n^{\epsilonsingleiteration} \) steps. (Lemma~\ref{lemma-e4})
	\item \(\eventfinit\)  is the event that \(\fsize(3)=4 \). (Lemma~\ref{lemma-event-finit})
	
	\item \(\eventfexp\) is the event that \(\fsize(i)\geq 1.5^{i} \) for all \(i\geq 3 \) for which the \((i-1)\)-st loop finished.\footnote{The value \(1.5\) here is arbitrary, and any \(a<2\) would work as well, but values closer to \(2\) would require larger \(i\)'s.} (Lemma~\ref{lemma-E-expsize})
	
		\item \(\eventresets\)  is the event that (Lemma~\ref{lemma-eventresets}) \[\gresets(i)\leq \frac{\fsize(i)\cdot \lfloor n^{1-\epsilonone} \rfloor }{n^{\espilonresets}} \] 

	\item \(\eventsmallprobsingle\) is the event, that within the first \(n^{\epsilonfour}\) loops, any single iteration of a multi-component \(\bx_j \cdot \lfloor n^{1-\epsilonone} \rfloor \) has effect \(\hat{\bv} \) on the probabilistic part with \(|\hat{\bv}(c)|\leq n^{1-\epsilonthree} \) for any \(c\in \countersset(\A)\). (Lemma~\ref{lemma-eventsmallprobsingle})

	\item \(\eventdoublereset\)  is the event that within the first \(n^{\epsilonfour} \) loops, whenever an iteration of a multi-component \(\bx_j \cdot \lfloor n^{1-\epsilonone} \rfloor+\bz \) finishes the \(\bx_j\)-bin contains counters vector \(\lfloor\frac{\vec{n}}{\bincount}\rfloor + \bv \) where \(|\bv(\Czero)|< n^{1-\epsilonthree} \). (Lemma~\ref{lemma-eventdoublereset})
	
		\item \(\eventbzbound\)  is the event that, within the first \(n^{\epsilonfour} \) loops, whenever a multi-component \(\bx_j \cdot \lfloor n^{1-\epsilonone} \rfloor+\bz \) is to be iterated it holds \(|\bz|\leq n^{\epsilonbigO} \). (Lemma~\ref{lemma-eventbzbound})

	
	\item 	\(\eventbigcounter\) is the event that, within the first \(n^{\epsilonfour} \) loops,  the main-bin never becomes negative on any counter, and every counter \(c\in C_+ \) has, in the main-bin, when the \(i \)-th loop finishes, value at least \(2\cdot \fsize(i) \cdot \lfloor n^{1-\epsilonone} \rfloor \).  (Lemma~\ref{lemma-E-bigC})
\end{itemize}
 
 Now consider the intersection of all these events. First note that from the lemmas~\ref{lemma-event-looping}-\ref{lemma-E-bigC} it holds  
\begin{multline*}
	\lim_{n\rightarrow\infty} \prob_{p\vec{n}}^\strategyDemon[\eventloops\cap \eventshortsucc\cap \eventfinit\cap \eventfexp \cap\\\cap \eventresets\cap \eventsmallprobsingle\cap \eventdoublereset\cap \eventbzbound\cap \eventbigcounter]=1
\end{multline*}

 These events also imply no bin gets negative on any counter within the first \(n^{\epsilonfour} \) loops since:
 \begin{itemize}
 	\setlength{\itemsep}{1pt}
 	\item looping-bin: follows directly from \(\eventloops\).

 		\item main-bin:  follows directly from \(\eventbigcounter\).
 	
 	\item \(\bx_i \)-bins: these bins never change the value of \(c\in C_+\). For \(c\in \Czero\), from \(\eventbzbound\) we have that  \(|\bv(\Czero)|\geq  n^{1-\epsilonthree} \) implies  \(|\bv(\Czero)|\leq n^{1-\epsilonthree}+a\cdot n^{\epsilonbigO} \) for some constant \(a\) that depends only on \(\A\). But this value is always lower than the initial counters vector in this bin since \(\epsilonbigO<1 \). And \(c\in\Czero\) can only become negative in the  \(\bx_i \)-bin when \(|\bv(\Czero)|\geq \frac{n}{\bincount}\).

 	\item resetting-bin: Whenever this bin changes it gets restored right away, and from \(\eventbzbound\) the change is by at most \(n^{1-\epsilonthree}+a\cdot n^{\epsilonbigO}<\lfloor\frac{n}{\bincount}\rfloor  \). 	
 \end{itemize}
 
 Hence the intersection of all these events also implies that the first \(n^{\epsilonfour} \) loops all finish. From \(\eventfexp \) it holds \(\fsize(\lfloor n^{\epsilonfour}\rfloor)\geq 1.5^{\lfloor n^{\epsilonfour}\rfloor} \).  Therefore the intersection also implies that each \(t\in \transition(\A) \) with \(\bx(t)>0 \) gets iterated at least \(1.5^{\lfloor n^{\epsilonfour}\rfloor}\) times (from \(\eventshortsucc\)) and that  the computation reaches a configuration \(p\bv \) with \(\bv(c)\geq 1.5^{\lfloor n^{\epsilonfour}\rfloor} \) for each \(c\in C_+ \) (from \(\eventbigcounter\)).
 
 In Section~\ref{sec-eps-assign} we show that for all \(0<\epsilon<\frac{1}{2} \) there exists an assignment for the \(\epsilon[*]\) variables with \(\epsilonfour=\frac{1}{2}-\epsilon  \) for which all the restrictions put upon these are valid. Hence this implies Lemma~\ref{lemma-strategy-witnesses}.
\end{proof}

 One could ask whether we could also put \(\epsilonfour\geq \frac{1}{2}\) and thus obtain a higher lower asymptotic estimate. Unfortunately, as per the following lemma any such value would break the restrictions  we put on the \(\epsilon[*] \) variables  and hence would break our proof. It remains   open whether this is due to an inefficiency of our proof, or whether no better lower asymptotic estimate can be obtained in the general case. 
 
 
\begin{restatable}{lemma}{lemmasqrtmustbe}\label{lemma-sqrt-mustbe}
	All  restrictions placed upon the \(\epsilon[*] \) variables in this paper imply \(\epsilonfour<\frac{1}{2} \).
\end{restatable}
We list all the restrictions on the \(\epsilon[*] \) variables in Table~\ref{Table-eps-equations} (Section~\ref{sec-eps-assign}). The proof of Lemma~\ref{lemma-sqrt-mustbe} is then purely technical and can be found in Section~\ref{sec-lemma-sqrt-mustbe-proof}. 

Specifically, setting \(\epsilonfour\geq \frac{1}{2} \)  would break both \(\eventdoublereset\) and \(\eventshortsucc\) happening with very large probability. 
\begin{itemize}
	\item In \(\eventshortsucc\) the larger number of iterations of simple components would mean we would start seeing iterations whose length is longer than our upper bound \(n^{\epsilonsingleiteration} \). This would then give us too large upper bounds on the number of steps per single iteration of a multi-component thus invalidating multiple other parts of the proof.
	\item In \(\eventdoublereset\) the larger number of iterations of multi-components \(\bx_j \cdot \lfloor n^{1-\epsilonone} \rfloor+\bz\) would mean we would start seeing such multi-components being iterated right after another iteration of \(\bx_j \cdot \lfloor n^{1-\epsilonone} \rfloor+\bz\). This would break our upper bound on the size of \(|\bv(\Czero)|\) which in turn would break our proof that no counter becomes negative in any \(\bx_j\)-bin. 
\end{itemize}

 We now proceed with proving that \(\strategyDemon \) satisfies each of the  individual events with very high probability. In order to increase readability the full proofs have been moved to Section~\ref{sec-proofs-of-full-somehhh}.

 \begin{restatable}{lemma}{lemmaeventlooping}\label{lemma-event-looping}
 	\(\lim_{n\rightarrow\infty} \prob_{p\vec{n}}^\strategyDemon[\eventloops]=1\)
\end{restatable}
 The expected number of steps per single loop is constant and as the number of loops is sub-linear we obtain Lemma~\ref{lemma-event-looping} by applying Markov inequality. A Full proof can be found in Section~\ref{app-lemma-event-looping-proof}. \qed

\begin{restatable}{lemma}{lemmae}\label{lemma-e4}
	\(\lim_{n\rightarrow\infty} \prob_{p\vec{n}}^\strategyDemon[\eventshortsucc]=1\)
\end{restatable}
The probability of avoiding some state/transition for \(n^{\epsilonsingleiteration} \) steps decreases exponentially fast in \(n^{\epsilonsingleiteration}\). Since \(\exp({n^{\epsilonsingleiteration}}) \) grows faster than \(2^{n^{\epsilonfour}} \) the probability of avoidance for \(2^{n^{\epsilonfour}} \) steps still decreases in \(n\). A full proof of Lemma~\ref{lemma-e4} can be found in Section~\ref{sec-proof-of-lemma-component-iterating-exp-not-too-long-prob}.\qed

\begin{restatable}{lemma}{lemmaeventfinit}\label{lemma-event-finit}
	\(\lim_{n\rightarrow\infty} \prob_{p\vec{n}}^\strategyDemon[\eventfinit\mid \eventshortsucc]=1\)
\end{restatable}\label{lemma-E1}
\(\gresets(1)=\gresets(2)=0 \) implies  \(\fsize(2)=2 \) and \(\fsize(3)=4\). Hence it suffices to show that with very high probability the probabilistic part of the computation does not change the counter vector by more than \(n^{1-\epsilonthree}\) within the  first two iterations of the multi-component \(\bx_j\cdot  \lfloor n^{1-\epsilonone} \rfloor \). This follows from the results about 1-dim VASS MDPs from \cite{AKCONCUR23} which give a lower asymptotic estimate of \(n^2\) before the probabilistic part changes a counter by more than \(n\). A full proof of Lemma~\ref{lemma-event-finit} can be found in the Section~\ref{app-lemma-event-finit-proof}.\qed

\begin{restatable}{lemma}{lemmaEexpsize}\label{lemma-E-expsize}
	\(\prob_{p\vec{n}}^\strategyDemon[\eventfexp\mid \eventfinit]=1\)
\end{restatable}
\(\gresets(i)\leq w\cdot \fsize(i)\) for our choice of the \(\epsilon[*]\) variables it holds \(\fsize(i)\geq \lfloor 1.9\cdot \fsize(i-1) \rfloor\) for all \(i\geq 2\), hence the proof is purely technical. A full proof of Lemma~\ref{lemma-E-expsize} can be found in Section~\ref{sec-lemma-E-expsize-proof}.\qed

\begin{restatable}{lemma}{lemmaeventresets}\label{lemma-eventresets}
	\(\lim_{n\rightarrow\infty} \prob_{p\vec{n}}^\strategyDemon[\eventresets\mid \eventshortsucc]=1\)
\end{restatable}

Using Lemma~\ref{lemma-random-part-prob-bound} we upper bound the probability of multiple consecutive iterations of multi-components exceeding effect \(n^{1-\epsilonthree} \) in the probabilistic part on any counter. This probability is small enough that we can safely assume each increase  of \(\gresets(i) \) requires too many iterations of multi-components which gives an upper bound on \(\gresets(i) \). A full proof of Lemma~\ref{lemma-eventresets} can be found in Section~\ref{sec-lemma-eventresets-proof}.\qed

 \begin{restatable}{lemma}{lemmaeventsmallprobsingle}\label{lemma-eventsmallprobsingle}
 	\(\lim_{n\rightarrow\infty} \prob_{p\vec{n}}^\strategyDemon[\eventsmallprobsingle\mid \eventshortsucc]=1\)
 \end{restatable}
 Proof of Lemma~\ref{lemma-eventsmallprobsingle} is a purely technical application of Lemma~\ref{lemma-random-part-prob-bound}. The full proof can be found in Section~\ref{sec-lemma-eventsmallprobsingle-proof}.\qed

\begin{restatable}{lemma}{lemmaeventdoublereset}\label{lemma-eventdoublereset}
	\(\lim_{n\rightarrow\infty} \prob_{p\vec{n}}^\strategyDemon[\eventdoublereset\mid \eventshortsucc,\eventsmallprobsingle]=1\)
\end{restatable}
We use Lemma~\ref{lemma-random-part-prob-bound} to upper bound the probability of any single iteration of a multi-component \( \bx_j \cdot \lfloor n^{1-\epsilonone} \rfloor+\bz\) having large enough effect on the counters in the probabilistic part to have the next multi-component iterated be also some \(\bx_j \cdot \lfloor n^{1-\epsilonone} \rfloor+\bz'\). A Full proof of Lemma~\ref{lemma-eventdoublereset} can be found in Section~\ref{app-lemma-eventdoublereset-proof}.\qed



\begin{restatable}{lemma}{lemmaeventbzbound}\label{lemma-eventbzbound}
	\(\eventbzbound \subseteq \eventdoublereset\cap \eventsmallprobsingle\)
\end{restatable}
Using \(\eventshortsucc\cap \eventdoublereset\) we can upper bound the size of \(|\bv(\Czero)|\) which then gives us an upper bound on \(|\bz|\) from basic properties of vector spaces. A full proof of Lemma~\ref{lemma-eventbzbound} can be found in Section~\ref{sec-lemma-eventbzbound-proof}.\qed

\begin{restatable}{lemma}{lemmaEbigC}\label{lemma-E-bigC}
	\(\lim_{n\rightarrow\infty} \prob_{p\vec{n}}^\strategyDemon[\eventbigcounter\mid \eventshortsucc,\eventdoublereset,\eventresets,\eventbzbound]=1\)
\end{restatable}
The effect of each multi-component iteration is zero on counters from \(\Czero\) and the individual iterations are too short to ever deplete these. Additionally, each finished loop effectively iterates \(\bx\) a total of \(\fsize(i)\cdot \lfloor n^{1-\epsilonone} \rfloor \) times, and the effect of \(\bx \) on the deterministic part is sufficiently large on \(C_+ \) to compensate for possible fluctuations. A full proof of Lemma~\ref{lemma-E-bigC} can be found in Section~\ref{sec-lemma-E-bigC-proof}.\qed

	\section{Dag-like VASS MDP}
\label{sec-dag-like}
Let \(\A \) be a DAG-like VASS MDP. If \(\A \) contains any MEC \(\MEC \) that has lower asymptotic estimate \(2^{\sqrt{n}}\) for \(\F_\MEC \) then this trivially gives same lower asymptotic estimate also for \(\F_\A \). If on the other hand every MEC of \(\A \) has a polynomial upper asymptotic estimate then  also \(\F_\A \) has a polynomial upper asymptotic estimate. Intuitively, this is because in DAG-like VASS MDP each MEC can be entered/exited at most once, and since everything ``inside'' a MEC is polynomial whenever a new MEC is reached all the counters are still only polynomial. We formalize this in the following lemma.

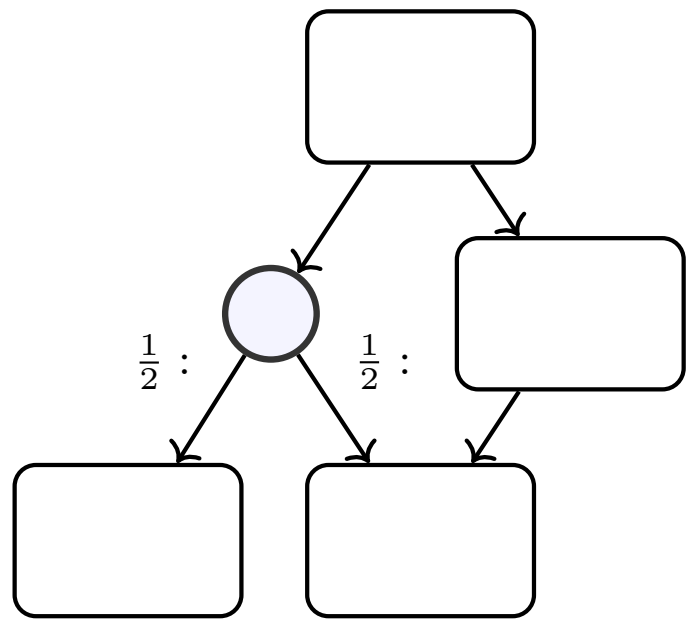
\begin{figure}[t]\centering
	\begin{tikzpicture}[x=1.5cm, y=1.5cm, font=\footnotesize]
	
		\begin{scope}[shift={(4,0)}]
			\node (A) at (0,0) [draw,thick,minimum width=1.5cm,minimum height=1cm, rounded corners] {};
			\node (B) at (0.66,-1) [draw,thick,minimum width=1.5cm,minimum height=1cm, rounded corners] {};
			\node[random] (C) at (-0.66,-1) {};
			\node (D) at (-1.29,-2) [draw,thick,minimum width=1.5cm,minimum height=1cm, rounded corners] {};
			\node (E) at (0,-2) [draw,thick,minimum width=1.5cm,minimum height=1cm, rounded corners] {};
			
			\draw [->, thick] (A) -- (C);
			\draw [->, thick] (C) -- node[above left]  {\(\frac{1}{2}: \)} (D);
			

			\draw [->, thick] (C) --  node[above right]  {\(\frac{1}{2}: \)} (E);
			
			\draw [->, thick] (A) -- (B);
			
			
			\draw [->, thick] (B) -- (E);
		\end{scope}
		
	\end{tikzpicture}
	
	\caption{Example of a MEC-decomposition of a DAG-like VASS MDP.}
	\label{fig-VASS-dsdsds}
\end{figure}

\begin{lemma}\label{lemma-daglike-superpolynomial-upper-bound}
	Let \(\A \) be a DAG-like  VASS MDP with MECs \(\MEC_1,\dots,\MEC_w\) such that each \(\MEC_i \) has an upper asymptotic estimate \(n^{k[i]} \) for \(\calC_{\MEC_i}[c]\) for all \(c\in\countersset(\A) \) and \(n^{2\cdot k[i]} \) for each \(\F_{\MEC_i}\in\{\calL_{\MEC_i},\calT_{\MEC_i}[t] \} \), where  \[k[i]= 3^{|\countersset(\A)|+ |\transitions(|\MEC_i)|} \] Then each \(\calC_\A[c] \) has an upper asymptotic estimate \(n^k\) and each \(\F_\A\in \{\calL_\A,\calT_\A[t] \} \) has an upper asymptotic estimate \(n^{2\cdot k} \),  where \[k=3^{w\cdot |\countersset(\A)|+|\transitions(\A)|} \]
\end{lemma}
\begin{proof}
	Since \(\A\) is DAG-like each computation can visit each MEC at most once. Let \(\bv_i \) be the counters vector when the computation either exits the \(i\)-th MEC visited, or upon termination if it terminates inside of the \(i\)-th MEC (or \(\bv_{i}=\bv_{i-1} \) if it enters less than \(i \) MECs).  We will show that for each \(i\leq w\), each \(\epsilon>0 \), each \(p\in \states(\A)\), and each strategy \(\strategyDemon \) it holds for all \(c\in \countersset(\A) \) \[\liminf_{n \to \infty} \prob_{p\vec{n}}^\strategyDemon[ \bv_i(c)\geq n^{(1+i\cdot \epsilon)\cdot a[i]} ]=0 \] where \[a[i]=3^{i\cdot |\countersset(\A)|+|\transitions(\MEC_1)|+\ldots+|\transitions(\MEC_i)|} \] where \(\MEC_j \) is the \(j\)-th MEC visited by the computation (or \(|\transitions(\MEC_j)|=0\) if \( \MEC_j\) does not exist).
	
	Note that this would prove that \(\calC_\A[c] \) has an upper asymptotic estimate \(n^k \) since \(a[w]\leq k \). Similarly, this would also prove that each \(\F_\A\in \{\calL_\A,\calT_\A[t] \} \) has an upper asymptotic estimate \(n^{2\cdot k} \) since the last MEC \(\MEC_j \) visited has an upper asymptotic estimate \(n^{2\cdot k[i]} \) for each \(\F_{\MEC_j}\in \{\calL_{\MEC_j},\calT_{\MEC_j}[t] \} \). And the counters vector upon reaching \(\MEC_j \) is at most \(n^{a[w-1]} \) and \[a[w-1]\leq  3^{(w-1)\cdot |\countersset(\A)|+|\transitions(\A)|-|\transitions(\MEC_j)|}  \] Therefore  \(\F_\A \) has an upper asymptotic estimate \[ (n^{a[w-1]})^{2\cdot k[i]}=n^{a[w-1]\cdot2\cdot  k[i]}\]
	and it holds \begin{align*}
a[w-1]\cdot2\cdot  k[i] 
&\leq 3^{(w-1)\cdot |\countersset(\A)|+|\transitions(\A)|-|\transitions(\MEC_j)|} \cdot2\cdot  3^{|\countersset(\A)|+ |\transitions(|\MEC_i)|}
\\& = 2\cdot  3^{w\cdot |\countersset(\A)|+|\transitions(\A)|} 
= 2\cdot k \end{align*}
	
	We proceed with an induction  over \(i\).
	
	\textbf{Base case \(i=1\):} this follows directly from the assumption that \(n^{k[1]}=n^{a[1]}\) is an upper asymptotic estimate of \(\calC_{\MEC_1}[c] \) for each \(c\in\countersset(\A) \).
	
\textbf{Induction step:} Assume this holds for \(i\), we will show it holds for \(i+1\) as well. If no \((i+1) \)-st MEC was reached during the computation then this holds trivially since \(\bv_{i+1}=\bv_i \) and \(a[i+1]>a[i] \). 
	Assume thus that the computation has reached the \(i\)-th MEC. Let \(E_\epsilon\) be the event that the \(i \)-th MEC has been reached with a counters vector \(\bv_i'\) such that \(\bv_i'(c)\leq n^{\big(1+(i-1)\cdot \epsilon\big)\cdot a[i-1]}\) for all \(c\in\countersset(\A) \). From the assumption that \(n^{k[i]}\) is an upper asymptotic estimate of \(\calC_{\MEC_i}[c] \) for each \(c\in\countersset(\A) \) we obtain that for each \(\epsilon>0 \), each \(p\in \states(\A)\), and each strategy \(\strategyDemon \) it holds for all \(c\in\countersset(\A) \) that \[\liminf_{n \to \infty} \prob_{p\vec{n}}^\strategyDemon\big[ \bv_i(c)\geq \big(n^{\big(1+(i-1)\cdot \epsilon\big)\cdot a[i-1]}\big)^{k[i]+\epsilon'} \mid E_\epsilon\big]
	=0 \] 
	
	where \[\epsilon'=\frac{k[i]\cdot \epsilon}{1+(i-1)\cdot \epsilon} \]
	
	We chose this value for \(\epsilon' \) as it satisfies the following technical lemma.
	
	\begin{restatable}{lemma}{lemmatechnicalepsprimeee}\label{lemma-technical-epsprime}
		It holds \[\big(1+(i-1)\cdot \epsilon\big)\cdot a[i-1]\cdot (k[i]+\epsilon')=(1+i\cdot \epsilon)\cdot a[i] \]
	\end{restatable}
	The proof of Lemma~\ref{lemma-technical-epsprime} is purely technical and can be found in Section~\ref{sec-lemma-technical-epsprime-proof}.

	Hence using Lemma~\ref{lemma-technical-epsprime} it holds 
	\begin{gather*}
		0=
		\liminf_{n \to \infty} \prob_{p\vec{n}}^\strategyDemon\big[ \bv_i(c)\geq \big(n^{\big(1+(i-1)\cdot \epsilon\big)\cdot a[i-1]}\big)^{k[i]+\epsilon'} \mid E_\epsilon\big]
		=\\=
		\liminf_{n \to \infty} \prob_{p\vec{n}}^\strategyDemon\big[ \bv_i(c)\geq n^{\big(1+(i-1)\cdot \epsilon\big)\cdot a[i-1]\cdot (k[i]+\epsilon')} \mid E_\epsilon\big]
		=
		\liminf_{n \to \infty} \prob_{p\vec{n}}^\strategyDemon\big[ \bv_i(c)\geq n^{(1+i\cdot \epsilon)\cdot a[i]} \mid E_\epsilon\big]
	\end{gather*} 
	
	From the law of total probability we get 
	\begin{multline*}
		\prob_{p\vec{n}}^\strategyDemon\big[ \bv_i(c)\geq n^{(1+i\cdot \epsilon)\cdot a[i]}\big]
		=\\=
		\prob_{p\vec{n}}^\strategyDemon\big[ \bv_i(c)\geq n^{(1+i\cdot \epsilon)\cdot a[i]} \mid E_\epsilon\big] 
		\cdot
		\prob_{p\vec{n}}^\strategyDemon\big[ E_\epsilon\big]
		+
		\prob_{p\vec{n}}^\strategyDemon\big[ \bv_i(c)\geq n^{(1+i\cdot \epsilon)\cdot a[i]} \mid \hat{E}_\epsilon\big] 
		\cdot
		\prob_{p\vec{n}}^\strategyDemon\big[ \hat{E}_\epsilon\big]
	\end{multline*} 
	where \(\hat{E}_\epsilon \) is the complement event to \(E_\epsilon\). From the induction assumption we have 
	\[\liminf_{n \to \infty} \prob_{p\vec{n}}^\strategyDemon[E_\epsilon]=1
	\]
	
	Hence 
	\begin{gather*}
		\liminf_{n \to \infty} 	\prob_{p\vec{n}}^\strategyDemon\big[ \bv_i(c)\geq n^{(1+i\cdot \epsilon)\cdot a[i]}\big]
		=\\=
		\liminf_{n \to \infty} 	\big(	\prob_{p\vec{n}}^\strategyDemon\big[ \bv_i(c)\geq n^{(1+i\cdot \epsilon)\cdot a[i]} \mid E_\epsilon\big] 
		\cdot
		\prob_{p\vec{n}}^\strategyDemon\big[ E_\epsilon\big]
		+
		\prob_{p\vec{n}}^\strategyDemon\big[ \bv_i(c)\geq n^{(1+i\cdot \epsilon)\cdot a[i]} \mid \hat{E}_\epsilon\big] 
		\cdot
		\prob_{p\vec{n}}^\strategyDemon\big[ \hat{E}_\epsilon\big]
		\big)
		\\\leq
		\liminf_{n \to \infty} 	\big(
		\prob_{p\vec{n}}^\strategyDemon\big[ \bv_i(c)\geq n^{(1+i\cdot \epsilon)\cdot a[i]} \mid E_\epsilon\big] 
		\cdot
		\prob_{p\vec{n}}^\strategyDemon\big[ E_\epsilon\big]
		+
		1
		\cdot
		\prob_{p\vec{n}}^\strategyDemon\big[ \hat{E}_\epsilon\big]
		\big)
		=
		0
	\end{gather*}
\end{proof}

	\section{Non DAG-like VASS MDPs}
\label{sec-non-dag}
Let \(\A \) be a non-DAG-like VASS MDP. If \(\A \) contains any MEC \(\MEC \) that has lower asymptotic estimate \(2^{\sqrt{n}}\) for \(\F_\MEC \) then this trivially gives same lower asymptotic estimate also for \(\F_\A \). If on the other hand every MEC of \(\A \) has a polynomial upper asymptotic estimate then every superpolynomial function is an upper asymptotic estimate of  \(\F_\A \) as per the Lemma~\ref{lemma-nondaglike-superpolynomial-pper-bound}. Then in Lemma~\ref{lemma-non-dag-example} we show that this does not necessarily imply that there exists any polynomial that is an upper upper asymptotic estimate of \(\F_\A \).

\begin{lemma}\label{lemma-nondaglike-superpolynomial-pper-bound}
	Let \(\A \) be a VASS MDP whose each MEC \(\MEC\) has an upper asymptotic estimate \(n^k \) where \[k=2\cdot 3^{|\countersset(\A)|+ |\transitions(\A)|} \] for each of \(\F_\MEC\in \{\calL_\MEC,\calC_\MEC[c],\calT_\MEC[t] \} \). Then each superpolynomial function \(f\) is an upper asymptotic estimate for each of  \(\F_\A\in \{\calL_\A,\calC_\A[c],\calT_\A[t] \} \).
\end{lemma}
\begin{proof}
	We utilize the notion of \emph{types} defined in \cite{AKCONCUR23}. Formally, for every computation $\pi$ of $\A$, let $\mecs(\pi)$ be the unique sequence of MECs visited by~$\pi$. 	A \emph{type} is then a finite sequence $\beta$ of MECs such that $\mecs(\pi) = \beta$ for some  computation~$\pi$. Observe that $\mecs(\pi)$ disregards the states that do not belong to any MEC; intuitively, this is because the transitions executed in such states do not influence the asymptotic growth of~$\F$. Observe that the length of $\mecs(\pi)$, denoted by $\mathit{len}\big(\mecs(\pi)\big)$, can be finite or infinite. The first possibility corresponds to the situation when an infinite suffix of $\pi$ stays within the same MEC.

	Types have two important properties. First is that each type \(\beta \) of finite length  can be seen as a DAG-like VASS MDP. Furthermore, from our assumption on \(\A\), from Lemma~\ref{lemma-daglike-superpolynomial-upper-bound} this DAG-like VASS MDP  has an upper asymptotic estimate \(n^{k} \) for each \(\F\in\{\calL,\calC[c],\calT[t] \} \) where \(k=2\cdot 3^{\length(\beta)\cdot ( |\countersset(\A)|+ |\transitions(\A)|)}  \).
	
	The second property is that for all strategies $\sigma $ and all $p \in \states(\A)$ it holds $\prob^\sigma_{p}[\mathit{len}(\mecs) = \infty] =0$ and  $\prob^\sigma_{p}[\mathit{len}(\mecs) \geq k]\leq a^k$ for some constant \(0<a<1\) that depends only on \(\A \). 
	
	Let \(f:\R\rightarrow\R\) be an arbitrary superpolynomial function.  Let \(g:\N\rightarrow\N \) be such that \(g(n) \) is the largest integer for which it holds \[f(n')\geq  n^{2\cdot 3^{g(n)\cdot ( |\countersset(\A)|\cdot |\transitions(\A)|)} }\] for all \(n'\geq n \). Since \(f \) is superpolynomial it holds  \(\liminf_{n\rightarrow\infty} g(n)=\infty\). 
	
	%
	%
	Then for every $\varepsilon>0$, every $p \in \states(\A)$, and every strategy $\sigma$ it holds
	\begin{gather*}
		\limsup_{n \to \infty} \prob^{\sigma}_{p \vec{n}}[\F \geq f(n^{1+\varepsilon}) ] 
		\leq 
		\limsup_{n \to \infty} \prob^{\sigma}_{p \vec{n}}[\F \geq n^{2\cdot 3^{g(n)\cdot ( |\countersset(\A)|\cdot |\transitions(\A)|)}} ]
		\leq \\\leq 
		\limsup_{n \to \infty} \prob^{\sigma}_{p \vec{n}}[\length(\mecs)\geq g(n) ] 
		\leq 
		\limsup_{n \to \infty} a^{g(n)} 
		=0
	\end{gather*}
	
	Hence \(f\) is an upper asymptotic estimate of \(\A\).
\end{proof}

We proceed with showing an example of a VASS MDP that has no polynomial upper asymptotic for \(\F \) but has all superpolynomial functions as an upper asymptotic estimate for \(\F \).

\begin{lemma}\label{lemma-non-dag-example}\label{lemma-nondag-example}
	There exists a VASS MDP \(\A\) such that
	\begin{itemize}
		\setlength{\itemsep}{0.2pt}
		\item  \(\calL\) has no polynomial upper asymptotic estimate;
		\item every superpolynomial function is an upper asymptotic estimate of \(\calL \);
	\end{itemize} 
\end{lemma}
\begin{proof}
	
	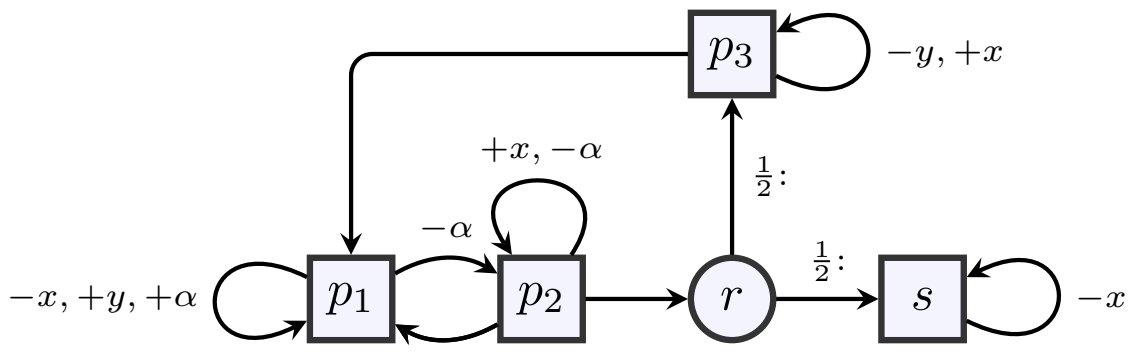
\begin{figure}[tb]
		\centering
		\lstset{basicstyle=\footnotesize}
		\begin{tikzpicture}[scale=.9, every node/.style={scale=0.9}, x=1.4cm, y=1.2cm, font=\scriptsize]
			\node [stoch] (s0) at (0,0)  {\(p_1 \)};
			\node [stoch] (s1) at (1,0)  {\(p_2 \)};
			
			\node [stoch] (d) at (2,1.5)  {\(p_3 \)};

			\node [random] (r) at (2,0)  {\(r \)};
			\node [stoch] (t) at (3,0)  {\(s\)};
			
			\draw [tran,loop left] (s0) to node[left] {$-x,+ y,+\alpha$ } (s0);   
			\draw [tran,loop above] (s1) to node[above] {$+x,-\alpha$ } (s1);
			
			\draw [tran,bend left] (s0) to node[above] {$-\alpha$ } (s1);
			
			\draw [tran,bend left] (s1) to node[left] { } (s0);
			
			\draw [tran,bend left] (s1) to node[left] { } (s0);
			
			\draw [tran] (r) to node[right] {\(\frac{1}{2} \): } (d);
			
			\draw [tran] (r) to node[above] {\(\frac{1}{2} \): } (t);
			
			\draw [tran,loop right] (t) to node[right] {$-x$ } (t);
			
			\draw [tran,loop right] (d) to node[right] {$-y,+x$ } (d);

			\draw [tran,rounded corners] (d) -- (-0,1.5) --    (s0);

			\draw [tran] (s1) to node[left] { } (r);
		\end{tikzpicture}
		\caption{A non-DAG-like VASS MDP from Lemma~\ref{lemma-non-dag-example}.}
		\label{figure-nondag-pvassexample}
	\end{figure}
	Consider the VASS MDP \(\A\) of fig.~\ref{figure-nondag-pvassexample}. Notice this VASS MDP is not DAG-like since \(p_1\) and \(p_3 \) are each contained in a different MEC yet one is reachable from the other.
	
	It is easy to see that the MECs containing \(p_3\) and \(s\) have polynomial upper asymptotic estimate as per Theorem~\ref{theorem-poly-or-exp-k}. The last MEC, which contains \(p_1\) and \(p_2\) is also polynomial as per Theorem~\ref{theorem-poly-or-exp-k} since there is no simple component containing either of the transitions \((p_1,-\alpha,p_2) \) or \((p_2,\vec{0},p_1) \), and hence the VASS MDP \(\A_1 \) computed during the analysis (see Section~\ref{sec-something-some}) contains only the self-loops over \(p_1\) and \(p_3\), which cannot be contained in any solution of \hyperref[fig-systems]{(I)} for \(\A_1 \). 	Hence \(\A_2 \) is empty which implies this MEC also has a polynomial upper asymptotic estimate from  Theorem~\ref{theorem-poly-or-exp-k}.
	
	Hence from Lemma~\ref{lemma-nondaglike-superpolynomial-pper-bound} we have that every superpolynomial function is an upper asymptotic estimate of \(\calL \).

	It remains to show that there is no upper asymptotic estimate \(n^k \) for  \(\calL \) for \( k\in \mathbb{N}\).		Assume the opposite. Then there exists \(k\in \mathbb{N}\) such that \(n^k \) is an upper asymptotic estimate of \(\calL \). 	In the following we describe the counters vectors of \(\A\) using vectors  \((a,b,c) \) where the order of the counters is \((x,y,z) \).

	Consider the following computation started in \(p_1(n,0,0)\). Repeating the loop over \(p_1\) we can reach  \(p_1(0,n,n) \). Then by going to \(p_2\) and utilizing the loop there we can reach \(p_2(n-1,n,0) \). From there, by returning to \(p_1\) and then using the loop  over \(p_1\) again we can reach \(p_1(0,n+ n-1,n-1) \). It is easy to see that by repeating this  \(n\) times we can reach the configuration \(p_2\big(0,\sum_{i=1}^n(n-i),0\big)=p_2\big(0,\frac{1}{2}\cdot n\cdot (n-1),0\big)\). Since the counters being higher can only increase \(\calL\) let us assume for simplicity we reached \(p_2\) with the counters vector being instead \( \big(0,(\frac{n}{2})^2,0\big) \geq (0,\frac{1}{2}\cdot n\cdot (n-1),0) \).
	
	Going through \(r\) we can, with probability \(\frac{1}{2}\), reach \( p_3\big(0,(\frac{n}{2})^2,0\big)\) and then by using the loop over \(p_3\) we can reach \(p_1\big((\frac{n}{2})^2,0,0\big)\). But notice we just moved from \(p_1(n,0,0) \) to \(p_1\big((\frac{n}{2})^2,0,0\big)\) with probability \(\frac{1}{2} \). Hence by repeating this procedure we can with probability \(\frac{1}{2^i} \) reach \(p_1\big((\frac{n}{2})^{2^i},0,0\big)\).
	
	But this means that there exists a strategy \(\sigma\) such that for all \(\epsilon>0 \) it holds
	\[
	\limsup_{n \to \infty} 	 \prob^{\sigma}_{p_1 \vec{n}}[\calL \geq n^{k+\varepsilon} ] 
	\geq 
	\limsup_{n \to \infty} 	 \prob^{\sigma}_{p_1 \vec{n}}[\calL \geq (\frac{n}{2})^{\lceil k+2\cdot \varepsilon\rceil} ] 
	\geq 
	\frac{1}{\lceil k+2\cdot  \varepsilon\rceil}>0
	\]	a contradiction with \(k\) being an upper asymptotic estimate of \(\calL \).
\end{proof}

		\section{Constraint Systems \hyperref[fig-systems]{(I)} and \hyperref[fig-systems]{(II)} - Details }
\label{sec-systems}
In this section we discuss in detail the properties of the Constraint Systems \hyperref[fig-systems]{(I)} and \hyperref[fig-systems]{(II)} (Fig.~\ref{fig-systems}).

\begin{figure}[H]
	\begin{tabular}{|c|c|}
		\hline
		{\begin{minipage}[c]{0.35\textwidth}\small
				\vspace{-1.26cm}
				Constraint system~(I):
				
				\vspace{0.18cm}
				Find $\bx \in \mathbb{Q}^{\transitions(\A)}$ such that\vspace{-0.245cm}
				\begin{align}
					\bx & \ge \vec{0} \nonumber\\
					\Delta(\bx)  & \ge \vec{0} \nonumber
				\end{align}
				
				\vspace{-0.2cm}and for each $\stateVass\in \states(\A)$\vspace{-0.22cm}
				\begin{align}
					\sum_{t\in \tout(\stateVass)}\bx(t)&=\sum_{t \in \tin(\stateVass)} \bx(t)\nonumber
				\end{align}
				%
				%
				%
				
				\vspace{-0.17cm}and for each $\stateVass\in \statesProb(\A)$, $t\in \tout(\stateVass)$\vspace{-0.21cm}
				\begin{align}
					\bx(t) & =P(t) \cdot \sum_{t'\in \tout(\stateVass)} \bx(t') \nonumber
				\end{align}
				\vspace{-0.05cm}
				\textbf{Objective:} \emph{Maximize}
				\begin{itemize}
					\item \vspace{-0.1cm}the number of valid inequalities of the form \vspace{-0.1cm}
					\[\sum_{t \in \transitions(\A)} \bx(t)\cdot \bu_t(c) > 0,\] 
					
					\vspace{-0.32cm}where \(c\in \countersset(\A)\).
					\item \vspace{-0.045cm}the number of valid inequalities of the form $\bx(t) > 0$.
				\end{itemize}
		\end{minipage}}
		&
		{\begin{minipage}[c]{0.62999\textwidth}\small
				\vspace{0.17cm}
				Constraint system (II):
				
				\vspace{0.18cm}
				Find $\ba \in \mathbb{Q}^{\countersset(\A)},\bb \in \mathbb{Q}^{\states(\A)}$ such that \vspace{-0.245cm}
				\begin{align}
					\ba,\bb & \ge  \vec{0} \nonumber
				\end{align}
				
				\vspace{-0.2cm}and for each $(\stateVass,\bu,\stateVassalt)\in \transitions(\A)$ where $\stateVass\in \statesDemon(\A)$ 	\vspace{-0.22cm}		
				\begin{equation}\label{eq-II-nondet-states}\bb(\stateVassalt)-\bb(\stateVass)+\sum_{c\in\countersset(\A)} \bu(c)\cdot \ba(c) \leq 0 \vspace{-0.29cm} \end{equation}
				
				\vspace{-0.17cm}and for each $\stateVass\in \statesProb(\A)$  \vspace{-0.28cm}
				\begin{multline}\label{eq-II-prob-states}
					\sum_{t= (\stateVass,\bu,\stateVassalt) \in \tout(\stateVass)}P(t)\cdot \big(\bb(\stateVassalt)-\bb(\stateVass)+
					\sum_{c\in \countersset(\A)}  \bu(c)\cdot \ba(c)\big)\leq 0 
				\end{multline}
				
				\textbf{Objective:}
				\emph{Maximize}
				\begin{itemize}
					\item \vspace{-0.15cm}the number of valid inequalities of the form $\ba(c) > 0$,
					\item\vspace{-0.155cm} the number of transitions $t=(\stateVass,\bu,\stateVassalt)\in \transitions(\A)$ such that $\stateVass\in \statesDemon(\A)$ and \vspace{-0.145cm}\vspace{-0.20cm} \[\bb(\stateVassalt)-\bb(\stateVass)+\sum_{c\in\countersset(\A)} \bu(c)\cdot \ba(c)<0,\] \vspace{-0.66cm}
					\item \vspace{-0.01cm}the number of states $\stateVass\in \statesProb(\A)$ such that \vspace{-0.5cm}\vspace{-0.13cm}
				\end{itemize}
				\begin{multline*}
					\sum_{t = (\stateVass,\bu,\stateVassalt) \in \tout(\stateVass)}P(t)\cdot \big(\bb(\stateVassalt)-\bb(\stateVass)+
					\sum_{c\in \countersset(\A)}  \bu(c)\cdot \ba(c)\big)< 0 
				\end{multline*}
				\begin{itemize}
					
					\item \vspace{-0.215cm}the number of states $\stateVass\in \statesProb(\A)$ for which  there exists a transition \((p,\bu,q)\in \tout(p) \) with\vspace{-0.28cm}
					\[\bb(\stateVassalt)-\bb(\stateVass)+\sum_{c\in\countersset(\A)} \bu(c)\cdot \ba(c)<0,\]
				\end{itemize}
		\end{minipage}}\\
		\hline
	\end{tabular}
	\caption{Constraint systems defined for a given VASS MDP $\A$.}
	\label{fig-systems}
\end{figure}

 These systems are an important  tool for VASS asymptotic analysis, including the analysis of VASS MDPs in this paper, and were first used in \cite{Zuleger:VASS-polynomial} for strongly connected VASS, and then adapted to VASS MDPs in \cite{AKCONCUR23}. Our version is the one from  from \cite{AKCONCUR23} where we add one more objective to \hyperref[fig-systems]{(II)} (the last listed objective).

Observe that both~\hyperref[fig-systems]{(I)} and~\hyperref[fig-systems]{(II)} are always satisfiable (set all coefficients to zero) and that the solutions of both constraint systems are closed under addition. Furthermore, all but the last objective of \hyperref[fig-systems]{(II)} are preserved under addition, and the last objective of  \hyperref[fig-systems]{(II)} is preserved under addition for almost all ratios as per the following lemma.

\begin{restatable}{lemma}{lemmaIIsolutionbadonlyonce}
	\label{lemma-II-solution-bad-only-once}
	Let \(p\in \statesProb(\A) \). Let \(\ba_1,\bb_1 \) and \(\ba_2,\bb_2 \) be solutions of \hyperref[fig-systems]{(II)} such that  \(\ba_1,\bb_1 \)  satisfies the last objective of \hyperref[fig-systems]{(II)} for \(p\). Then for all but possibly single \(0<k\) we have that \(\ba_1+k\cdot \ba_2,\bb_1+k\cdot \bb_2 \) is a solution of \hyperref[fig-systems]{(II)} which satisfies the last objective of \hyperref[fig-systems]{(II)} for \(p\).
\end{restatable}

The proof of Lemma~\ref{lemma-II-solution-bad-only-once} is purely technical and can be found in Section~\ref{sec-proof-of-lemma-II-solution-bad-only-once}.



Hence for both~\hyperref[fig-systems]{(I)} and~\hyperref[fig-systems]{(II)}, the set of inequalities for which the maximization objective is satisfied is the same for every \emph{maximal} solution. Furthermore, we can find a maximal solution in polynomial time using linear programming as per the following example.

\begin{example}\label{example-linear-programming-polynomial}
	Let \(X \) be one objective of \hyperref[fig-systems]{(I)} (say $X\equiv \bx(t) > 0$ for some \(t\in \transitions(\A)\)), then we can consider an extended linear system obtained by adding \(X\) to \hyperref[fig-systems]{(I)}. We can then use linear programming over rationales to decide whether this extended linear system has a rational  solution, and if it does then we can compute it in polynomial time.  Since there are only polynomially many objectives, we can do this for each single objective of \hyperref[fig-systems]{(I)} and then sum up all the solutions we obtained for the extended systems to obtain a  solution of \hyperref[fig-systems]{(I)} that is guaranteed to be maximal. The same can be done for \hyperref[fig-systems]{(II)}.
\end{example}



The key tool for the analysis is then the following dichotomy which follows from the duality of \hyperref[fig-systems]{(I)} and~\hyperref[fig-systems]{(II)}.

\begin{restatable}{lemma}{lemmadichotomy}
	\label{lemma:dichotomy}
	[Lemma~7 of \cite{AKCONCUR23}]
	Let $\bx$ be a maximal solution to the constraint system~\hyperref[fig-systems]{(I)} and $\ba,\bb$ be a maximal solution to the constraint system~\hyperref[fig-systems]{(II)}. Then, for each counter $c\in \countersset(\A)$  exactly one of these holds:	\begin{itemize}
		\item $\Delta(\bx)(c)>0$
		\item $\ba(c)>0$
		
	\end{itemize}  and for each transition $t = (\stateVass,\bu,\stateVassalt)\in \transitions(\A)$:
	\begin{itemize}
		\item[$\triangleright$] if $\stateVass\in \statesDemon(\A)$ then  exactly one of these holds
		\begin{itemize}
			\item[$\bullet$] $\bx(t)>0$ 
			\item[$\bullet$] \(\bb(\stateVassalt)-\bb(\stateVass)+\sum_{c\in\countersset(\A)} \bu(c)\cdot \ba(c)<0\)
		\end{itemize} 
		\item[$\triangleright$] if $\stateVass\in \statesProb(\A)$ then exactly one of these holds
		\begin{itemize}
			\item[$\bullet$] $\bx(t)>0$
			\item[$\bullet$] \(\sum_{t'=(\stateVass,\bu',\stateVassalt') \in \tout(\stateVass)}P(t')\cdot \big(\bb(\stateVassalt')-\bb(\stateVass)+\sum_{c\in\countersset(\A)}  \bu'(c)\cdot \ba(c)\big)< 0\)
		\end{itemize} 
	\end{itemize}
\end{restatable}

\begin{note}While our version of Lemma~\ref{lemma:dichotomy} is stronger than the version in \cite{AKCONCUR23} (as we use ``exactly one of'' instead of only ``one of''), the proof of Lemma~7 in \cite{AKCONCUR23} actually directly proves this stronger version (in fact the full version of \cite{AKCONCUR23} already uses the ``exactly one of'' in the proof in the Lemmas~16~and~17. It is just omitted in the statement of Lemma~7). 
\end{note}

\section{Pointing at Simple Components}
\label{sec-pointing}
In this section we formalize the notion of which simple component is being pointed at a given step (see also  Section~\ref{sec-det-prob-decompositions}).

We begin by recalling the pointing decomposition from \cite{AKCONCUR23}.


A \emph{path} in \(\A \) is a sequence \(p_0,\bu_1,p_1,\dots,p_k  \) where \(p_i\in \states(\A) \) and \((p_{i},\bu_{i+1},p_{i+1})\in \transitions(\A) \) for all \(i\). Each path, together with an initial counters vector \(\bv \) define uniquely a corresponding computation, and each computation defines uniquely a corresponding path in the obvious way. 

Given two paths \(\alpha=p_0,\bu_1,p_1,\dots,p_k \) and \(\beta=q_0,\bv_1,q_1,\dots,q_\ell \) we define \[\alpha \circ \beta= p_0,\bu_1,p_1,\dots,p_k,\bv_1,q_1,\bv_2,q_2,\dots,q_\ell \] if \(\alpha\neq \epsilon \), and \(\alpha\circ\beta=\beta \) if \(\alpha=\epsilon \). Note that if \(p_k=q_0\) then \(\alpha\circ \beta \) is a path. We say that a path \(\alpha=p_0,\dots,p_k \) is \emph{compatible} with a strategy \(\sigma \) if \(\prob_{p_0}^\sigma(\alpha)>0 \). Given a path \(\alpha=p_1,\dots,p_k \), we use \(\length(\alpha)=k \) to denote the length of \(\alpha\), and \(\alpha_{..i}=p_0,\dots,p_i \) to denote the prefix of \(\alpha \) of length \(i\).

\begin{definition}
	Let $\A$ be a VASS MDP, $\pi_1,\dots, \pi_k \in \cMDstrats{\A}$, and $p_1,\dots,p_k \in \states(\A)$. A \mbox{\emph{cMD-decomposition}} of a path $\alpha = s_1,\ldots,s_m$ under  
	$\pi_1,\dots, \pi_k$ and $p_1,\dots,p_k$ is a decomposition of $\alpha$ into finitely many paths $\alpha = \gamma_1^1 \circ \cdots\circ\gamma_1^k \; \circ\; \gamma_2^1\circ\cdots\circ\gamma_2^k \; \circ \; \cdots \; \circ \; \gamma_\ell^1\circ\cdots\circ\gamma_\ell^k$ satisfying the following conditions:
	\begin{itemize}
		\item for all $i < \ell$ and $j\leq k$, the last state of $\gamma_i^j$ is the same as the first state of $\gamma_{i+1}^j$;
		\item for every \(j\leq k \), $\gamma_1^j \circ \cdots \circ \gamma_\ell^j$ is a path that begins with \(p_j \) and is compatible with $\pi_j$.   
	\end{itemize} 
\end{definition}
Note that $\pi_1,\dots, \pi_k$ and $p_1,\dots,p_k$ are not necessarily pairwise different, and the length of $\gamma_{i}^j$ can be zero. Also note that the same $\alpha$ may have several cMD-decompositions. 

The next lemma shows that for every VASS MDP $\A$, one can \emph{fix} cMD strategies $\pi_1,\ldots,\pi_k$ and states $p_1,\ldots,p_k$ such that \emph{every} path $\alpha$ in $\A$ has a cMD-decomposition under $\pi_1,\ldots,\pi_k$ and $p_1,\ldots,p_k$. Furthermore, such a decomposition is constructible \emph{online} as $\alpha$ is read from left to right.

\begin{lemma}\label{lemma-MD-decompose}
	[Lemma~14 of \cite{AKCONCUR23}]
	For every VASS MDP \(\A\), there exist \(\pi_1,\dots,\pi_k \in \cMDstrats{\A}\),   \(p_1,\dots,p_k \in \states(\A)\), and a function \(\Decompose_\A \) such that the following conditions are satisfied for every finite path $\alpha$:
	\begin{itemize}
		\item \(\Decompose_\A(\alpha) \) returns a cMD-decomposition of $\alpha$ under $\pi_1,\dots,\pi_k$ and $p_1,\ldots,p_k$;
		\item $\Decompose_\A(\alpha)=\Decompose_\A(\alpha_{..\length(\alpha)-1})\; \circ\; \gamma^1\circ\cdots\circ\gamma^k$, where exactly one of 
		$\gamma^i$ has positive length (the $i$ is called the \emph{\(\Decompose_\A\)-mode} of $\alpha$);
		\item If the last state of \(\alpha_{..\length(\alpha)-1}\) is probabilistic, then the \(\Decompose_\A\)-mode of $\alpha$ does not depend on the last transition of $\alpha$. 
	\end{itemize}
\end{lemma}

Lemma~\ref{lemma-MD-decompose} effectively allows us to assign to each computational step in \(\A\) a cMD strategy/state  pair \((\sigma,p)\)  in such a way that we can see the computation on \(\A\) as simply an interweaving of the individual computations on \(\A_\sigma\). Hence we can see a strategy in \(\A\) as if, instead of choosing the next transition in non-deterministic states, it were \emph{pointing} at which cMD strategy/state  pair to follow next.

An important observation is that the computation for a given cMD strategy/state  pair \((\sigma,p)\) is a computation on a VASS Markov chain \(\A_\sigma\), initiated in \(p\). And each such Markov chain reaches a MEC in constant expected time. Furthermore, each MEC of \(\A_\sigma \) can be seen a simple component (say centered in the first state of the given MEC reached). Hence towards the asymptotic analysis of VASS MDPs, we can effectively view each of the cMD strategy/state  pairs as if doing nothing but iterating some simple component of \(\A \).  This allows us to  effectively view any computation on any VASS MDP as an interweaving of ``repeated iterations of simple components'' (plus some asymptotically irrelevant part).

Let $\alpha = \gamma_1^1 \circ \gamma_1^2\circ\ldots\circ\gamma_1^k\circ\gamma_2^1\circ\ldots\circ\gamma_2^k\circ\gamma_3^1\circ\dots\circ\gamma_\ell^k$ be the cMD-decomposition of \(\alpha \) produced by Lemma~\ref{lemma-MD-decompose}. For every \(j\leq k \), we define the \emph{pointing tuple}  \(\pointpair_j \) as either \(\emptypointpair \) if  $\beta_j=\gamma_1^j \circ \cdots \circ \gamma_\ell^j$ did not yet reach a MEC of \(\A_{\pi_j} \) , or  \(\pointpair_j=(\by_j,q_j,j) \) where \(q_j \) is the first state along \(\beta_j\) in a MEC of \(\A_{\pi_j} \) and \(\by_j \) is the simple component of \(\A_{\pi_j} \) centered in \(q_j\). Note that \(\by_j \) is defined uniquely as per Lemma~\ref{lemma-simple-components-unique-per-cetnering}.

We define \(\DecomposeComponent_\A(\alpha)=\pointpair_j  \) where \( j\) is the \(\Decompose_\A(\alpha)\)-mode of \(\alpha \). We call \(\DecomposeComponent_\A(\alpha) \) the \emph{pointing tuple} pointed at in the last step of \(\alpha \), and we say that the last step of \(\alpha \) \emph{pointed at} \(\DecomposeComponent_\A(\alpha)\). Additionally, if \(\DecomposeComponent_\A(\alpha)=(\by,p,i)\) we say the simple component pointed at in the last step of \(\alpha \) is \(\by \).

In the following Lemma we show that the total effect of all the computational steps which point at \(\emptypointpair \)  is asymptotically negligible.
\begin{restatable}{lemma}{lemmanonpointersirrelevant}
	\label{lemma-non-pointers-irrelevant}
	Let \(f\in \omega(1) \) be an increasing unbounded function, and let \(\bv\) be the sum of the effects of all computational steps in \(\A\) for which \(\DecomposeComponent_\A(\alpha)=\emptypointpair\). Then for each \(p\in\states(\A)\), and strategy \(\sigma\)     it holds \(\liminf_{n\rightarrow\infty}\prob_{p\vec{n}}^\sigma[\max_{c\in\countersset(\A)}\bv(c)<f(n)]=1 \).  
\end{restatable}
\begin{proof}
	This follows from the probability of not reaching a MEC within \(x\) steps decreases exponentially fast in \(x\).
	
	As per Lemma~\ref{lemma-MD-decompose}  the function \(\Decompose_\A(\alpha)  \) can return only \(k\) many different cMD strategy/state pairs \((\pi,p) \), where \(k\) depends only on \(\A \) (and hence is a constant in \(n\)). Furthermore, there exists a probability \(p>0 \) that depends only on \(\A \) such that every at most \(\states(\A) \) steps any computation in any \(\A_\pi \) reaches a MEC. Hence the probability of not having reached a MEC within \(i \) steps in \(\A_\pi \) is at most \[(1-p)^{\lfloor\frac{i}{\states(\A)}\rfloor} \] Effect of each transition is likewise bounded by a constant \(u \). Hence if the computation in each \(\A_\pi \) has length at most \(\frac{f(n)}{k\cdot u} \) then this implies that \(|\bv(c)|<k\cdot u\cdot \frac{f(n)}{k\cdot u}= f(n)  \). But the probability of any such computation being longer than \(\frac{f(n)}{k\cdot u}\) is upper bounded by \[k\cdot (1-p)^{\lfloor\frac{f(n)}{k\cdot u\cdot \states(\A)}\rfloor} \] which converges to \(0\) as \(n\) tends towards \(\infty\). Lemma holds.  
\end{proof}

\section{Conclusion}

We have also shown that for VASS MDPs either every superpolynomial function is an upper asymptotic estimate of termination/\(c\)-counter complexity or the complexity has \(2^{\sqrt{n}} \) as a lower asymptotic estimate,and it is decidable in polynomial time in the size of the VASS MDP which case holds. We have also shown that the former case does not necessarily imply the existence of a polynomial upper asymptotic estimate unless the VASS MDP is DAG-like. For DAG-like VASS MDPs we have shown that the former case does imply a polynomial upper asymptotic estimate for the termination/\(c\)-counter complexity. In doing these we have formalized the concepts of multi-components, deterministic/probabilistic decomposition, and we have shown surprising results regarding the geometry of VASS MDPs.

Additionally we have shown an alternative definition of asymptotic estimates using the natural notion of failure-time function.

The analogous results for VASS from \cite{Leroux:Polynomial-termination-VASS} have been extended in \cite{Zuleger:VASS-polynomial} by showing that in the polynomial case one can compute an exact degree of the polynomial that gives a tight asymptotic bound. A natural next step would thus be to see whether one could similarly compute in the polynomial case for DAG-like VASS MDPs, an exact degree of a polynomial that gives a tight asymptotic estimate for the termination/\(c\)-counter complexity. Another direction would be to further study asymptotic estimates in respect to their close relation to the failure-time function to see where exactly this type of asymptotic analysis of failure-time function would be applicable.

\section{Conclusion}

We have shown that for VASS MDPs, either every superpolynomial function is an upper asymptotic estimate of the termination/\(c\)-counter complexity, or this complexity admits \(2^{\sqrt{n}}\) as a lower asymptotic estimate, and that it is decidable in polynomial time in the size of the VASS MDP which of these two cases holds. We have further shown that the former case does not necessarily imply the existence of a polynomial upper asymptotic estimate, unless the VASS MDP is DAG-like, in which case it does. Along the way, we formalized the notions of multi-components, introduced a technique of deterministic/probabilistic decomposition, and established surprising results on the geometry of VASS MDPs. We have also given an alternative characterization of asymptotic estimates in terms of the natural notion of a failure-time function.

The analogous results for VASS from \cite{Leroux:Polynomial-termination-VASS} have been extended in \cite{Zuleger:VASS-polynomial} by showing that in the polynomial case one can compute an exact degree of the polynomial that gives a tight asymptotic bound. A natural next step would thus be to see whether one could similarly compute in the polynomial case for DAG-like VASS MDPs, an exact degree of a polynomial that gives a tight asymptotic estimate for the termination/\(c\)-counter complexity. Another direction would be to further study asymptotic estimates in respect to their close relation to the failure-time function to see where exactly this type of asymptotic analysis of failure-time function would be applicable.

	
	\bibliography{citate,used,cituse}
	
	\appendix

\section{Asymptotic Estimates - Basic Properties}
\label{sec-asest-basic-prop}
In this section, we list some properties and give examples of asymptotic estimates, defined in Section~\ref{sec-comp-VASS-runs}.

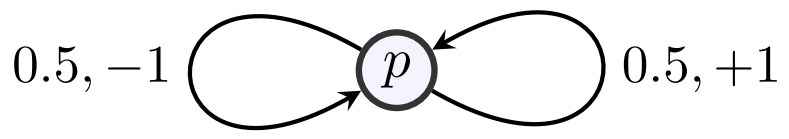
\begin{figure}[h]
	\parbox[c]{.5\textwidth}{
		\begin{tabbing}
			\hspace*{1em} \= \hspace*{2em} \= \hspace*{4em} \= \hspace*{4em} \= \kill
			\> \textbf{input} $N$\\
			\> \textbf{repeat}\\
			\>\> \textbf{random choice:}\\
			\>\>\> $0.5:\ \ N:=N+1;$\\
			\>\>\> $0.5:\ \ N:=N-1;$\\
			\> \textbf{until} $N=0$
	\end{tabbing}}\hspace*{4em}
	\parbox[c]{.5\textwidth}{
		\begin{tikzpicture}[x=2cm, y=2cm]
			\node [stoch,label={[shift={(0,-1)}]$\A$}] (r) at (3,0)  {$p$};
			\draw [tran,loop right] (r) to node[right] {$0.5, +1$} (r); 
			\draw [tran,loop left]  (r) to node[left] {$0.5, -1$} (r); 
		\end{tikzpicture}
	}
	\caption{A probabilistic program with infinite expected running time for every $N\geq 1$, and its $1$-dimensional VASS MDP model~$\A$.}
	\label{fig-prob-prg}
\end{figure}

Let us start with a simple motivating example from \cite{AKCONCUR23}. Consider the simple probabilistic program of Fig.~\ref{fig-prob-prg}. The program inputs a positive integer $N$ and then repeatedly increments/decrements~$N$ with probability $0.5$ until $N=0$. One can easily show that for every $N \geq 1$, the program terminates with probability one, and the expected termination time is \emph{infinite}. Based on this, one may conclude that the execution takes a very long time, independently of the initial value of~$N$. However, this conclusion is \emph{not} consistent with practical experience gained from trial runs (see e.g., \cite{Ajdarow2025}). The program tends to terminate ``relatively quickly'' for small~$N$, and the termination time \emph{does} depend on~$N$. 
Hence, the function assigning $\infty$ to every $N \geq 1$ is \emph{not} a faithful characterization of the asymptotic growth of termination time.  Asymptotic estimates are motivated  by the following observations:
\begin{itemize}
	\item For every $\varepsilon >0$, the probability of all runs terminating after more than $n^{2+\varepsilon}$ steps (where $n$ is the initial value of $N$) approaches \emph{zero} as $n \to \infty$.
	\item For every $\varepsilon >0$, the probability of all runs terminating after more than $n^{2-\varepsilon}$ steps (where $n$ is the initial value of $N$) approaches \emph{one} as $n \to \infty$.
\end{itemize}
Since the execution time is ``squeezed'' between $n^{2-\varepsilon}$ and $n^{2+\varepsilon}$ for an arbitrarily small $\varepsilon > 0$ as $n \to \infty$, it can be characterized as ``asymptotically quadratic''. This analysis is in accordance with experimental outcomes. One can easily show that this program indeed has a tight asymptotic estimate \(n^2 \) for its running time.

Other basic properties of asymptotic estimates include: \begin{itemize}
	\item having a finite upper asymptotic estimate is equivalent to almost-sure finiteness \cite{Ajdarow2025};
	\item asymptotic estimates are compositional (unlike the asymptotics of the expectation) \cite{Ajdarow2025};
	\item upper bounds on the expectation carry over to asymptotic estimates \cite{AKCONCUR23} (i.e., if \(\E[\F]\in \calO(f) \) then \(f\) is an upper asymptotic estimate of \(\F \));
	\item lower asymptotic estimates carry over to the expectation \cite{AKCONCUR23}.
\end{itemize}

We refer to \cite{Ajdarow2025} for more details regarding asymptotic estimates.

\section{Multi-Components - Intuition}
\label{sec-multi-components-intuition}
Several of the techniques used in this paper require us to extend the notions of multi-cycles and simple cycles from non-probabilistic systems (such as VASS) onto probabilistic systems (such as VASS MDP) in a way that captures the same notions about strategies. While the mathematical objects of a multi-cycle and a simple cycle  is properly defined also for VASS MDPs, it no longer necessarily  corresponds to strategies in VASS MDPs in same way as in VASS, as illustrated in the following example.

\begin{figure}[H]
	\centering
	\begin{tikzpicture}[scale=.85, every node/.style={scale=0.85}, x=2.1cm, y=2.1cm, font=\footnotesize]
		\node[demonic] (s1) at (0,0) {$\stateVass$};
		\node[demonic] (s2) at (1,0.5) {$\stateVass_1$};
		\node[demonic] (s3) at (1,-0.5) {$\stateVass_2$};

		\draw [tran, bend left] (s1) to  (s2);
		\draw [tran, bend right] (s1) to  (s3);
		\draw [tran] (s2) to (s1);
		\draw [tran] (s3) to (s1);
		
%
		
		\begin{scope}[shift={(2.5,0)}]
				\node[random] (s1) at (0,0) {$\stateVass$};
			\node[demonic] (s2) at (1,0.5) {$\stateVass_1$};
			\node[demonic] (s3) at (1,-0.5) {$\stateVass_2$};

			\draw [tran, bend left] (s1)  to node[above] {\parbox{5em}{\centering \(\frac{1}{2}:\)}} (s2);
			\draw [tran, bend right] (s1) to node[below] {\parbox{5em}{\centering \(\frac{1}{2}:\)}}   (s3);
			\draw [tran] (s2) to (s1);
			\draw [tran] (s3) to (s1);
		\end{scope}
		
%
	\end{tikzpicture}
	\caption{An example of a demonic VASS \(\A_1\) (left) and a demonic pVASS \(\A_2\) (right). }
	\label{figure-fdsfsdvass-from-theorem-expectation-vassgame-locking-not-enofdfsfdsfsdugh}
\end{figure}
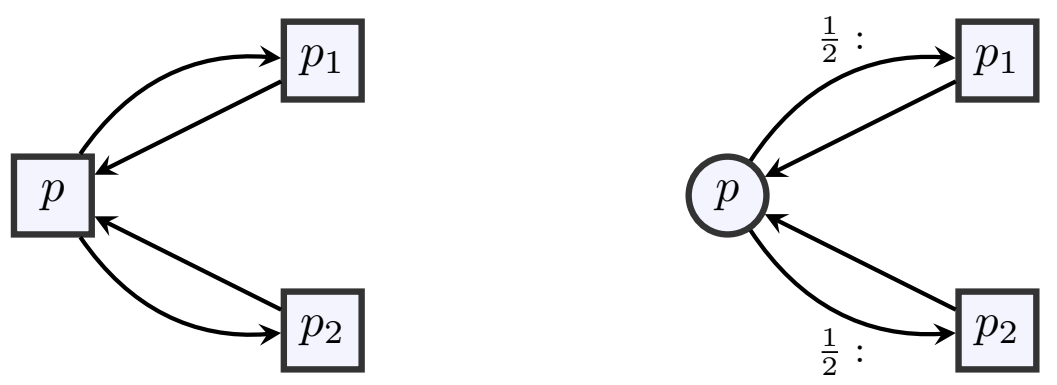

\begin{example}\label{examosdsd}
	Consider the VASS \(\A_1\) of Fig.~\ref{figure-fdsfsdvass-from-theorem-expectation-vassgame-locking-not-enofdfsfdsfsdugh} (left) and the VASS MDP \(\A_2\) of Fig.~\ref{figure-fdsfsdvass-from-theorem-expectation-vassgame-locking-not-enofdfsfdsfsdugh} (right). Note that \(\pi=p,(p,\vec{0},p_1),p_1,(p_1,\vec{0},p),p \) is a simple cycle in both, however only in \(\A_1 \) does there exist a strategy that iterates \(\pi\), as any strategy in \(\A_2\) will with probability \(1\) eventually reach \(p_2\) which does not lie on \(\pi \).	
\end{example}

In non-probabilistic setting, a multi-cycle can be seen as a conical sum of simple cycles, and each simple cycle in a VASS \(\A \) can be ``equated'' with a MEC of \(\A_\strategyDemon \) for some cMD strategy \(\strategyDemon \). Hence in non-probabilistic setting a multi-cycle can be seen as a conical sum of MECs of various \(\A_\strategyDemon \) with \(\strategyDemon\in \cMDstrats{\A} \). In probabilistic setting this is no longer true, as in VASS MDPs simple cycles no longer necessarily correspond to MECs of \(\A_\strategyDemon \) with \(\strategyDemon\in \cMDstrats{\A} \) (see Example~\ref{examosdsd}). Hence we need to consider a different object, one that corresponds to conical sums of MECs of \(\A_\strategyDemon \) with \(\strategyDemon\in \cMDstrats{\A} \). As such MECs can be seen as the most elementary component of any strategy (one could even call it a \emph{simple component}) we call the conical combination of these ``simple components''  a \emph{multi-component}. For our purposes, representing multi-components using their Parikh images will prove highly convenient. While the Parikh image of a multi-cycle is well defined, it is a weighted sum of the Parikh images of the individual simple cycles (and a Parikh image of a simple cycle is a vector \(\by\in \{0,1\}^{\transitions(\A)} \) such that \(\by(t)=1 \) if \(t\) lies on the simple cycle and \(\by(t)=0 \) otherwise), it is not immediately clear how to define a Parikh image of a multi-component.  Using classical results about Parikh images, one can define a multi-cycle of \(\A\) as a conical sum of Parikh images of simple cycles, i.e., as a solution to the following system of linear equations	\begin{center}\begin{tabular}{|cc|}
		\hline
		{\begin{minipage}[c]{0.38\textwidth}\small
		\vspace{0.2cm}
				$\bx \in \mathbb{Q}^{\transitions(\A)}$ such that
				\begin{align}
					\bx & \ge \vec{0} \nonumber
				\end{align}
				and for all $\stateVass\in \states(\A)$
				\begin{align}
					\sum_{t\in \tout(\stateVass)}\bx(t)&=\sum_{t \in \tin(\stateVass)} \bx(t)\nonumber
				\end{align}
				%
				%
				%
			
		\end{minipage}}&\\\hline
	\end{tabular}\end{center}

As per this definition a multi-cycle \(\bx \) can be interpreted as assigning to each transition of \(\A\) a non-negative flow such that for each state what flows in also flows out. 

In VASS MDPs the outgoing flow from probabilistic states is controlled probabilistically, so in multi-components, for each probabilistic state \(\stateVass\) the flow distribution on \(\tout(p) \) must have the same ratios as the probability distribution on \(\tout(p) \) given by \(P\). This gives us the following definition of a multi-component.

\begin{definition}\label{def-multicomponent}
	Let \(\A \) be a VASS MDP. A \emph{multi-component} of \(\A \) is a solution to the following system of linear equations	\begin{center}\begin{tabular}{|cc|}
			
			\hline
			
			{\begin{minipage}[c]{0.42\textwidth}\small
					
					\vspace{0.2cm}
					
					\vspace{0.2cm}
					$\bx \in \mathbb{Q}^{\transitions(\A)}$ such that
					\begin{align}
						\bx & \ge \vec{0} \nonumber
					\end{align}
				and	for all $\stateVass\in \states(\A)$
					\begin{align}
						\sum_{t\in \tout(\stateVass)}\bx(t)&=\sum_{t \in \tin(\stateVass)} \bx(t)\nonumber
					\end{align}
					%
					%
					%
					and for all $\stateVass\in \states_p(\A)$, $t\in \tout(\stateVass)$
									\begin{align}
											\bx(t) & =P(t) \cdot \sum_{t'\in \tout(\stateVass)} \bx(t') \nonumber
										\end{align}
					
			\end{minipage}}&\\\hline
	\end{tabular}\end{center}

\end{definition}

Note that this is exactly the definition of a multi-component used in Section~\ref{sec-multicomponents}. Note that every multi-component is also a multi-cycle, but not the other way around. 

\begin{figure}[H]
	\centering
	\begin{tikzpicture}[scale=.85, every node/.style={scale=0.85}, x=2.1cm, y=2.1cm, font=\footnotesize]
			
			
			\node[] (X) at (1cm,-3.1cm-1em) {$\A$};
			\node[random] (t1) at (0,0) {$\stateVass_1$};
			\node[stoch] (t2) [below of=t1,node distance=1.9cm] {$\stateVass_2$};
			\node[stoch] (t3) [right of=t1,node distance=1.9cm] {$\stateVass_3$};
			\node[stoch] (t4) [below of=t3,node distance=1.9cm] {$\stateVass_4$};

			\node[] () at (-0.94cm,0.75cm) {{\parbox{5em}{\centering $\frac{1}{5}:$ {\boldmath$1$} }}};
			\draw [tran,loop left] (t1) to (t1);

			\node[] () at (0.9cm,0.75cm) {{\parbox{5em}{\centering $\frac{1}{10}:${\boldmath$\frac{1}{2}$}}}};
			\draw [tran,loop right] (t1) to (t1);
			
			\draw [tran,loop left] (t2) to node[left] {{\parbox{2em}{\centering{\boldmath$3$}}}} (t2);
			
			\draw [tran,loop right] (t3) to node[right] {{\parbox{2em}{\centering{\boldmath$\frac{1}{2}$}}}} (t3);
			
			\draw [tran,loop right] (t4) to node[right] {{\parbox{2em}{\centering {\boldmath$0$}}}} (t4);

			\draw [tran,bend right] (t1) to node[ left] {\parbox{3.48em}{\centering \(\frac{7}{10}:\){\boldmath$\frac{35}{10}$}}} (t2);
			\draw [tran, bend right] (t2) to node[ right] {\parbox{1em}{{\boldmath$5$}}} (t1);
			
			\draw [tran, bend right] (t3)  to node[ left] {\parbox{1em}{{\boldmath$\frac{2}{3}$}}} (t4);
			\draw [tran, bend right] (t4) to node[ right] {\parbox{1em}{{\boldmath$\frac{2}{3}$}}} (t3);
			
			\draw [tran, bend right] (t2) to node[below] {\parbox{1em}{{\boldmath$0$}}} (t4);
			
			\draw [tran, bend right] (t4) to node[above] {\parbox{1em}{{\boldmath$0$}}} (t2);

	\end{tikzpicture}
	\caption{An example of a multi-component \(\bx \). The bold values next to transitions \(t\) denote the value \(\bx(t) \).  }
	\label{fidssgure-fdsfsdvass-from-theorem-expectation-vassgame-locking-not-enofdfsfdsfsdugh}
\end{figure}
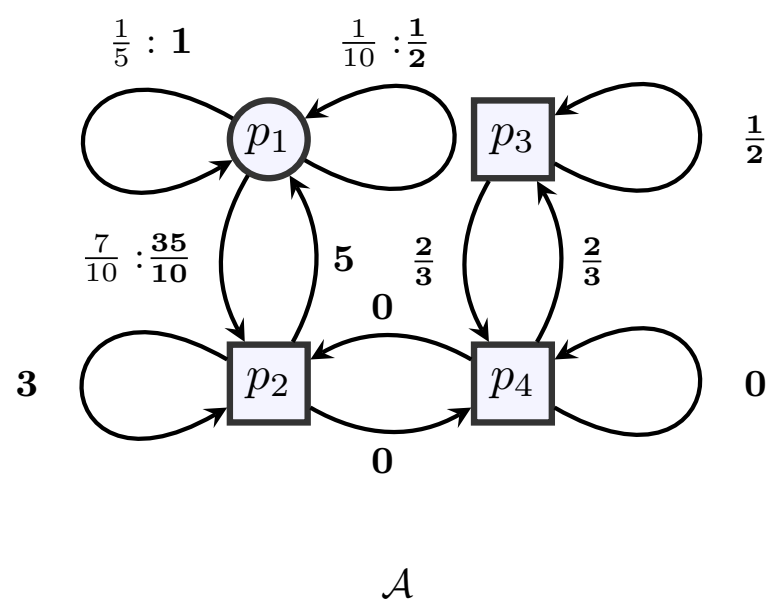

The next two technical Lemmas~\ref{lemma-component-recentering}~and~\ref{lemma-simple-components-unique-per-cetnering} show that simple components correspond to MECs of \(\A_\sigma \) for \(\sigma\in \cMD(\A) \) (up to a multiplication by positive constant). Exactly as we wanted. Note also that multi-components are closed under addition, multiplication by non-negative scalar, and subtraction assuming the result is non-negative.


\begin{restatable}{lemma}{lemmacomponentrecentering}
	\label{lemma-component-recentering}
	Let \(\by \) be a simple component of \(\A \) centered in \(\stateVass\). Let \(\stateVassalt \) be a state of \(\A_\by \). Then there exists \(a>0 \) such that \(a\cdot \by \) is a simple component of \(\A \) centered in \(\stateVassalt \).
\end{restatable} 
\begin{proof}\label{proof-lemma-component-recentering}
	Let \begin{equation*}
		a=\frac{1}{\sum_{t \in \tout(q)} \by(t)}
	\end{equation*} Note that \(a>0\) as \(\sum_{t \in \tout(q)} \by(t)>0 \) from the definitions of \(\by \) and \(q\).  It holds \[\sum_{t \in \tout(q)} a\cdot \by(t) =a\cdot \sum_{t \in \tout(q)}  \by(t) =\frac{1}{\sum_{t \in \tout(q)} \by(t)}\cdot \sum_{t \in \tout(q)} \by(t)=1 \] Hence \(a\cdot \by \) is 
	centered in \(q \). Since \(\by(t)>0 \) iff \(a\cdot \by(t)>0 \) we have   \(\A_\by=\A_{a\cdot \by} \). Therefore \(a\cdot \by \) is a simple component of \(\A\) centered in \(q\).
\end{proof}

\begin{restatable}{lemma}{lemmasimplecomponentsuniquepercetnering}
	\label{lemma-simple-components-unique-per-cetnering}
	Let \(\by_1\) and \(\by_2 \) be simple components both centered in \(\stateVass\) and such that \(\A_{\by_1}=\A_{\by_2}\). Then it holds \(\by_1=\by_2 \).
\end{restatable}
\begin{proof}\label{proof-lemma-simple-components-unique-per-cetnering}
	For \(t\notin \transitions(\A_{\by_1}) \) it holds \(\by_1(t)=0\) by the definition of \(\A_{\by_1} \) and \(\by_2(t)=0\) by our assumption on \(\A_{\by_2} \).
	
	Let 	\[b_1=\frac{1}{\sum_{t\in \transitions(\A)} \by_1(t)} \] \[b_2=\frac{1}{\sum_{t\in \transitions(\A)} \by_2(t)} \]  and  \[\overline{\by}_1=b_1\cdot \by_1\] \[\overline{\by}_2=b_2\cdot \by_2\] be the normalized vectors of \(\by_1 \) and \(\by_2 \).
	
	Since \(\A_{\by_1}=\A_{\by_2} \) is an irreducible VASS Markov chain it contains a single unique invariant distribution, hence from  Lemma~\ref{app-lemma-solution-I-startegy} it holds for each state \(q\in \states(\A)\) that \[\bp_1(q)=\sum_{t\in \tout(q)} \overline{\by}_1(t)=\sum_{t\in \tout(q)} \overline{\by}_2(t)=\bp_2(q) \] 
	
	Hence for each \(q\in \statesDemon(\A) \), since there exists only single \(t\in\{t\in \transitions(\A) \mid\by_1(t)>0 \}\cap \tout(p)   \) it holds for each \(t\in \tout(q)\) that \( \bar{\by}_1(t) = \bar{\by}_2(t) \). 
	
	And for each \(q\in \statesProb(\A) \) it holds \[\sum_{t\in \tout(q)} P(t)\cdot \sum_{t'\in \tout(q)}\bar{\by}_1=\sum_{t\in \tout(q)} \bar{\by}_1 = \sum_{t\in \tout(q)} \bar{\by}_2 = \sum_{t\in \tout(q)} P(t)\cdot \sum_{t'\in \tout(q)}\bar{\by}_2 \] which gives \(\sum_{t\in \tout(q)}\bar{\by}_1=\sum_{t\in \tout(q)}\bar{\by}_2 \). Remember also that for each \(t\in \tout(q)\) it holds \[\bar{\by}_1(t)=P(t)\cdot \sum_{t\in \tout(q)}\bar{\by}_1 = P(t)\cdot\sum_{t\in \tout(q)}\bar{\by}_2 = \bar{\by}_2(t) \]
	
	Hence it holds \(\bar{\by}_1=\bar{\by}_2 \) which means that \[b_1\cdot \by_1=\bar{\by}_1=\bar{\by}_2=b_2\cdot \by_2 \]
	and hence \[\by_1=\frac{b_2}{b_1}\cdot \by_2 \]
	
	Since both \(\by_1 \) and \(\by_2 \) are centered in \(p\) it therefore holds
	\[1=\sum_{t\in\tout(p)}\by_1(t)=\sum_{t\in\tout(p)}\frac{b_2}{b_1}\cdot \by_2(t)=\frac{b_2}{b_1}\cdot 1 \] which gives \(b_1=b_2=1 \). Hence \(\by_1 =\by_2 \).
\end{proof}

\section{Full Proofs}

This section contains all the missing proofs from the rest of the paper.

\subsection{Proof of Theorem~\ref{observation-f-p-f}}
\label{app-observation-f-p-f}

\observationfpf*


The \(\Rightarrow \) directions follow directly from \cite{Ajdarow2025} (Lemmas~1~and~2). The remaining \(\Leftarrow \) directions then follow from the following two Lemmas, each proving the \(\Leftarrow \) direction for one of the bullet points.

\begin{lemma}
	Let \(f:\mathbb{R}\rightarrow\mathbb{R} \) be such that \(\lim_{n\rightarrow\infty} \frac{f( n)}{f( n^{1+\epsilon})}=0\) for every \(\epsilon>0 \).
	If for every \(\epsilon>0 \) and every \(p<1 \) it holds \(f_p^\F\in \bigO\big(f(n^{1+\epsilon})\big) \) then \(f\) is an upper asymptotic estimate of \(\F \).
\end{lemma}

\begin{proof}
	For each \(p<1 \) and each \(\epsilon>0 \) there exists  \(a_{p,\epsilon}\in \mathbb{N} \) and \(n_{p,\epsilon}\in \mathbb{N}\) such that  \[f_p^\F(n)\leq a_{p,\epsilon} \cdot f( n^{1+\epsilon} )\leq \cdot f( n^{1+2\cdot \epsilon} )\] for all \(n\geq n_{p,\epsilon} \). Hence for all \(n\geq n_{p,\epsilon} \), every $q \in \states(\A)$, and every strategy $\sigma$ it holds \[\Prob_{q\vec{n}}^\sigma[\F\leq  f( n^{1+2\epsilon} )]  \geq\Prob_{q\vec{n}}^\sigma[\F\leq  a_{p,\epsilon}\cdot f( n^{1+\epsilon} )] \geq \Prob_{q\vec{n}}^\sigma[\F\leq  f_p^\F(n)]\geq  p\]
	
	Hence 
	\begin{gather*}	
		\limsup_{n\rightarrow\infty} \Prob_{q\vec{n}}^\sigma[\F\geq f( n^{1+3\cdot\epsilon} )] 
		= \limsup_{n\rightarrow\infty} \big(1-\Prob_{q\vec{n}}^\sigma[\F< f( n^{1+3\epsilon} )]\big) 
		\leq\\\leq
		\limsup_{n\rightarrow\infty}\big( 1-\Prob_{q\vec{n}}^\sigma[\F\leq f( n^{1+2\epsilon} )]\big)
		\leq  \limsup_{n\rightarrow\infty} \big(1-\max \{p<1 \mid n> n_{p,\epsilon} \}\big)=0 \end{gather*}
	
	Hence \(f\) is an upper asymptotic estimate of \(\F \).
\end{proof}

\begin{lemma}
	
	Let \(f:\mathbb{R}\rightarrow\mathbb{R} \) be such that \(\lim_{n\rightarrow\infty} \frac{f( n)}{f( n^{1+\epsilon})}=0\) for every \(\epsilon>0 \).
	If for every \(\epsilon>0 \) and every \(p<1 \) it holds \(f_p^\F\in \Omega\big(f(n^{1-\epsilon})\big) \) then \(f\) is a lower asymptotic estimate of \(\F \).
	
\end{lemma}

\begin{proof}
	For each \(p<1 \) and each \(\epsilon>0 \) there exists  \(a_{p,\epsilon}\in \mathbb{N} \) and \(n_{p,\epsilon}\in \mathbb{N}\) such that \[f_p^\F(n)\geq a_{p,\epsilon}\cdot  f( n^{1-\epsilon} )> f( n^{1-2\cdot \epsilon} )\] for all \(n\geq n_{p,\epsilon} \). 	Hence for each \(p<1\) and each \(\epsilon>0 \) it holds for each $q \in \states(\A)$ and each strategy $\sigma$ \[ \Prob_{q\vec{n}}^\sigma[\F\geq  f( n^{1-2\cdot \epsilon} )]\geq \Prob_{q\vec{n}}^\sigma[\F> a_{p,\epsilon} \cdot f( n^{1-\epsilon} )]\geq \Prob_{q\vec{n}}^\sigma[\F>  f^\F_p(n)]=1-\Prob_{q\vec{n}}^\sigma[\F\leq   f^\F_p(n)] \geq 1-p \]  for all \(n\geq n_{p,\epsilon}\).  Hence
	\begin{gather*}
		\liminf_{n\rightarrow\infty} \Prob_{q\vec{n}}^\sigma[\F\geq f( n^{1-2\cdot \epsilon} )]
		\geq 
		\liminf_{n\rightarrow\infty} \big( 1-\min \{p<1\mid n> n_{p,\epsilon} \}\big) = 1
	\end{gather*}
	\(f\) is a lower asymptotic estimate of \(\F \).
\end{proof}

\subsection{Proof of Lemma~\ref{lemma-polynomial-upper}}
\label{sec-lemma-polynomial-upper-proof}\label{sec-polynomial-upper}
\lemmapolynomialupper*

%
\begin{proof}
	Assume towards contradiction this does not hold. Let \(\variablenotk\in \mathbb{N}_0 \) be the smallest index  for which there exists \(x\in C_{\variablenotk+1} \) (or \(x\in T_{\variablenotk+1} \)) such that \(\calC[x]\) (or \(\calT[x] \)) does not have an upper asymptotic estimate \(n^{3^{\variablenotk}}  \) (or \(n^{2\cdot 3^{\variablenotk}}  \)).
	

	\textbf{Case \(\variablenotk=0 \):} If \(x\) is a counter then as per the definition of \(C_1 \)  it holds \(\Delta(\bx_0)(x)=0 \). Therefore from Lemma~\ref{lemma:dichotomy} it holds \(\ba_0(x)>0 \). Hence we obtain an upper asymptotic estimate \(n=n^{3^0}\) for \(\calC[x] \) from the following lemma.

	\begin{lemma}[Theorem~8 of \cite{AKCONCUR23}]\label{lemma-C-limPlinear-if-no-+}
		For each counter $c$, if $\ba(c)>0$ then $n$ is a tight asymptotic estimate of $\calC[c]$.  Otherwise, i.e., when $\ba(c)=0$, the function $n^2$ is a lower asymptotic estimate of $\calC[c]$.
	\end{lemma}

	In the case that \(x \) is a transition we get from the definition of \(T_1=M_0\) that every simple component \(\by\) of \(\A\) with \(\by(x)>0\) contains a transition \(t\) with \(u_{t}(r_{0})< 0 \). As per the following lemma this implies that every simple component of \(\A \) is either decreasing or zero-unbounded on \(r_0\).
	
	\begin{restatable}{lemma}{lemmaMidecreasingorunbounded}\label{lemma-Mi-decreasing-or-unbounded}
		Let \(\by \) be a simple component of \(A_i \). There exists \(t\in M_i\) with \(\by(t)>0 \) iff \(\by\) is either decreasing or zero-unbounded on \(r_{i} \).
	\end{restatable}
	This follows from the expected effect of every single computational step on \(r_i \) being always non-positive. A full proof of Lemma~\ref{lemma-Mi-decreasing-or-unbounded} can be found in Section~\ref{sec-lemma-Mi-decreasing-or-unbounded-proof}.
	
	Furthermore, since every computational step of \(\A\) is non-increasing on average on \(r \) (see Section~\ref{sec-systems}) there exists no simple component that is increasing on \(r \) as per the following lemma.
	\begin{restatable}{lemma}{dimlemmacomponentrealeffektiseffecfft}
		\label{1dim-lemma-component-realeffekt-is-effect}
		Let \(\by \) be a simple component and \(c\) a counter. Let \(\bw \) be the long-run average effect of \(\A_{\by} \). Then there exist \(a,b>0 \) such that \(\bw=a\cdot \E[\realeffect_\by]=b\cdot \Delta(\by) \).
	\end{restatable}
	The proof of Lemma~\ref{1dim-lemma-component-realeffekt-is-effect} is purely technical and can be found in Appendix~\ref{sec-proof-of-1dim-lemma-component-realeffekt-is-effect}.

	Hence we obtain an upper asymptotic estimate \(n^2=n^{2\cdot 3^0} \) for \(\calT[t] \) from the results about \(1\)-dimensional VASS MDPs from \cite{AKCONCUR23} (third to last bullet point on page~12) which states that a transition \(t \) has an upper asymptotic estimate \(n^2\) for \(\calT[t] \) if there exists neither an increasing simple component nor a simple component that is zero-bounded while containing \(t\).

	
	\textbf{Case \(\variablenotk>0 \):}  Let \(rank_\variablenotk(p\bv)=\bb_\variablenotk(p)+\sum_{c\in\countersset(\A_\variablenotk)}  \ba_\variablenotk(c)\cdot\bv(c) \) be the ranking function defined by \(\ba_\variablenotk,\bb_\variablenotk \) on \(\A_\variablenotk \).  In the same way as in \(\cite{Zuleger:VASS-polynomial} \) we extend this ranking function onto \(\A \) as 	\[rank(p\bv)=\bb_\variablenotk(p)+\sum_{c\in\countersset(\A)} \ba_\variablenotk\big((c,\MEC_p)\big)\cdot \bv(c) \] where \(\MEC_p \) is the MEC of \(\A_\variablenotk \) which contains \(p\) (and if no such MEC exists  we put \(\ba_\variablenotk\big((c,\MEC_p)\big)=\bb_\variablenotk(p)=0 \)).  
	

	Notice that any transition \(t\in \transitions(\A_\variablenotk)\) has the same effect on both \(rank_\variablenotk \) as well as \(rank\), and that this effect is the effect on the counter \(r_{\variablenotk} \). Remember also that \(rank_\variablenotk\) is non-increasing on average at any computational step in \(\A_{\variablenotk} \) (see Section~\ref{sec-systems}).  Hence the only steps that can increase				 \(rank\) on average are those using transitions from \(\transitions(\A)\setminus\transitions(\A_l)= T_\variablenotk\).  But each \(t\in T_\variablenotk\) has an upper asymptotic estimate \(n^{2\cdot 3^{\variablenotk-1}} \) for \(\calT[t] \) from our assumption on \(\variablenotk\). The maximal possible increase of \(rank\) per single use of a transition \(t\in T_\variablenotk \) comes partly from the effect \(\bu_t \), but mainly  from changing weights assigned to the counters (i.e., when \(\ba_\variablenotk\big((c,\MEC_p)\big)\neq \ba_\variablenotk\big((c,\MEC_q)\big) \)). But   weights  can change only for counters \(c\in C_\variablenotk \), and these have an upper asymptotic estimate \(n^{ 3^{l-1}} \) for \(\calC[c] \) from our choice of \(\variablenotk\) as well.  Hence  the maximal possible increase of \(rank\) from transitions \(T_\variablenotk\) has an upper upper asymptotic estimate \(n^{2\cdot 3^{\variablenotk-1}} \cdot n^{3^{\variablenotk-1}} =n^{ 3^{\variablenotk}} \).

	
	Let \(P_0V_0,P_1V_1,\dots\) be the random variables denoting a computation in \(\A\).	We define \(\beta(n)=|T_\variablenotk|\cdot |C_\variablenotk|\cdot n^{3^{l}}\) and \(R_0,R_1,\dots\) as \[R_0=rank(P_0V_0)+\beta\big(\max_{c\in\countersset(\A)}V_0(c)\big) \] \[R_i=\begin{cases}R_{i-1}+rank(P_iV_i)-rank(P_{i-1}V_{i-1}) & \textit{ if } (P_{i-1},V_{i}-V_{i-1},P_{i})\notin T_\variablenotk \\
		R_{i-1} & \textit{ else } 
	\end{cases} \]  
	
	
	Notice that if \(R_\tau<0 \) then there exists \(c\in \countersset(\A) \) such that \(V_\tau(c)<0 \). Furthermore,  as per the following lemma	\(R_0,R_1,\dots\) is a supermartingale  which behaves in either ``decreasing'' or ``zero-unbounded'' manner eventually whenever a transition from \(T_{\variablenotk+1}\setminus T_{\variablenotk} \) is taken.
	
	\begin{restatable}{lemma}{lemmasupermartingale}\label{lemma_supermartingale}
		For  all initial configurations \(P_0V_0\) and all strategies \(\sigma\) it holds that 					\(R_0,R_1,\dots\) is a supermartingale. Furthermore, for each simple component \(\by\in \simplecompset(\A) \) with \(\{t\in \transitions(\A)\mid \by(t)>0 \}\cap (T_{\variablenotk+1}\setminus T_{\variablenotk})\neq \emptyset\) there exists a transition \(t\in \{t\in \transitions(\A)\mid \by(t)>0 \} \) for which 
  \[\prob^{\strategyDemon}_{p\bv}\big[R_i=\E^{\strategyDemon}_{p\bv}[R_{i-1}]\mid (P_{i-1},V_{i}-V_{i-1},P_{i})=t,P_0V_0,\dots,P_{i-1}V_{i-1}\big]\neq 1 \]
	\end{restatable}
	
	The proof of Lemma~\ref{lemma_supermartingale} is purely technical and can be found in Section~\ref{app-proof-of-lemma_supermartingale}.

	Consider now the 1-dimensional VASS MDP \(\A' \) created from \(\A\) by considering only a counter \(r\) encoding  \(R_0,R_1,\dots \) (i.e., effect of every transition \(t\) in \(\A' \) is the effect of \(t\) in \(\A \) on \(R_0,R_1,\dots\)). From Lemma~\ref{lemma_supermartingale} we have that there exists no strategy with positive long-term average effect on \(r\). Hence from Lemma~\ref{1dim-lemma-component-realeffekt-is-effect} there is no simple component increasing on \(r\) in \(\A' \).

	Therefore the following lemma gives a linear upper asymptotic estimate for \(\calC_{\A'}[r]\).
	
	\begin{lemma}\label{lemma-no-+-effect-on-C}[Lemma~29 of the full version of \cite{AKCONCUR23}]
		Given a 1-dimensional VASS MDP \(\A\), if \(\A\) contains no  increasing simple component, then \(n\) is an upper asymptotic estimate of \(\calC[c]\). 
	\end{lemma}
	
	But since the initial value of \(r \) is \(rank(P_0V_0)+\beta\big(\max_{c\in\countersset(\A)}V_0(c)\big)
	\leq
	a\cdot n+|T_\variablenotk|\cdot |C_\variablenotk|\cdot n^{3^{\variablenotk}}\in \calO(n^{3^\variablenotk}) \) this implies that any counter \(c\in\countersset(\A)\) with \(\ba_\variablenotk(c)>0 \) has an upper asymptotic estimate \(n^{3^\variablenotk} \). But if  \(x \) is a counter then we have from the definition of  \(C_{\variablenotk+1} \) and  Lemma~\ref{lemma:dichotomy}  that \(\ba_\variablenotk(x)=a >0 \). 
	
	Therefore it must hold that \(x\) is a transition. But from Lemma~\ref{lemma_supermartingale}  each simple component \(\by  \) of \(\A \), which contains \(t\), is either decreasing or zero-unbounded on \(r\). This gives us an upper asymptotic estimate \(n^2 \) for \(\calT_{\A'}[x] \) from the results about \(1\)-dimensional VASS MDPs from \cite{AKCONCUR23} (third to last bullet point on page~12). Same as in the case above this gives an upper asymptotic estimate \((n^{3^\variablenotk})^2=n^{2\cdot 3^\variablenotk} \) for \(\calT_{\A}[x] \). A contradiction.
	
	Lemma~\ref{lemma-polynomial-upper} holds.
\end{proof}

\subsubsection{Proof of Lemma~\ref{lemma-Mi-decreasing-or-unbounded}}\label{sec-lemma-Mi-decreasing-or-unbounded-proof}
\lemmaMidecreasingorunbounded*
\begin{proof}
	\(\Rightarrow \):
	Let \(t=(p,\bu,q)\in M_i\cap \{t\mid \by(t)>0 \}\). Remember that
	\begin{gather*}
		\Delta(\by)(r_i)
		=
		\sum_{t'\in \transitions(\A_i)} \bu_{t'}(r_i)\cdot \by(t')
		\\=
		\sum_{p'\in \statesDemon(\A_i)}\sum_{t'\in \tout(p')}\bu_{t'}(r_i)\cdot \by(t') + \sum_{p'\in \statesProb(\A_i)}\sum_{t'\in \tout(p')}\bu_{t'}(r_i)\cdot \by(t')
		\\=
		\sum_{p'\in \statesDemon(\A_i)}\sum_{t'\in \tout(p')}\bu_{t'}(r_i)\cdot \by(t') + \sum_{p'\in \statesProb(\A_i)}\sum_{t'\in \tout(p')}\bu_{t'}(r_i)\cdot P(t')\cdot   \sum_{t''\in \tout(p')} \by(t'')
		\\=
		\sum_{p'\in \statesDemon(\A_i)}\sum_{t'\in \tout(p')}\bu_{t'}(r_i)\cdot \by(t') + \sum_{p'\in \statesProb(\A_i)}\big(\sum_{t'\in \tout(p')} \by(t')\big)\cdot \sum_{t'\in \tout(p')}\bu_{t'}(r_i)\cdot P(t') \end{gather*} 
	where the second to last equality follows from \eqref{eq-multicomponent-prob-split}. From constraint~\eqref{eq-II-nondet-states} of \hyperref[fig-systems]{(II)} it holds for each \(p'\in \statesDemon(\A_i)\) and each \(t'\in \tout(p') \) that  \(
	\bu_{t'}(r_i)\leq 0\)
	and from constraint~\eqref{eq-II-prob-states} of  \hyperref[fig-systems]{(II)} it holds for each \(p'\in \statesProb(\A_i)\)
	\[				 \sum_{t'\in \tout(p')}\bu_{t'}(r_i)\cdot P(t') \leq 0
	\]
	
	Hence it holds \(\Delta(\by)(r_i)\leq 0\), hence \(\by\)  is either decreasing, zero-bounded, or zero-unbounded on \(r_i\).  Assume towards contradiction that \(\by\) is zero-bounded on \(r_i\). Then from Lemma~\ref{multicomp-lemma-zero-bounded-iff-zero-cycles} every cycle of \(\A_\by \) has effect exactly zero on \(r_i \). If every transition \(t'\in \{t\mid \by(t)>0 \}\) had effect exactly \( 0\) on \(r_{i} \) then this is a contradiction with \(t\in M_i \). Hence there exists a transition \((p',\bu',q')\in \{t\mid \by(t)>0 \} \) with \(\bu'(r_i)>0 \). 
	
	If \(p'\in \statesDemon(\A_i) \) then this is a contradiction with the constraint~\eqref{eq-II-nondet-states} of \hyperref[fig-systems]{(II)}. Hence \(p'\in \statesProb(\A_i) \). But every state from \(p''\in \statesProb(\A_i) \) that has an outgoing transition increasing \(r_i\) also has an outgoing transition decreasing \(r_i\) from the constraint~\eqref{eq-II-prob-states} of  \hyperref[fig-systems]{(II)}. Hence if we consider the computation which starts in \(p' \), takes this decreasing transition, and then in each state it reaches it picks an outgoing transition \(t' \) with \(\bu_{t'}(r_i)\leq 0 \) eventually we must witness a cycle whose effect is on \(r_i \) is negative, which is a contradiction with Lemma~\ref{multicomp-lemma-zero-bounded-iff-zero-cycles} below. 
	
	\(\Leftarrow\):
	We will show that if \(M_i\cap \{t\mid\by(t)>0\}=\emptyset \) then  \(\by(t)>0 \) implies \(\bu_t(r_i)=0 \). Note that if \(\by(t)>0 \) implies \(\bu_t(r_i)=0 \) then  \(\by \) is zero-bounded on \(r_i\) as per Lemma~\ref{multicomp-lemma-zero-bounded-iff-zero-cycles}.
	
	Assume  thus \(M_i\cap \{t\mid\by(t)>0\}=\emptyset \) and \(\by \) contains a \(t=(p,\bu,q)\in \{t\mid\by(t)>0\} \) with \(\bu(r_i)\neq 0 \). Since \(t\notin M_{i}\) we have \(\bu(r_i)\geq 0 \).  If \( p\in\statesDemon(\A_i)\) then from constraint~\eqref{eq-II-nondet-states} of \hyperref[fig-systems]{(II)} we have that \(\bu(r_i)\leq 0 \) and hence  \(\bu(r_i)= 0 \). Hence 	\(p\in\statesProb(\A_i)\), but then from constraint~\eqref{eq-II-prob-states} of \hyperref[fig-systems]{(II)} it holds  
	\[
	\sum_{t'= (\stateVass,\bu',\stateVassalt') \in \tout(\stateVass)}P(t)\cdot \bu_{t'}(r_i)=\sum_{t'= (\stateVass,\bu',\stateVassalt') \in \tout(\stateVass)}P(t)\cdot \big(\bb_i(\stateVassalt')-\bb_i(\stateVass)+\sum_{c\in \countersset(\A)}  \bu'(c)\cdot \ba_i(c)\big)\leq 0 
	\]
	From \eqref{eq-multicomponent-prob-split}  we also have \(\by(t')>0 \) for each \(t'\in \tout(p) \). Hence if \(\bu(r_i)> 0 \) then there exists \(t'\in \tout(p)\) with \(\bu_{t'}(r_i)< 0 \) and hence \(t'\in M_i \) which is a contradiction with \(M_i\cap \{t\mid\by(t)>0\}=\emptyset \). 		
\end{proof}

\begin{restatable}{lemma}{multicomplemmazeroboundediffzerocycles}
	\label{multicomp-lemma-zero-bounded-iff-zero-cycles}
	A  simple component \(\by \) is zero-bounded on \(c\) iff every cycle \(\pathm\) in \(\A_\by \) satisfies \(\Delta(\pathm)(c)=0 \).
	
\end{restatable}

%
%
\begin{proof}
	``$\Rightarrow$'': From the definition of zero-boundedness we have \(\prob[\realeffect_\by(c)=0]=1 \). Assume towards a contradiction that there exists a cycle \(\pathm=p\bv_0,\dots,p\bv_l \) in \(\A_\by \) with \(\Delta(\pathm)(c)\neq 0 \). Since \(\A_\by \) is strongly connected, there exists a computation \(\alpha=p_\by\bu_0,\dots,p\bu_a,\dots,p_\by\bu_b \). Furthermore since a single iteration of \(\by \) has positive probability of being equal to \(\alpha \)  it holds \(\Delta(\alpha)(c)=0 \). But this means that there also exists a computation  \(\alpha'=p_\by\bu_0,\dots,p\bu_a,\dots,p\big(\bu_a+\Delta(\pi)\big),\dots,p_\by\big(\bu_b+\Delta(\pi)\big) \) with \(\Delta(\alpha')(c)=\big(\bu_b+\Delta(\pi)\big)(c)-\bu_0(c)=\Delta(\alpha)(c)+\Delta(\pi)(c)=\Delta(\pi)(c)\neq 0 \). But a single iteration of \(\by\) has a positive probability of being equal to \(\alpha' \) as well this is a contradiction with \(\prob[\realeffect_\by(c)=0]=1 \).
	
	``$\Leftarrow$'': Since each iteration of \(\by \) is a cycle this means that effect of a single iteration of \(\by \) on \(c\) is always \(0\). Thus \(\prob[\realeffect_\by(c)=0]=1 \) which also implies \(\E[\realeffect_\by(c)]=0 \). But as per Lemma~\ref{1dim-lemma-component-realeffekt-is-effect} this gives us that \(\by \) is zero-bounded on \(c\).
\end{proof}

\subsubsection{Proof of Lemma~\ref{1dim-lemma-component-realeffekt-is-effect}}
\label{sec-proof-of-1dim-lemma-component-realeffekt-is-effect}
\dimlemmacomponentrealeffektiseffecfft*
\begin{proof}
	As we can view the computation on \(\A_\by \) as a series of iterations of \(\by \), and each iteration finishes with probability \(1\), it holds that \begin{itemize}
		\item \(\E^{\strategyDemon_\by}_{\stateVass_\by}[\realeffect_\by](c)>0 \) implies \(\bw(c)>0 \),
		\item  \(\E^{\strategyDemon_\by}_{\stateVass_\by}[\realeffect_\by](c)<0 \) implies \(\bw(c)<0 \),
		\item and \(\E^{\strategyDemon_\by}_{\stateVass_\by}[\realeffect_\by](c)=0 \) implies \(\bw(c)=0 \).
	\end{itemize}
	Hence there exists some \(a>0 \) such that \(\bw(c)=a\cdot \E^{\strategyDemon_\by}_{\stateVass_\by}[\realeffect_\by](c) \).

	Now let \[b=\frac{1}{\sum_{t\in \transitions(\A)} \by(t)} \] and let \(\overline{\by}=b\cdot \by\) be the normalized vector of \(\by \). Note that \(b>0\). A standard result regarding Markov chains is that an irreducible (i.e., strongly connected) Markov chain contains a single unique invariant distribution, and that the effect of a single computational step from this invariant distribution is the long-run average effect \(\bw \).\footnote{See e.g. \cite{Norris1997} (Section~1.10).} Hence from the following lemma it holds \(\bw=b\cdot \Delta(\by) \). 
\end{proof}

\begin{restatable}{lemma}{dimlemmacomponentrealeffektiseffect}
	\label{app-lemma-solution-I-startegy}
	Let \(\bp\in \mathbb{Z}^{\states(\A)} \) be such that \(\bp(p)=\sum_{t\in \tout(p)} \overline{\by}(t) \). Then \(\bp \) is an invariant distribution on \(\A_\by \). Also the expected effect of a single computational step in \(\A_\by \) taken from distribution \(\bp \) is equal to \(\Delta(\bar{\by})\).
\end{restatable} 
\begin{proof}
	Let us consider a single computation step in \(\A_\by \) taken from distribution \( \bp\), and let \(X \) be resulting distribution on transitions during this step. Then for each transition \(t=(p,\bu,q)\in \transitions(\A_\by) \) it holds:	 
	\begin{itemize}
		\item if \(p\in \statesProb(\A) \) then \( X(t)=P(t)\cdot \bp(p) =P(t)\cdot \sum_{t\in \tout(p)} \overline{\by}(t)=\overline{\by}(t) \) where the last equality holds from \eqref{eq-multicomponent-prob-split}.
		\item if \(p\in  \statesDemon(\A) \) then \( X(t)=
		\bp(p)
		=
		\sum_{t\in \tout(p)} \overline{\by}(t)
		=
		\overline{\by}(t)
		\) where the last equality holds from  \(p\) having only a single outgoing transition in \(\A_{\by} \).
	\end{itemize} 
	As the next distribution \(\bp' \) on states can be expressed as \[\bp'(p)=\sum_{t\in \tin(p)} X(t) = \sum_{t\in \tin(p)} \overline{\by}(t)=\sum_{t\in \tout(p)}\overline{\by}(t)=\bp(p) \] \(\bp \) is an invariant distribution on \( \A_\by\). Here the second to last equality follows from \eqref{eq-multicomponent-flow-cond}.

	The expected effect of a single computational step from \(\bp\) is \[\sum_{t\in\transitions(\A_\by)} X(t)\cdot \bu_t=\sum_{t\in\transitions(\A_\by)} \bar{\by}(t)\cdot \bu_t=\Delta(\bar{\by}) \]	Lemma holds.
\end{proof}

\subsubsection{Proof of Lemma~\ref{lemma_supermartingale}}
\label{app-proof-of-lemma_supermartingale}
\lemmasupermartingale*

We prove that \(R_0,R_1,\dots\) is a supermartingale in Lemma~\ref{local-lemma-martinglae}. For the claim regarding simple components follows directly from the definition of \(M_{l+1} \), as for for each \(\by \) there exists \(t\in \{t\in \transitions(\A)\mid \by(t)>0 \}\cap (T_{\variablenotk+1}\setminus T_{\variablenotk})\neq \emptyset\) with \(\bu_t(r_i)<0 \) and therefore \[\prob^{\strategyDemon}_{p\bv}\big[R_i>R_{i+1}[R_{i-1}]\mid (P_{i-1},V_{i}-V_{i-1},P_{i})=t,P_0V_0,\dots,P_{i-1}V_{i-1}\big]= 1 \]

\begin{lemma}\label{local-lemma-martinglae}
	For  all initial configurations \(P_0V_0\) and all strategies \(\sigma\) it holds that 					\(R_0,R_1,\dots\) is a supermartingale.
\end{lemma}
\begin{proof}
We consider two separate cases, either \(P_{i-1}\in \statesDemon(\A) \) or  \(P_{i-1}\in \statesProb(\A) \). 
	
	
	\textbf{Case \(P_{i-1}\in \statesDemon(\A)\):} then
	\begin{gather*}
		\E^{\strategyDemon}_{p\bv}[R_i\mid P_0V_0,\dots,P_{i-1}V_{i-1}]
		=\\=
		\sum_{t\in \tout(P_{i-1})\cap T_{\variablenotk}}\strategyDemon(P_0V_0,\dots,P_{i-1}V_{i-1})(t)\cdot R_{i-1} 
		+\\+
		\sum_{t=(P_i,\bu,q)\in \tout(P_{i-1})\setminus T_\variablenotk}\strategyDemon(P_0V_0,\dots,P_{i-1}V_{i-1})(t)\cdot \big(R_{i-1}+rank(qV_{i-1}+\bu)-rank(P_{i-1}V_{i-1})\big)
		\\\allowdisplaybreaks=
		R_{i-1} +
		\sum_{t=(P_i,\bu,q)\in \tout(P_{i-1})\setminus T_\variablenotk}\strategyDemon(P_0V_0,\dots,P_{i-1}V_{i-1})(t)\cdot \big(rank(qV_{i-1}+\bu)-rank(P_{i-1}V_{i-1})\big)
		\\\allowdisplaybreaks=
		R_{i-1} +\sum_{t=(P_{i-1},\bu,q)\in \tout(P_{i-1})\setminus T_\variablenotk}
		\strategyDemon(P_0V_0,\dots,P_{i-1}V_{i-1})(t) \cdot 
		\\ \cdot
		\Big(\bb_\variablenotk(q)+\sum_{c\in \countersset(\A)} \ba_\variablenotk\big((c,\MEC_q)\big)\cdot (V_{i-1}+\bu)(c) - \bb_\variablenotk(P_{i-1})-\sum_{c\in\countersset(\A)} \ba_\variablenotk\big((c,\MEC_{P_0})\big)\cdot V_{i-1}(c)   \Big)
		\\\allowdisplaybreaks=
		R_{i-1} +\sum_{t=(P_{i-1},\bu,q)\in \tout(P_{i-1})\setminus T_\variablenotk}
		\strategyDemon(P_0V_0,\dots,P_{i-1}V_{i-1})(t)
		\cdot\\\cdot\Big(\bb_\variablenotk(q)-\bb_\variablenotk(P_{i-1})+\sum_{c\in \countersset(\A)} \ba_\variablenotk\big((c,\MEC_{q})\big)\cdot \bu(c) \Big)
		\\\allowdisplaybreaks=
		R_{i-1} +\sum_{t=(P_{i-1},\bu,q)\in \tout(P_{i-1})\setminus T_\variablenotk}
		\strategyDemon(P_0V_0,\dots,P_{i-1}V_{i-1})(t)
		\cdot\big(\bb_\variablenotk(q)-\bb_\variablenotk(P_{i-1})+\sum_{c\in \countersset(\A_\variablenotk)} \ba_\variablenotk(c)\cdot \bu(c) \big)
	\end{gather*}
	where the second to last equality follows form \(\MEC_{P_0}=\MEC_q\) for each \((P_0,\bu,q)\in\transitions(\A_\variablenotk) \).
	
 Hence 
	\begin{gather*}
		R_{i-1} +\sum_{t=(P_{i-1},\bu,q)\in \tout(P_{i-1})\setminus T_\variablenotk}
		\strategyDemon(P_0V_0,\dots,P_{i-1}V_{i-1})(t)
		\cdot\big(\bb_\variablenotk(q)-\bb_\variablenotk(P_{i-1})+\sum_{c\in \countersset(\A_\variablenotk)} \ba_\variablenotk(c)\cdot \bu(c) \big)
		\leq\\\leq 
		R_{i-1} +0
	\end{gather*}
	

	\textbf{Case \(P_{i-1}\in \statesProb(\A) \):} by the definition of \(T_\variablenotk\), it holds either \(\tout(P_{i-1})\subseteq T_\variablenotk \) or \(\tout(P_{i-1})\cap T_\variablenotk=\emptyset \). 
	\begin{itemize}
		\item If  \(\tout(P_{i-1})\subseteq T_\variablenotk \) then by the definition of \(R_0,R_1,\dots \) it holds  \[\E^{\strategyDemon}_{p\bv}[R_i\mid P_0V_0,\dots,P_{i-1}V_{i-1}]=R_{i-1} \]
		\item If  \(\tout(P_{i-1})\cap T_\variablenotk=\emptyset \) then \begin{gather*}
			\E^{\strategyDemon}_{p\bv}[R_i\mid P_0V_0,\dots,P_{i-1}V_{i-1}]
			=\\=R_{i-1}+\sum_{t=(P_0,\bu,q)\in\tout(P_0)} P(t)\cdot \Big(\bb_\variablenotk(q)-\bb_\variablenotk(P_0)+\sum_{c\in\countersset(\A)}\ba_\variablenotk\big((c,\MEC_{P_0})\big)\cdot \bu(c)   \Big) 
			\\=R_{i-1}+\sum_{t=(P_0,\bu,q)\in\tout(P_0)} P(t)\cdot \big(\bb_\variablenotk(q)-\bb_\variablenotk(P_0)+\sum_{c\in\countersset(\A_\variablenotk)}\ba_\variablenotk(c)\cdot \bu(c)   \big) 
		\end{gather*}
		
				Hence \begin{gather*}
			R_{i-1}+\sum_{t=(P_0,\bu,q)\in\tout(P_0)} P(t)\cdot \big(\bb_\variablenotk(q)-\bb_\variablenotk(P_0)+\sum_{c\in\countersset(\A_\variablenotk)}\ba_\variablenotk(c)\cdot \bu(c)   \big) \leq R_{i-1}
		\end{gather*} 
	\end{itemize} 				
\end{proof}

\subsection{Proof of Lemma~\ref{lemma-ci-ti-ai-ptime-computable}}\label{sec-lemma-ci-ti-ai-ptime-computable-proof}
\lemmacitiaiptimecomputable*
\begin{proof}
	We can compute \(\bx_i,\ba_i \), and \(\bb_i \) in polynomial time as discussed in Section~\ref{sec-systems}. This  gives us \(C_{i+1} \) as well as allowing us to compute the value \(\bu_t (r_i)\) for each \(t=(p,\bu,q)\in \transitions(\A_i)\).  	
	
	From constraint~\eqref{eq-II-nondet-states} of \hyperref[fig-systems]{(II)} it holds  \(\bu_t (r_i)\leq 0\) whenever \(p\in \statesDemon(\A_i)\). And if \(p\in\statesProb(\A_i) \) then from constraint~\eqref{eq-II-prob-states} of \hyperref[fig-systems]{(II)} it holds  \(\sum_{t'\in\tout(p)}\bu_{t'}(r_i)\leq 0\). Hence the only transitions with \(\bu_t(r_i)>0 \) can be those from probabilistic states \(p\in \statesProb(\A_i)\) for which there exists another transition \(t'\in \tout(p)\) with \(\bu_{t'}(r_i)<0 \).  From \eqref{eq-multicomponent-prob-split} we thus have that any simple component \(\by \) of \(\A_i \) containing \(t\) with \(\bu_t(r_i) \) also contains \(t'\) with \(\bu_{t'}(r_i)<0 \). This gives us that \(t\notin M_i \) iff there exists a simple component \(\by\) of \(\A_i \) with \(\by(t)>0 \) that contains only transitions whose effect on \(r_i\) is exactly \(0\).
	
	Hence to compute \(T_{i+1} \) it suffices to remove each transition \(t \) with \(\bu_t(r_i)\neq 0 \) from \(\A_i \), and then iteratively remove all states \(p\) which either have no remaining incoming transition or
	\begin{itemize}
		\item if \(p\in\statesDemon(\A_i)\)  they have no  remaining outgoing transition;
		\item if \(p\in\statesProb(\A_i) \) any of their outgoing transition was removed.
	\end{itemize} 
	
	It is easy to see that any simple component of \(\A_i \) containing any of the removed transitions must contain a transition \(t\) with \(\bu_t(r_i)\neq 0 \) (this follows from \eqref{eq-multicomponent-flow-cond}). On the other hand, any transition \(t\) that is left is contained in some EC of the ``pruned'' VASS MDP and hence there exists a simple component \(\by \) of \(\A_i \) containing only transitions that are left while also containing \(t\). Clearly such ``pruning'' can be done in polynomial time. 
\end{proof}

\subsection{Proofs of Section~\ref{sec-exponential-lower}}

\subsubsection{Proof of Lemma~\ref{lemma-random-part-prob-bound}}
\label{sec-lemma-random-part-prob-bound-proof}

\lemmarandompartprobbound*

\begin{proof}
	Let \(P_0V_0,P_1V_1,\dots \) be random variables encoding a computation on \(\A \).
	
	
	For each \(c\in \countersset(\A)\) and each simple component \(\by \) of \(\A \) let \(\ba_\by^c,\bb_\by^c \) be a maximal solution of \hyperref[fig-systems]{(II)} for  \(\A_{\hat{\by}}^c \), where \(\A_{\hat{\by}}^c \) is the VASS MDP \(\A_{\hat{\by}} \) restricted only to the counter \(c\). 
	
	Since \(\Delta(\hat{\by})(c)=0 \) and \(\hat{\by} \) is the only simple component of \(\A_{\hat{\by}}^c \), it follows from Lemma~\ref{lemma-decompose-multicomponents-into-components} that \(\hat{\by} \) is a maximal solution of \hyperref[fig-systems]{(I)} for  \(\A_{\hat{\by}}^c \). Hence from Lemma~\ref{lemma:dichotomy} it holds \(\ba_\by^c(c)>0 \) and we can wlog. assume that \(\ba_\by^c(c)=1 \) (see Section~\ref{sec-systems}).
	
	We define \(R_0^c=\sum_{\by\in Y(\A)}\max_{p\in \states(\A_\by)}\bb_\by^c(p)  \) and \[R_i^c=\begin{cases}
		R_{i-1}^c+\hat{V}_i(c)-\hat{V}_{i-1}(c)+\bb_\by^c(P_i)-\bb_\by^c(P_{i-1}) & \textit{if } \DecomposeComponent_\A(P_0V_0,\dots,P_{i}V_{i})=(\by,p,j) \\ & \textit{for some } p,j \textit{ and }|R_i^c-R_0^c|< \xconstant \\
		R_{i-1}^c & \textit{else}
	\end{cases} \]
	
	Let \(\tau[c] \) be a stopping time such that \(|\hat{V}_{\tau[c]}(c)|\geq \xconstant \) for the first time.
	Notice that for any \(i\leq \tau[c]\) it holds  \(|R_i^c-\hat{V}_i(c)|\leq \sum_{\by\in \simplecompset(\A)}\max_{p\in \states(\A_\by)}\bb_\by^c(p) =R_0^c\). Hence \(\tau[c]\leq \yconstant \) implies \(|R_\yconstant^c-R_0^c|\geq \xconstant-R_0^c\). It thus suffices to bound the probability of \(|R_\yconstant^c-R_0^c|\geq \xconstant-R_0^c\).
	
	From Lemma~\ref{lemma-prob-part-martingaleonecounter} below we have that \(R_0^c,R_1^c,\dots \) is a martingale. Notice also that \(|R_{i}-R_{i-1}|\leq a \) for some constant \(a\) that depends only on \(\A \). Hence we can use Azuma inequality (\cite{azuma_ineq}) to obtain 
	\[\prob_p^\strategyDemon\big[|R_{\yconstant}^c-R_0^c|\geq \xconstant-R_0^c\big]
	\leq
	2\cdot \exp\big( \frac{-(\xconstant-R_0^c)^2
	}{
		2\cdot a^2\cdot \yconstant
	}\big)
	\]
	
	which gives \begin{gather*}
		\prob_p^\strategyDemon[\tau\leq y]
		\leq \sum_{c\in\countersset(\A)} \prob_p^\strategyDemon\big[\tau[c]\leq y\big] \leq 
		\sum_{c\in\countersset(\A)} \prob_p^\strategyDemon\big[|R_{\yconstant}^c-R_0^c|\geq \xconstant-R_0^c\big] 
		\leq \\\leq 
		|\countersset(\A)|\cdot  	2\cdot \exp\big( \frac{-(\xconstant-\max_{c\in \countersset(\A)}R_0^c)^2
		}{
			2\cdot a^2\cdot \yconstant
		}\big)
	\end{gather*}
	Since each of \(|\countersset(\A)|\cdot  	2\), \(2\cdot a^2\), and    \(\max_{c\in \countersset(\A)}R_0^c \) is a constant that depends only on \(\A \) the lemma follows.
\end{proof}

\begin{lemma}\label{lemma-prob-part-martingaleonecounter}
	\(R_0^c,R_1^c,\dots\) is a martingale.
\end{lemma}
\begin{proof}
	If \(\DecomposeComponent_\A(P_0V_0,\dots,P_{i}V_{i})=\emptypointpair\) or \(|R_{i-1}^c-R_0^c|\geq \xconstant\)  then per the definition of \(R_i \) it holds \[\E_p^\strategyDemon[R_i\mid P_0V_0,\dots,P_{i-1}V_{i-1}]=\E_p^\strategyDemon[R_{i-1}] \]
	
	Assume thus that \(\DecomposeComponent_\A(P_0V_0,\dots,P_{i}V_{i})\neq \emptypointpair\). We show this implies  
	\[\E_p^\strategyDemon[\hat{V}_i(c)-\hat{V}_{i-1}(c)+\bb_\by^c(P_i)-\bb_\by^c(P_{i-1})\mid P_0V_0,\dots,P_{i-1}V_{i-1} ]=0 \]
	
	We consider two separate cases depending on whether \(P_{i-1} \) is non-deterministic or probabilistic.
	
	\textbf{Case \(P_{i-1}\in \statesDemon(\A) \):} for each simple component \(\by \) containing \(P_i\), since \(\vec{1},\bb_\by^c \) is a maximal solution of \hyperref[fig-systems]{(II)} for  \(\A_{\hat{\by}}^c \) we have from \eqref{eq-II-nondet-states}  for the transition \((P_{i-1},\bu,q)\in \tout(P_{i-1})\cap \transitions(\A_\by) \) that \(\bb_\by^c(q)-\bb_\by^c(P_{i-1})+\bu_{\hat{\by}}(c)\leq 0 \) where \(\bu_{\hat{\by}} \) is the effect of this transition in \(\A_{\hat{\by}} \). Furthermore, since \(\Delta(\hat{\by})=\vec{0} \) it holds that \(\hat{\by} \) is a maximal solution of  of \hyperref[fig-systems]{(I)} for  \(\A_{\hat{\by}}^c \) and thus we get from  Lemma~\ref{lemma:dichotomy} that this is actually an equality \(\bb_\by^c(q)-\bb_\by^c(P_{i-1})+\bu_{\hat{\by}}(c)= 0 \). Hence
	\begin{gather*}
		\E_p^\strategyDemon[\hat{V}_i(c)-\hat{V}_{i-1}(c)+\bb_\by^c(P_i)-\bb_\by^c(P_{i-1})\mid P_0V_0,\dots,P_{i-1}V_{i-1} ]
		=\\=
		\sum_{(\by,p,j)} \prob_p^\strategyDemon[\DecomposeComponent_\A(P_0V_0,\dots,P_{i}V_{i})=(\by,p,j) \mid P_0V_0,\dots,P_{i-1}V_{i-1} ]  \cdot 
		\\\cdot \big(\bu_{\hat{\by}}(c)+\bb_\by^c(q)-\bb_\by^c(P_{i-1})\big)
		\\=
		\sum_{(\by,p,j)} \prob_p^\strategyDemon[\DecomposeComponent_\A(P_0V_0,\dots,P_{i}V_{i})=(\by,p,j) \mid P_0V_0,\dots,P_{i-1}V_{i-1} ]  \cdot  0 =0
	\end{gather*} 
	Hence this case satisfies the martingale property.
	
	\textbf{Case \(P_{i-1}\in \statesProb(\A) \):} Let  \(\DecomposeComponent_\A(P_0V_0,\dots,P_{i}V_{i})=(\by,p,j)\). From \(\vec{1},\bb_\by^c \) being a maximal solution of \hyperref[fig-systems]{(II)} for  \(\A_{\hat{\by}}^c \) we have from \eqref{eq-II-prob-states} that \[\sum_{t= (P_{i-1},\bu,\stateVassalt) \in \tout(\stateVass)}P(t)\cdot \big(\bb_\by^c(\stateVassalt)-\bb_\by^c(P_{i-1})+\bu_{\hat{\by}}(c)\big)\leq 0 \] where  \(\bu_{\hat{\by}} \) is the effect of this transition in \(\A_{\hat{\by}} \). Furthermore, from \(\hat{\by} \) being a maximal solution of \hyperref[fig-systems]{(I)} for  \(\A_{\hat{\by}}^c \) we get from  Lemma~\ref{lemma:dichotomy} that this is actually an equality \[\sum_{t= (P_{i-1},\bu,\stateVassalt) \in \tout(\stateVass)}P(t)\cdot \big(\bb_\by^c(\stateVassalt)-\bb_\by^c(P_{i-1})+\bu_{\hat{\by}}(c)\big)= 0 \] Hence
	\begin{gather*}
		\E_p^\strategyDemon[\hat{V}_i(c)-\hat{V}_{i-1}(c)+\bb_\by^c(P_i)-\bb_\by^c(P_{i-1})\mid P_0V_0,\dots,P_{i-1}V_{i-1} ]
		=\\=
		\sum_{t=(P_{i-1},\bu,q)\in\tout(P_{i-1})} P(t)\cdot \big(\bu_{\hat{\by}}(c)+\bb_\by^c(q)-\bb_\by^c(P_{i-1})\big) =0
	\end{gather*} 
	Hence this case also satisfies the martingale property.
\end{proof}

\subsection{Proof of Lemma~\ref{lemma-decompose-multicomponents-into-components}}
\label{sec-lemma-decompose-multicomponents-into-components-proof}
\lemmadecomposemulticomponentsintocomponents*
\begin{proof}
	Let \(\by_1,\dots,\by_k \) be all the simple components of \(\A \). Let \(\bx_0=\bx \), and for each \(1\leq i \leq k \) let \(\bx_i=\bx_{i-1}-a_i\cdot\by_i \) where \(a_i \) is the largest value such that \(\bx_{i-1}-a_i\cdot\by_i\geq \vec{0} \). Note that \(a_i\geq 0  \) since  \(\bx_{i-1}-0\cdot\by_i\geq \vec{0} \). Also note that such \(a_i\) always exists and is rational as it holds \(a_i=\min_{t\in \transitions(\A)} \frac{\bx_{i-1}(t)}{\by_i(t)} \) and \(\transitions(\A)\) is finite.

	If \(\bx_k=\vec{0} \) then it holds \(\bx=\sum_{i=1}^{k}a_i\cdot\by_i \) with \(a_i\geq 0  \) for all \(1\leq i\leq k\). Assume therefore that \(\bx_k\neq \vec{0} \).  As \(\bx_k\) is a multi-component of \(\A\) the VASS MDP \(\A_{\bx_k} \) must contain at least one MEC, and this MEC must contain at least one MEC of \(\A_\sigma \) for some \(\strategyDemon\in \cMDstrats{\A_{x_k}} \), and thus \(\A_{\bx_k} \) contains at least one simple component \(\by \) of \(\A_{\bx_k} \). Since \(\by\) is also a simple component of \(\A \) there exists \(1\leq i\leq k\) such that \(\by=\by_i \), and therefore \(\by_i(t)>0 \) implies \(\bx_k(t)>0 \) for each \(t\in \transitions(\A)\). Since \(\bx_i\geq \bx_k \) this also means that \(\by_i(t)>0 \) implies \(\bx_{i}(t)>0 \) for each \(t\in \transitions(\A)\). But this is a contradiction with how we chose \(a_i \) as this means there exists \(a_i'>a_i \) such that \(\bx_{i-1}-a_i'\cdot \by_i\geq \vec{0} \).  Hence  \(\bx_k=\vec{0} \), which finishes proof of the first part of the lemma.
	
	It remains to show that we can do this decomposition such that \(a_\by>0 \) iff \(\by \) is a simple component of \(\A_\bx \). It holds trivially that \(a_\by=0 \) if \(\by \) is not a simple component of \(\A_\bx \) as that means there exists a transition \(t\) with \(\by(t)>0\) while \(\bx(t)=0 \). Hence it suffices to show that we can do this decomposition in such a way that \(a_\by>0 \) for every simple component \(\by \) of \(\A_\bx \). 
	
	We will show that if there exists a decomposition \(\bx=\sum_{\by} a_\by\cdot \by\) for which there are exactly \(i\geq 1 \) simple components \(\by \) of \(\A_\bx \) with \(a_\by=0 \) then there also exists a decomposition \(\bx=\sum_{\by} a_\by' \cdot\by\) such that there are at most \(i-1 \) simple components \(\by \) of \(\A_\bx \) with \(a_\by'=0 \).
	
	Note that this would finish the proof of the lemma, as there are only finitely many  simple components in \(\A_\bx \) and we have already shown that at least one decomposition  exists.
	
	Let \(\bx=\sum_{\by} a_\by\cdot \by\) be such that there are exactly \(i\geq 1 \) simple components \(\by \) of \(\A_\bx \) with \(a_\by=0 \). Let \(\by' \) be a simple component of \(\A_{\bx} \) such that \(a_{\by'}=0 \).
	
	It holds \(\frac{\bx}{2}=\sum_{\by} \frac{a_\by}{2}\cdot \by \). Furthermore, Since \(\by' \) is a simple component of \(\A_{\bx} \) it holds for every \(t\in \transitions(\A)\)  that \(\by'(t)>0 \) implies \(\frac{\bx}{2}(t)>0 \). Hence there exists \(b>0 \) such that \(\frac{\bx}{2}-b\cdot \by'\geq \vec{0}  \). Hence  \(\frac{\bx}{2}-b\cdot \by' \) is also a multi-component of \(\A \). Therefore we can apply the first part of this lemma to obtain a decomposition \(\frac{\bx}{2}-b\cdot \by' =\sum_{\by} b_\by \cdot\by \)  with \(b_\by\geq 0 \) for all \(\by \). We can then decompose  \(\bx \) as \[\bx= \frac{\bx}{2}+(\frac{\bx}{2}-b\cdot \by') + b\cdot \by'=\sum_{\by} \frac{a_\by}{2}\cdot \by +\sum_{\by} b_\by\cdot  \by  + b\cdot \by'=\sum_{\by} a_\by'\cdot \by\] where \(a_{\by}'=\frac{a_\by}{2}+b_\by \) if \(\by\neq \by' \) and \(a_{\by'}'=\frac{a_{\by'}}{2}+b_{\by'}+b \). Clearly \(a_\by>0 \) implies \(a_{\by}'>0 \) while \(a_{\by'}'>0 \). Hence there are at most \(i-1\) components \(\by \) of \(\A_\bx \) for which it holds \(a_{\by}'=0 \). Lemma Holds.
\end{proof}

\subsection{Proof of Lemma~\ref{lemma-vecspaces}}\label{sec-lemma-vecspaces-proof}
\lemmavecspaces*

We begin by showing that \(\spanCone(\hat{\by})(\Czero)\) is a vector space (Lemma~\ref{lemma-span-hat-y-vector-space}), then we show that \(\Delta\big(\multicompset(\MEC)\big)(\Czero)\) is also a vector space (Lemma~\ref{lemma-multicomponents-inverse}), and then finally we show that \(\spanCone(\hat{\by})(\Czero)\subseteq \Delta\big(\multicompset(\MEC)\big)(\Czero)\) (Lemma~\ref{lemma-deviations-resetable}).

\begin{restatable}{lemma}{lemmaspanhatyvectorspace}
	\label{lemma-span-hat-y-vector-space}
	Let \(\by \) be a simple component of \(\A_k \). Then \(\spanCone(\hat{\by})(\Czero)\) is a vector space.
\end{restatable}
\begin{proof}
	Since \(\spanCone(\hat{\by})(C_0)\) is a cone it suffices to show \(\spanCone(\hat{\by})(C_0)\) is closed under multiplication by \(-1\).  Let \(\bu\in \spanCone(\hat{\by})(C_0) \) be arbitrary. 
	
	From the definition of \(\spanCone(\hat{\by})(C_0) \) we can write \(\bu=\sum_{\bv\in \support(\hat\by)(C_0)} \bv\cdot a_\bv\) for some constants \(a_\bv\geq 0 \) where only finitely many constants are non-zero. Also, since \(\Delta(\hat{\by})(C_0)=\vec{0} \) it holds \(\sum_{\bv\in  \support(\hat\by)(C_0)}\bv\cdot \prob[\realeffect_{\hat{\by}}(C_0)=\bv]=\vec{0} \). This gives us for each \(l\) that \[\sum_{\bv\in  \support(\hat\by)(C_0)}\bv\cdot \big(\prob[\realeffect_{\hat{\by}}(C_0)=\bv]-l\cdot a_\bv\big)=-l\cdot \bu \]
	
	Let \[a=\max_{\bv\in \support(\hat\by)(C_0)} a_\bv \] \[b=\min_{\bv\in \support(\hat\by)(C_0); a_\bv>0}\prob[\realeffect_{\hat{\by}}(C_0)=\bv] \] Then it holds \(\prob[\realeffect_{\hat{\by}}(C_0)=\bv]-\frac{b}{a}\cdot a_\bv\geq 0 \) for each \(\bv\in \support(\hat\by)(C_0) \) and hence \(-\frac{b}{a}\cdot \bu  \) lies in the convex hull of \(\support(\hat\by)(C_0) \). Thus as per Carathéodory's theorem it holds \(-\bu\in \spanCone(\hat{\by})(C_0) \). 
\end{proof}

The following lemma shows that  \(\Delta\big(\multicompset(\MEC)\big)(\Czero)\) is also a vector space.

\begin{restatable}{lemma}{lemmamulticomponentsinverse}\label{lemma-multicomponents-inverse}
	Let \(\MEC \) be a MEC of \(\A_k \). Let \(\bu\in \Delta\big(\multicompset(\MEC)\big)(\Czero) \). Then it holds \(-\bu\in \Delta\big(\multicompset(\MEC)\big)(\Czero) \).
\end{restatable}		
\begin{proof}
	
	Let \(\bz\in \multicompset(\MEC) \) be such that \(\bu=\Delta(\bz)(\Czero)\). Let \(\bx_\MEC \) be the restriction of \(\bx \) onto \(\MEC \) (i.e., \(\bx_\MEC(t)=\bx(t) \) if \(t\) is in \(\MEC \) and \(\bx_\MEC(t)=0 \) otherwise). From Lemma~\ref{lemma-k-converged} we have that every transition of \(\MEC \) is included in \(\bx_\MEC \). Hence we can wlog. assume that \(\bx_\MEC>\bz \) (as otherwise we can simply take a multiple of \(\bx_B\)). Thus \(\bx_\MEC-\bz \) is a multi-component of \(\MEC \) with \(\Delta(\bx_\MEC-\bz)(\Czero)=\Delta(\bx_\MEC)(\Czero)-\Delta(\bz)(\Czero)=\vec{0}-\bu \). Hence \(-\bu\in\Delta\big(\multicompset(\MEC)\big)(\Czero) \).
\end{proof}

The core of our lower bound comes from the following lemma, which allows us to ``reset'' the probabilistic part of the computation in \(\A_k \) whenever it threatens to kill the run.
\begin{restatable}{lemma}{lemmadeviationsresetable}\label{lemma-deviations-resetable}
	Let \(\MEC \) be a MEC of \(\A_k \) and let \(\by \) be a simple component of \(\MEC \). Then it holds \(\spanCone(\hat{\by})(\Czero)\subseteq \Delta\big(\multicompset(\MEC)\big)(\Czero)\).	
\end{restatable}

\begin{proof}
	Both \(\spanCone(\hat{\by})(\Czero)\) and \(\Delta\big(\multicompset(\MEC)\big)(\Czero)\) are vector spaces as per Lemmas~\ref{lemma-span-hat-y-vector-space}~and~\ref{lemma-multicomponents-inverse}, respectively.  Hence if \(\spanCone(\hat{\by})(\Czero)\nsubseteq \Delta\big(\multicompset(\MEC)\big)(\Czero)\) then there exists a vector \(\bv\in \mathbb{R}^{\Czero} \) that is orthogonal to \(\Delta\big(\multicompset(\MEC)\big)(\Czero)\) but not to \( \spanCone(\hat{\by})(\Czero)\). Notice also that if 	\(\spanCone(\hat{\by})(\Czero)=\{\vec{0} \}\) this lemma holds trivially, hence we can assume \(\hat{\by}\) is zero-unbounded on some \(c\in \Czero\).
	
	For each \(\bu\in \mathbb{R}^{\countersset(\A)} \), let \(\Pi(\bu) \) be the orthogonal projection of \(\bu \) onto \(\bv\).
	
	Let \(\MEC_\bv \) be the \(1\)-dimensional VASS MDP created from  \(\MEC \) by replacing every transition effect vector \(\bu_t \) with \(\Pi(\bu_t) \).  Let \(c_\bv\) be the only counter of \(\MEC_\bv \).
	
	Since \(\bv \) is orthogonal to \(\Delta\big(\multicompset(\MEC)\big)(\Czero)\) it holds for every multi-component \(\bx\in \multicompset(\MEC) \) that \(\Delta(\bx)(c_\bv)=0 \). Hence from Lemma~\ref{lemma:dichotomy} the maximal solution \(\ba_\bv,\bb_\bv \) of \hyperref[fig-systems]{(II)} for \(\MEC_\bv \) has \(\ba_\bv(c_\bv)>0 \) (and we can wlog. assume \(\ba_\bv(c_\bv)=1 \)).  But since \(\hat{\by} \) is zero-unbounded on \(c_\bv\) there exists a transition \(t=(p,\bu,q)\) such that \[
	\bb_\bv(\stateVassalt)-\bb_\bv(\stateVass)+  \bu_t(c_\bv)\cdot \ba_\bv(c_\bv)
	=
	\bb_\bv(\stateVassalt)-\bb_\bv(\stateVass)+  \Pi(\bu_t)
	< 0
	\]
	That is because if no such transition were to exist then each cycle over \(p_\by \) would have effect exactly \(0\) on the ranking function \(rank_\bv(p_\by\bu)=\bb_\bv(p_\by)+\bu(c_\bv) \), but that means the effect of each such cycle  on \(c_\bv\) is \(0\) and thus \(\hat{\by} \) would be zero-bounded on \(c_\bv\), which it is not.
	
	Since orthogonal projection is a linear transformation, we can express \(\Pi(\bu)=\sum_{c\in \countersset(\A)} a_c\cdot \bu(c) \) for some constants \(a_c\in \mathbb{R} \) which do not depend on \(\bu \) and such that \(a_c=0 \) for each \(c\notin \Czero \).
	
	Hence we can represent the solution \(\ba_\bv,\bb_\bv \) in terms of a \(\A_k \) using \(\ba'(c)=a_c  \) as \begin{gather*}
		\bb_\bv(\stateVassalt)-\bb_\bv(\stateVass)+  \Pi(\bu_t)
		=
		\bb_\bv(\stateVassalt)-\bb_\bv(\stateVass)+  \sum_{c\in \countersset(\A)} a_c\cdot \bu_t(c)
		=\\=
		\bb_\bv(\stateVassalt)-\bb_\bv(\stateVass)+  \sum_{c\in \countersset(\A)} \ba'(c)\cdot \bu_t(c)
		< 0
	\end{gather*}
	
	Notice that \(\ba',\bb_\bv \) can be seen as a solution of \hyperref[fig-systems]{(II)} for \(\A_k \) except it may not hold \(\ba'\geq \vec{0} \).
	
	Remember that \(\ba,\bb \) is some maximal solution of \hyperref[fig-systems]{(II)} for \(\A_k \). From Lemma~\ref{lemma:dichotomy} it holds \(\ba(\Czero)>\vec{0} \). Since these solutions are closed under addition (see Section~\ref{sec-systems}), we can wlog.  assume \(\ba>\ba' \) as well as \(\ba>-\ba' \). But this means that \(\ba+\ba'\geq \vec{0} \).
	
	Hence \((\ba+\ba'),(\bb+\bb_\bv)\) is a solution of \hyperref[fig-systems]{(II)} for \(\A_k \), and it holds \[
	(\bb+\bb_\bv)(\stateVassalt)-(\bb+\bb_\bv)(\stateVass)+ \sum_{c\in \countersset(\A)} (\ba+\ba')(c)\cdot \bu_t(c) < 0
	\]
	
	If \(t=(p,\bu,q) \) where \(p\in \statesDemon(\A) \) then this gives us from Lemma~\ref{lemma:dichotomy} that the maximal solution \(\bx \) of \hyperref[fig-systems]{(I)} for \(\A_k \) satisfies \(\bx(t)=0 \) and thus from the definition of \(\A_{k+1}\) it holds \(\A_k\neq \A_{k+1} \), a contradiction as per Lemma~\ref{lemma-k-converged}.
	
	Assume thus that \(p\in \statesProb(\A) \). But this means that for the counter \(r_k=rank_k \) defined by some maximal solution of \hyperref[fig-systems]{(II)} for \(\A_k \) there exists a transition 	\(t'\in \tout(p) \) with	 \(\bu_{t'}(r_k)<0 \), and thus \(t'\in M_{k+1}\) and therefore from the definition of \(\A_{k+1} \) it holds \(t'\notin \transitions(\A_{k+1}) \). But this is a contradiction as per Lemma~\ref{lemma-k-converged}  it holds \(\A_k=\A_{k+1}\).
\end{proof}

\subsection{Proofs of Section~\ref{sec-sigma-analysis}}
\label{sec-proofs-of-full-somehhh}

\subsubsection{Proof of Lemma~\ref{lemma-event-looping}}
\label{app-lemma-event-looping-proof}
\lemmaeventlooping*
\begin{proof}
	Let \(\kappa(i) \) represent the number of steps taken while looping during the first \(i\) loops (i.e., the number of steps whose effect is added to the looping-bin). Notice that the expected number of steps per single loop is constant in \(n\). Thus from Markov inequality we obtain \[\prob_{p\vec{n}}^\strategyDemon\big[\kappa(n^{\epsilonfour})\geq \lfloor\frac{n}{ \bincount}\rfloor\cdot \frac{1}{\maxupdate{\A}}\big]\leq \frac{a\cdot n^{\epsilonfour}}{n} \]  for some constant \(a\), where \(\maxupdate{\A}\) is the largest possible change of counter per transition in \(\A\).  Which from \(	\epsilonfour<1 \) converges to \(0 \) as \(n \) goes towards infinity. Since each step modifies each  counter by at most \(\maxupdate{\A} \) the lemma follows.
\end{proof}

\subsubsection{Proof of Lemma~\ref{lemma-e4}}\label{sec-proof-of-lemma-component-iterating-exp-not-too-long-prob}
\lemmae*


This follows directly from the next lemma and from \begin{equation}\label{eq_eps_1}
	\epsilonfour<\epsilonsingleiteration
\end{equation}
\vspace{-\belowdisplayskip}\vspace{-\belowdisplayskip}
\begin{restatable}{lemma}{dfdsfdsfsdfsdeeesss}
	\label{lemma-component-iterating-exp-not-too-long-prob}
	Let \(\A\) be a VASS MDP, \(\by\in\simplecompset(\A)\), \(k\in \N\), and \(\epsilonfour,\epsilonsingleiteration>0\) be such that \(\epsilonfour<\epsilonsingleiteration \). For each \(n\in \N\) let \(E_n \) be the set of all computations \(\pi \)  in \(\A_\by \) from \(\stateVass_\by \), such that the prefix of \(\pi \) of length \(2^{n^{\epsilonfour}}\) does not contain any sub-computation of length at most \(n^{\epsilonsingleiteration} \) that does not contain some transition or state of \(\A_\by \). Then it holds \(\lim_{n\rightarrow\infty}\prob^{\strategyDemon_\by}_{\stateVass_\by}[E_n]=1 \).
\end{restatable}

\begin{proof}
	Let us fix some transition \(t=(p,\bu,q)\). From Lemma~\ref{lemma-component-recentering} we can  wlog. assume \(p=p_\by \), and that \(t\) is the only outgoing transition of \(p\) (as otherwise we can split the transition \(t\) into two by adding a new non-deterministic state \(p'\) along with a new transition \((p,\vec{0},p') \) and replacing \(t=(p,\bu,q) \) with \((p',\bu,q \)).	 
	
	Notice it suffices to prove a version of this lemma for this \(t\) only (as opposed to all transitions and all states), as this implies the full version of the lemma.
	
	Hence it suffices to show that while iterating \(\by \) not a single iteration of \(\by \) takes more than \(n^{\epsilonsingleiteration} \) steps within the first \(2^{n^{\epsilonfour}} \) total steps with very high probability. Formally, let \(E_n' \) be the set of all computations in \(\A_\by \) of length at most \(2^{n^{\epsilonfour}} \) that do not contain a sub-computation of length at least \(n^{\epsilonsingleiteration} \) which does not contain \(p \). We show that \(\lim_{n \rightarrow \infty}\prob_{p}^{\strategyDemon_\by}[E_n']=1 \).

	Notice that for each state \(r \) of \(\A_\by \) there exists a simple path \(\pi \) from \(r \)  to \(\stateVass_\by \) with \(\prob_{r}^{\strategyDemon_\by}[\pi]\geq |\states(\A)|\cdot\min_{t\in\transitions(\A)}P(t)\geq a  \) for some constant \(a>0\) (here we use \(P(t)=1\) for transition from non-deterministic states). Hence the probability of a single iteration of \(\by \) taking more than \(n^{\epsilontwo}\) steps can be upper bounded by \[(1-a)^{\lfloor\frac{n^{\epsilonsingleiteration}}{|\states(\A)|}\rfloor} \]
	
	Thus within the first \(2^{n^{\epsilonfour}}\) iterations of \(\by \) it holds \[\prob^{\strategyDemon_\by}[E_n']\geq 1-2^{n^{\epsilonfour}} \cdot  (1-a)^{\lfloor\frac{n^{\epsilonsingleiteration}}{|\states(\A)|}\rfloor} \]

	Let \(k \) be such that \(2\cdot (1-a)^{k}<1 \) (the existence of \(k\) follows from \(1-a<1\)). It holds
	\begin{gather*}
		2^{n^{\epsilonfour}}\cdot (1-a)^{\lfloor\frac{n^{\epsilonsingleiteration}}{|\states(\A)|}\rfloor}
		=
		(2\cdot (1-a)^{k})^{n^{\epsilonfour}} \cdot (1-a)^{\lfloor\frac{n^{\epsilonsingleiteration}}{|\states(\A)|}\rfloor - k\cdot n^{\epsilonfour}}
	\end{gather*}
	from \(\epsilonfour<\epsilonsingleiteration \) it holds \[\lim_{n\rightarrow\infty}\lfloor\frac{n^{\epsilonsingleiteration}}{|\states(\A)|}\rfloor - k\cdot n^{\epsilonfour}= \infty \] 
	Hence
	\begin{gather*}
		\lim_{n\rightarrow\infty}\prob_{\stateVass}^{\strategyDemon_\by}[E_n']
		\geq 
		\lim_{n\rightarrow\infty}\big( 1-2^{n^{\epsilonfour}} \cdot  (1-a)^{\lfloor\frac{n^{\epsilonsingleiteration}}{|\states(\A)|}\rfloor}\big)
		= \\=
		\lim_{n\rightarrow\infty}\Big( 1-\big(2\cdot (1-a)^{k}\big)^{n^{\epsilonfour}} \cdot (1-a)^{\lfloor\frac{n^{\epsilonsingleiteration}}{|\states(\A)|}\rfloor - k\cdot n^{\epsilonfour}}\Big)
		=1-0=1	
	\end{gather*}
\end{proof}

\subsubsection{Proof of Lemma~\ref{lemma-event-finit}}
\label{app-lemma-event-finit-proof}
\lemmaeventfinit*
\newcommand{\eventginit}{E_{\gresets}}
\begin{proof}	
	Let \(	\eventginit	\) be the event that \(\gresets(1)=\gresets(2)=0 \). Notice that  \(\eventfinit\subseteq\eventginit \) since \(\eventginit\) implies \(\fsize(1)=1\), \(\fsize(2)=2 \), and \(\fsize(3)=4\) from \eqref{eq-fsize-def}.

	\(	\eventginit	\) is equivalent to the first two loops performing, for each \(j\in\{1,\dots,w\} \), exactly three iterations of the multi-component \(\bx_j \cdot \lfloor n^{1-\epsilonone} \rfloor \).  But this happens exactly when, for each \(j\in\{1,\dots,w\} \), the probabilistic part of the first two iterations of \(\bx_j \cdot \lfloor n^{1-\epsilonone} \rfloor \) does not at any point exceed the effect \(n^{1-\epsilonthree} \) on any \(c\in\Czero \).
	
	Furthermore, conditioned on \(\eventshortsucc\), these iterations of all the \(\bx_j \cdot \lfloor n^{1-\epsilonone} \rfloor \), summed over all \(j\in\{1,\dots,w\}\), take at most \(2\cdot \sum_{\by\in\simplecompset(\A_k)}a_\by\cdot \lfloor n^{1-\epsilonone} \rfloor\cdot  n^{\epsilonsingleiteration}  \leq n^{\epsilonbigY} \) steps, where we use \begin{equation}\tag{\ref{eq-eps-new-3}}
		\epsilonbigY>1-\epsilonone+\epsilonsingleiteration
	\end{equation}
	
	For a given counter \(c\in \Czero\) let \(E_n^c \) be the event that the probabilistic part of the computation does not change the counter \(c \) by more than \(n^{\epsilonbigZ} \) within the first \(n^{\epsilonbigY}\) steps. 	Since the probabilistic part of every simple component is either zero-bounded or zero-unbounded on each \(c\in\countersset(\A)\), and any computation on \(\A\) can be seen as an interweaving of iterations of simple components (see Section~\ref{sec-pointing}), using the results about 1-dim VASS MDPs from \cite{AKCONCUR23} we get that as long as 
	\begin{equation}\label{eq-eps-new-xtwo}
		\epsilonbigZ<2\cdot \epsilonbigY
	\end{equation}
	it holds \[\liminf_{n\rightarrow\infty} \prob_{p\vec{n}}^\strategyDemon[E_n^c]=1 \]  
	
	
	But since \begin{equation}\label{eq-eps-new-xthree}
		\epsilonbigZ<1-\epsilonthree
	\end{equation} this implies that \[\liminf_{n\rightarrow\infty} \prob_{p\vec{n}}^\strategyDemon[\eventfinit\mid\eventshortsucc]\geq \liminf_{n\rightarrow\infty} \prob_{p\vec{n}}^\strategyDemon[\eventginit\mid\eventshortsucc]\geq \liminf_{n\rightarrow\infty} \prob_{p\vec{n}}^\strategyDemon[\bigcap_{c\in \Czero} E_n^c]=1\]
\end{proof}
\subsubsection{Proof of Lemma~\ref{lemma-E-expsize}}\label{sec-lemma-E-expsize-proof}
\lemmaEexpsize*
\begin{proof}
	Notice that \(\gresets(i)\leq w\cdot \fsize(i)\). Hence for all \(i\geq 2\), for which the \((i-1) \)-st loop finished, it holds from~\eqref{eq-fsize-def} that
	\begin{align*}
		\fsize(i)
		&\geq
		\big\lfloor\frac{2\cdot \fsize(i-1)\cdot \lfloor n^{1-\epsilontwo} \rfloor - w\cdot \fsize(i-i)\cdot n^{1-\epsilonthree+\epsilonX}}{\lfloor n^{1-\epsilontwo} \rfloor}\big\rfloor
		\\&\geq
		\lfloor 2\cdot \fsize(i-1) - \frac{1}{10} \cdot \fsize(i-i)\rfloor
		\\&=\lfloor 1.9\cdot \fsize(i-1)\rfloor  \end{align*}
	
	where the second inequality follows from \begin{equation}\label{eq-eps-new-1}
		1-\epsilontwo>1-\epsilonthree+\epsilonX
	\end{equation}
	
	We  proceed with induction over \(i\geq 3 \) to prove that \(\fsize(i)\geq 1.5^i \) conditioned on \(\eventfinit\).
	
	\textbf{Base case \(i=3\):} From the conditioning on \(\eventfinit\) it holds \(\fsize(3)=4\geq 1.5^3=3.375 \), base case holds.
	
	\textbf{Induction step:} Assume \(i\geq 4\), \(\fsize(i-1)\geq 1.5^{i-1} \), and  the  \((i-1) \)-st loop finished. Then 
	\[
	\fsize(i)
	\geq
	\lfloor 1.9\cdot \fsize(i-1)\rfloor
	\geq   
	\lfloor 1.9\cdot 1.5^{i-1}\rfloor
	=
	\lfloor 0.4\cdot 1.5^{i-1} + 1.5^i \rfloor				
	\]
	
	Since \( 0.4\cdot 1.5^{i-1}\geq  0.4\cdot 1.5^{3}= 1.35\) it  holds 
	\[
	\fsize(i)	
	\geq 
	\lfloor 0.4\cdot 1.5^{i-1}+1.5^i \rfloor		\geq \lfloor1 + 1.5^i \rfloor \geq 1.5^i
	\] 
	Induction step holds.
\end{proof}

\subsubsection{Proof of Lemma~\ref{lemma-eventresets}}\label{sec-lemma-eventresets-proof}
\lemmaeventresets*
\begin{proof}
	Notice that the following implies \(\eventresets\): if the probabilistic part from any  \(n^{\espilonresets+\epsilonANYSMALL}\)  iterations of any \(\bx_j \cdot \lfloor n^{1-\epsilonone} \rfloor\) does not at any point exceed effect  \(n^{1-\epsilonthree} \) on any counter, within the first \(n^{\epsilonfour} \) loops.

	Conditioned on \(\eventshortsucc\), each single iteration of \(\bx_j \cdot \lfloor n^{1-\epsilonone} \rfloor\) consists of  \(\sum_{\by\in \simplecompset(\A_k)}a_\by \) iterations of simple components,and  each iteration of any simple component takes at most \(n^{\epsilonsingleiteration} \) steps. Hence \(n^{\espilonresets+\epsilonANYSMALL}\) iterations of multi-components \(\bx_j \cdot \lfloor n^{1-\epsilonone} \rfloor\) take at most \[n^{\espilonresets+\epsilonANYSMALL}\cdot \sum_{\by\in\simplecompset(\A_k)}a_\by \cdot  n^{\epsilonsingleiteration}\leq n^{\epsilonTECHNICALHERE-\epsilonFIXINGDIFFERENTSMALL}\] steps where the ineqaulity follows from \begin{equation}\label{eq-eps-technical}
		\espilonresets+\epsilonANYSMALL+\epsilonsingleiteration<\epsilonTECHNICALHERE-\epsilonFIXINGDIFFERENTSMALL
	\end{equation}
	
	From Lemma~\ref{lemma-random-part-prob-bound} the probability of this many steps exceeding the effect \(n^{1-\epsilonthree-\epsilonANYSMALLDIFFERENT} \) on any counter is upper bounded by\begin{gather*}
		f\cdot \exp\big( \frac{-(n^{1-\epsilonthree-\epsilonANYSMALLDIFFERENT}-g)^2
		}{
			h\cdot n^{\epsilonTECHNICALHERE-\epsilonFIXINGDIFFERENTSMALL}
		}\big) 
		\leq 
		f\cdot \exp\big( \frac{-n^{2\cdot (1-\epsilonthree-2\cdot \epsilonANYSMALLDIFFERENT)}
		}{
			n^{\epsilonTECHNICALHERE}
		}\big) 
		=\\=
		f\cdot \exp\big(-n^{\frac{1}{2}-2\cdot \epsilonthree-4\cdot \epsilonANYSMALLDIFFERENT)}
		\big) 
	\end{gather*} for some constants \(f,g,h\) that depend only on \(\A\).

	During the \(i\)-th loop each multi-component \(\bx_j \cdot \lfloor n^{1-\epsilonone} \rfloor\) is iterated at most \(\fsize(i)\leq 2^i\) times. Hence within the first \(n^{\epsilonfour} \) loops the number of multi-component iterations is at most \(w\cdot \sum_{i=1}^{n^{\epsilonfour}}2^i\leq w\cdot 2\cdot 2^{n^{\epsilonfour}} \).  This thus gives  \[\prob_{p\vec{n}}^\strategyDemon[\eventresets\mid \eventshortsucc] \geq 1- w\cdot 2\cdot 2^{n^{\epsilonfour}}  \cdot f\cdot \exp\big( -n^{\frac{1}{2}-2\cdot \epsilonthree-4\cdot \epsilonANYSMALLDIFFERENT }
	\big) \]
	
	which converges to \(1 \) as \(n\rightarrow\infty\) from \begin{equation}\label{eq-eps-technicaldif}
		\epsilonfour <\frac{1}{2}-2\cdot \epsilonthree-4\cdot \epsilonANYSMALLDIFFERENT
	\end{equation}
\end{proof}

\subsubsection{Proof of Lemma~\ref{lemma-eventsmallprobsingle}}\label{sec-lemma-eventsmallprobsingle-proof}
\lemmaeventsmallprobsingle*
\begin{proof}
	Conditioned on  \(\eventshortsucc\) a single iteration of any \(\bx_j \cdot \lfloor n^{1-\epsilonone} \rfloor \) takes at most \[ n^{\epsilonsingleiteration}\cdot\sum_{\by\in \simplecompset(\A_k)}a_\by \cdot\lfloor n^{1-\epsilonone} \rfloor \leq n^{\epsilonNEWFIXSECOND}\] steps where we use 	\begin{equation}\label{eq-eps-new tesitng}
		\epsilonsingleiteration+1-\epsilonone<\epsilonNEWFIXSECOND
	\end{equation} From Lemma~\ref{lemma-random-part-prob-bound} we get an upper bound on the probability of this many steps exceeding the effect \(n^{1-\epsilonthree}\) on any counter in the probabilistic part of
	\[
	f\cdot \exp\big( \frac{-(n^{1-\epsilonthree}-g)^2
	}{
		h\cdot n^{\epsilonNEWFIXSECOND}
	}\big) 
	\leq 
	f\cdot \exp\big( \frac{-n^{2-2\cdot \epsilonthree-3\cdot \epsilonANYSMALLDIFFERENT}
	}{
		n^{\epsilonNEWFIXSECOND+\epsilonANYSMALLDIFFERENT}
	}\big)
	= 
	f\cdot \exp( -n^{0.5-2\cdot \epsilonthree-4\cdot \epsilonANYSMALLDIFFERENT})
	\] for some constants  \(f,g\), and \(h\) that depend only on \(\A \).

	During the \(i\)-th loop each such multi-component is iterated at most \(\fsize(i)\cdot w\leq 2^i\cdot w \) times. Hence during the first \(n^{\epsilonfour}\) loops the maximal number of multi-component iterations is \[\sum_{i=1}^{n^{\epsilonfour}} 2^i\cdot w = 2\cdot (2^{n^{\epsilonfour}}-1)\leq 2^{n^{\epsilonfour}+1} \]
	
	Therefore the probability of there being at least one iteration of any \(\bx_j \cdot \lfloor n^{1-\epsilonone} \rfloor \) withing the first \(n^{\epsilonfour}\) loops that has \(|\hat{\bv}(c)|> n^{1-\epsilonthree} \) on any counter is upper bounded by \[2^{n^{\epsilonfour}+1}\cdot 	f\cdot \exp( -n^{0.5-2\cdot \epsilonthree-4\cdot \epsilonANYSMALLDIFFERENT}) \]
	which converges to \(0\) as \(n\) tends to \(\infty \) from \begin{equation}\tag{\ref{eq-eps-technicaldif}}
		\epsilonfour<\frac{1}{2}-2\cdot \epsilonthree-4\cdot \epsilonANYSMALLDIFFERENT
	\end{equation}
\end{proof}

\subsubsection{Proof of Lemma~\ref{lemma-eventdoublereset}}
\label{app-lemma-eventdoublereset-proof}  
\lemmaeventdoublereset*
\begin{proof}
	It suffices to show the effect of the probabilistic part from any single iteration of any \(\bx_j \cdot \lfloor n^{1-\epsilonone} \rfloor+\bz\) is less than \(n^{1-\epsilonthree} \) on each counter, within the first \(n^{\epsilonfour} \) loops.
	
	
	
	Let \(E_i \) be the event that the effect on counters  in the probabilistic part does not exceed \(n^{1-\epsilonthree} \) at any point during a single iteration of any \(\bx_j \cdot \lfloor n^{1-\epsilonone} \rfloor+\bz\), within the first \(i \) iterations of the multi-components \(\bx_j \cdot \lfloor n^{1-\epsilonone} \rfloor+\bz\) (not counting iterations of just \(\bx_j \cdot \lfloor n^{1-\epsilonone} \rfloor\))
	

	Conditioned on \(E_i \) when the condition \(|\bv(\Czero)|\geq  n^{1-\epsilonthree} \) holds for the \((i+1) \)-st time and  some \(\bx_j \cdot \lfloor n^{1-\epsilonone} \rfloor+\bz\) is to be iterated for the \((i+1)\)-st time, this condition was exceeded by iterating \(\bx_j \cdot \lfloor n^{1-\epsilonone} \rfloor\). Hence, conditioned on \(\eventsmallprobsingle\), at that point it holds \[|\bv(\Czero)|\leq n^{1-\epsilonthree}+n^{1-\epsilonthree} \]  From basic properties of vector spaces we thus have that \[|\bz|\leq a\cdot |\bv(\Czero)| \leq a\cdot n^{1-\epsilonthree} \leq n^{\epsilonbigO}    \] for some constant \(a\) where the last inequality holds from \begin{equation}
		\epsilonbigO>1-\epsilonthree
		\label{eq-eps-new-3}
	\end{equation}
	
	Hence  one such iteration of \(\bx_j \cdot \lfloor n^{1-\epsilonone} \rfloor+\bz\) involves at most \(\sum_{\by\in \simplecompset(\A_k)}(a_\by+n^{\epsilonbigO}) \)	iterations of simple components, and from conditioning on \(\eventshortsucc\) this gives at most \[n^{\epsilonsingleiteration}\cdot \sum_{\by\in \simplecompset(\A_k)}(a_\by+n^{\epsilonbigO})\leq n^{\epsilondoublebigOO}  \] steps in total, where we use \begin{equation}\label{eq-eps-newly}
		\epsilonsingleiteration+ \epsilonbigO<\epsilondoublebigOO
	\end{equation}
	
	Hence from Lemma~\ref{lemma-random-part-prob-bound} we get an upper bound on the probability of this number of steps exceeding effect \(n^{1-\epsilonthree} \) on any counter, which gives us
	\begin{gather*}	
		\prob_{p\vec{n}}^\strategyDemon[E_{i+1}\mid E_{i},\eventshortsucc]
		\geq
		1- f\cdot \exp\big( \frac{-(n^{1-\epsilonthree}-g)^2
		}{
			h\cdot n^{\epsilondoublebigOO}
		}\big) 
		\geq\\\geq 
		1- f\cdot \exp\big( \frac{-n^{2-2\cdot \epsilonthree-3\cdot \epsilonFIXINGDIFFERENTSMALL}
		}{
			n^{\epsilondoublebigOO+\epsilonFIXINGDIFFERENTSMALL}
		}\big) 
		==
		1- f\cdot \exp\big( -n^{0.5-2\cdot \epsilonthree-4\cdot \epsilonFIXINGDIFFERENTSMALL}
		\big)
	\end{gather*}
	where \(f,g\), and \(h \) are some  constants that depend only on \(\A \).
	
	Since \(\eventdoublereset\subseteq E_{\fsize(n^{\epsilonfour})} \) it thus holds \begin{gather*}
		\prob_{p\vec{n}}^\strategyDemon[\eventdoublereset\mid \eventshortsucc] 
		\geq
		\prod_{i=1}^{\fsize(n^{\epsilonfour})}  \prob_{p\vec{n}}^\strategyDemon[E_{i+1}\mid E_{i},\eventshortsucc]
		\geq\\\geq
		\prod_{i=1}^{\fsize(n^{\epsilonfour})} \big(
		1- f\cdot \exp( -n^{0.5-2\cdot \epsilonthree-4\cdot \epsilonFIXINGDIFFERENTSMALL} )
		\\\allowdisplaybreaks\geq 
		1-\fsize(n^{\epsilonfour})\cdot  f\cdot \exp( -n^{0.5-2\cdot \epsilonthree-4\cdot \epsilonFIXINGDIFFERENTSMALL} )
		\\\allowdisplaybreaks\geq
		1-2^{n^{\epsilonfour}}\cdot  f\cdot \exp( -n^{0.5-2\cdot \epsilonthree-4\cdot \epsilonFIXINGDIFFERENTSMALL} )
	\end{gather*}
	where the last inequality follows from  \(\fsize(n^{\epsilonfour})\leq 2^{n^{\epsilonfour}}\). But from 
	\begin{equation}\tag{\ref{eq-eps-technicaldif}}
		\epsilonfour<\frac{1}{2}-2\cdot \epsilonthree-4\cdot \epsilonANYSMALLDIFFERENT
	\end{equation}
	it holds \[ \lim_{n\rightarrow\infty}\prob_{p\vec{n}}^\strategyDemon[\eventdoublereset\mid \eventshortsucc]  \geq \lim_{n\rightarrow\infty}\big( 1-2^{n^{\epsilonfour}}\cdot  f\cdot \exp( -n^{0.5-2\cdot \epsilonthree-2\cdot \epsilonFIXINGDIFFERENTSMALL} )\big)=1 \]
\end{proof}

\subsubsection{Proof of Lemma~\ref{lemma-eventbzbound}}\label{sec-lemma-eventbzbound-proof}
\lemmaeventbzbound*
\begin{proof}
	From  \(\eventdoublereset \) whenever some multi-component \(\bx_j \cdot \lfloor n^{1-\epsilonone} \rfloor+\bz\) is to be iterated, the bound \(|\bv(\Czero)|\geq n^{1-\epsilonthree}\)  was exceeded after iterating \(\bx_j \cdot \lfloor n^{1-\epsilonone}\rfloor \). Hence, from \(\eventsmallprobsingle\) at that point it holds \(|\bv(\Czero)|\leq n^{1-\epsilonthree}+n^{1-\epsilonthree} \). From basic properties of vector spaces we thus have \begin{equation}\label{eq-weird-biC-internal}
		|\bz|\leq a\cdot |\bv(\Czero)| \leq a\cdot n^{1-\epsilonthree}\leq n^{\epsilonbigO}
		\tag{*}    \end{equation} for some constant \(a\) where the last inequality holds from 
	\begin{equation}\tag{\ref{eq-eps-new-3}}
		\epsilonbigO>1-\epsilonthree
	\end{equation}
	
\end{proof}

\subsubsection{Proof of Lemma~\ref{lemma-E-bigC}}\label{sec-lemma-E-bigC-proof}
\lemmaEbigC*
\begin{proof}	
	Let \(E_i \) be the event that during the first \(i\) loops, the main-bin never becomes negative on any counter, and every counter \(c\in C_+ \) has, in the main-bin, when the \(j \)-th loops finishes value at least \(2\cdot \fsize(j) \cdot \lfloor n^{1-\epsilonone} \rfloor \). Note that for \(\eventbigcounter=E_{\lfloor n^{\epsilonfour}\rfloor}  \). 
	
	From the following Lemma~\ref{lemma-sommmm} it thus holds
	\begin{gather*}
		\prob_{p\vec{n}}^\strategyDemon[\eventbigcounter\mid \eventshortsucc,\eventdoublereset,\eventresets,\eventbzbound]
		\geq \allowdisplaybreaks\\\geq
		\prod_{i=1}^{\lfloor n^{\epsilonfour}\rfloor} \prob_{p\vec{n}}^\strategyDemon[E_i\mid E_{i-1}, \eventshortsucc,\eventdoublereset,\eventresets,\eventbzbound]
		\allowdisplaybreaks\\\geq 
		\prod_{i=1}^{\lfloor n^{\epsilonfour}\rfloor}  \Big( 1-f\cdot \exp\big(- \fsize(i)\cdot n^{0.5-2\cdot \epsilonone-4\cdot \epsilonFIXINGDIFFERENTSMALL} 
		\big)
		\Big)
		\allowdisplaybreaks\\\geq 
		1-\sum_{i=1}^{\lfloor n^{\epsilonfour}\rfloor}  f\cdot \exp\big(- \fsize(i)\cdot n^{0.5-2\cdot \epsilonone-4\cdot \epsilonFIXINGDIFFERENTSMALL} 
		\big)
		\allowdisplaybreaks\\\geq 
		1- \lfloor n^{\epsilonfour}\rfloor\cdot  f\cdot \exp\big(-  n^{0.5-2\cdot \epsilonone-4\cdot \epsilonFIXINGDIFFERENTSMALL} 
		\big)
	\end{gather*}
	where the last inequality follows from \(\fsize(i)\geq 1 \). This converges to \(1\) as \(n\) tends to $\infty$ from \begin{equation}\label{eq-eps-harddd}
		\frac{1}{2}-2\cdot \epsilonone-4\cdot \epsilonFIXINGDIFFERENTSMALL>0
	\end{equation}
\end{proof}

\begin{lemma}\label{lemma-sommmm}
	For each \(1\leq i\) it holds \[ \prob_{p\vec{n}}^\strategyDemon[E_i\mid E_{i-1}, \eventshortsucc,\eventdoublereset,\eventresets,\eventbzbound]
	\geq
	1-f\cdot \exp\big(- \fsize(i)\cdot n^{0.5-2\cdot \epsilonone-4\cdot \epsilonFIXINGDIFFERENTSMALL} 
	\big)\] where \(f\) is some constant that depends only on \( \A\).
\end{lemma}
\begin{proof}

	We do separately the case \(i=1 \) and \(i>1 \).
	
	\textbf{Case \(i=1\):} If the first loop finishes then from the conditioning on  \(\eventshortsucc\) the sum of all the multi-components iterated by \(\sigma\) during the \(1\)-st loop is exactly \(\bx\cdot \lfloor n^{1-\epsilonone} \rfloor \). Hence the effect from the deterministic part of the computation on the main bin for is \(\sum_{\by\in\simplecompset(\A_k)} a_\by\cdot \E[\realeffect_\by] \cdot \lfloor n^{1-\epsilonone} \rfloor \). It holds \(\E[\realeffect_\by](\Czero)=\vec{0} \) and for each \(c\in C_+\) it holds  from \eqref{eq-multi-comp-it-big} that \(\sum_{\by\in\simplecompset(\A_k)} a_\by\cdot \E[\realeffect_\by](c) \cdot \lfloor n^{1-\epsilonone} \rfloor\geq 3\cdot \lfloor n^{1-\epsilonone} \rfloor \). Therefore when the \(1\)-st loop finishes the main-bin contains the counters vector \(\bv+\hat{\bv}\) with \((\bv+\hat{\bv})(\Czero)=\frac{\vec{n}}{\bincount} \) and \(\bv(C_+)\geq \frac{\vec{n}}{\bincount} + \vec{3}\cdot \lfloor n^{1-\epsilonone}\rfloor\) where \(\hat{\bv}\) is the effect of these iterations of multi-components on the probabilistic part of the computation. 
	
	From \(\eventshortsucc\), iterating all these multi-components takes at most \[ n^{\epsilonsingleiteration}\cdot \sum_{\by\in\simplecompset(\A_k)} a_\by\cdot \lfloor n^{1-\epsilonone} \rfloor \leq n^{\epsilonNEWFIXSECOND}\] steps where we use 
	\begin{equation}\tag{\ref{eq-eps-new tesitng}}
		\epsilonsingleiteration+1-\epsilonone<\epsilonNEWFIXSECOND
	\end{equation}
	
	Hence we can apply
	Lemma~\ref{lemma-random-part-prob-bound} to obtain an upper bound on \(|\hat{\bv}|\geq \lfloor n^{1-\epsilonone} \rfloor \) within this number of steps
	\begin{gather*}
		\prob_{p\vec{n}}^\strategyDemon\big[|\hat{\bv}|\geq \lfloor n^{1-\epsilonone} \rfloor\big]
		\leq
		f\cdot \exp\big( \frac{-(\lfloor n^{1-\epsilonone} \rfloor-g)^2
		}{
			h\cdot n^{\epsilonNEWFIXSECOND}
		}\big)
		\leq 
		f\cdot \exp\big( \frac{- n^{2-2\cdot \epsilonone-3\epsilonFIXINGDIFFERENTSMALL} 
		}{
			n^{\epsilonNEWFIXSECOND+\epsilonFIXINGDIFFERENTSMALL}
		}\big)
		=\\=
		f\cdot \exp( - n^{0.5-2\cdot \epsilonone-4\cdot \epsilonFIXINGDIFFERENTSMALL} 
		)
	\end{gather*} for some constants \(f,g,h \) that depend only on \(\A \). Note that  \(|\hat{\bv}|\geq \lfloor n^{1-\epsilonone} \rfloor\) implies 	\[(\bv+\hat\bv)(C_+)\geq \frac{\vec{n}}{\bincount} + \vec{3}\cdot \lfloor n^{1-\epsilonone}\rfloor + \hat\bv \geq 2 \cdot \lfloor n^{1-\epsilonone} \rfloor =2\cdot \fsize(i) \cdot \lfloor n^{1-\epsilonone} \rfloor \] hence \[\prob_{p\vec{n}}^\strategyDemon[E_1\mid \eventshortsucc,\eventdoublereset,\eventresets,\eventbzbound ]\geq 1-f\cdot \exp( - n^{0.5-2\cdot \epsilonone-4\cdot \epsilonFIXINGDIFFERENTSMALL} 
	) \]

	

	\textbf{Case \(i>1\):} 		The number of multi-component iterations during the \(i\)-th loop is exactly \(w\cdot \fsize(i) \cdot \lfloor n^{1-\epsilonone} \rfloor \). Out of these, exactly \(\gresets(i) \) are iterations of some \(\bx_j \cdot \lfloor n^{1-\epsilonone} \rfloor+\bz\) and the rest are iterations of \(\bx_j \cdot \lfloor n^{1-\epsilonone} \rfloor\). Conditioned on \(\eventbzbound\) it holds \(|\bz|\leq n^{\epsilonbigO} \). This gives at most \[w\cdot \fsize(i) \cdot \lfloor n^{1-\epsilonone} \rfloor \sum_{\by\in\simplecompset(\A_k)} a_\by  + \gresets(i)\cdot |\simplecompset(\A_k)| \cdot  n^{\epsilonbigO} \] simple component iterations during the \(i\)-th loop. Since \(\gresets(i)\leq w\cdot \fsize(i) \) this can be rewritten as \begin{gather*}
		w\cdot \fsize(i) \cdot \lfloor n^{1-\epsilonone} \rfloor \sum_{\by\in\simplecompset(\A_k)} a_\by  + \gresets(i)\cdot |\simplecompset(\A_k)| \cdot  n^{\epsilonbigO} 
		\leq \\\leq
		w\cdot \fsize(i) \cdot \big(\lfloor n^{1-\epsilonone} \rfloor \sum_{\by\in\simplecompset(\A_k)} a_\by  + 	|\simplecompset(\A_k)| \cdot  n^{\epsilonbigO} \big)
	\end{gather*} Conditioned on \(\eventshortsucc \) this gives at most \[n^{\epsilonsingleiteration}\cdot 	w\cdot \fsize(i) \cdot \big(\lfloor n^{1-\epsilonone} \rfloor \sum_{\by\in\simplecompset(\A_k)} a_\by  + 	|\simplecompset(\A_k)| \cdot  n^{\epsilonbigO} \big) \leq \fsize(i)\cdot  n^{\epsilonSOMETHINGNEW} \] steps where we use 
	\begin{equation}
		\tag{\ref{eq-eps-new tesitng}}
		\epsilonsingleiteration+1-\epsilonone<\epsilonSOMETHINGNEW
	\end{equation}
	\begin{equation}\tag{\ref{eq-eps-newly}}
		\epsilonsingleiteration+ \epsilonbigO<\epsilondoublebigOO
	\end{equation}
	Let \(\hat{F} \) be the event that at any point the probabilistic part of the computation during these iterations of multi-components modifies any counter by more than \( \fsize(i)\cdot  n^{1-\epsilonone-\epsilonFIXINGDIFFERENTSMALL} \) in total. From	Lemma~\ref{lemma-random-part-prob-bound} we have 
	\begin{gather*}
		\prob_{p\vec{n}}^\strategyDemon[\hat{F}] 
		\leq
		f\cdot \exp\big( \frac{-(\fsize(i)\cdot n^{1-\epsilonone-\epsilonFIXINGDIFFERENTSMALL} -g)^2
		}{
			h\cdot \fsize(i)\cdot  n^{\epsilonSOMETHINGNEW}
		}\big)
		\leq \\\leq
		f\cdot \exp\big( \frac{- \fsize^2(i)\cdot n^{2-2\cdot \epsilonone-3\cdot \epsilonFIXINGDIFFERENTSMALL} 
		}{
			\fsize(i)\cdot n^{\epsilonSOMETHINGNEW+\epsilonFIXINGDIFFERENTSMALL}
		}\big)
		=
		f\cdot \exp\big(- \fsize(i)\cdot n^{0.5-2\cdot \epsilonone-4\cdot \epsilonFIXINGDIFFERENTSMALL} 
		\big)
	\end{gather*} for some constants \(f,g,h \) that depend only on \(\A \). 
	
	We now proceed to show that conditioned on \(\hat{F} \) it is not possible for any counter to become negative in the main-bin during the \(i\)-th loop. We prove this separately for the counters from from \(\Czero \) and \(C_+ \).

	
	\textbf{Case \(c\in \Czero\):} Remember that for these counters only the effect on the deterministic part of the computation is put into the main-bin. 
	
	Furthermore, whenever any multi-component iteration finishes the effect on the counter \(\Czero \) in the main-bin is exactly zero since \(\Delta(\bx_j)(\Czero)=\vec{0} \) and the effect of \(\bz \) on these counters is perfectly  compensated for. Therefore if any of these counters were to become negative it has to happen during a single multi-component iteration.
	
	Whenever any multi-component \(\bx_j \cdot \lfloor n^{1-\epsilonone} \rfloor+\bz\) is to be iterated it holds from the conditioning on \(\eventbzbound\) that \(|\bz|< n^{\epsilonbigO} \). Therefore the number of simple-component iterations per single multi-component iteration is at most \(\sum_{\by\in\simplecompset(\A_k)}a_\by \cdot \lfloor n^{1-\epsilonone} \rfloor  +   n^{\epsilonbigO}\). This gives a maximal possible decrease during single multi-component iteration as \[\big(\sum_{\by\in\simplecompset(\A_k)}a_\by \cdot \lfloor n^{1-\epsilonone} \rfloor  +   n^{\epsilonbigO}\big)\cdot \max_{\by\in\simplecompset(\A_k)} \max_{c\in\countersset(\A)}\big|\E[\realeffect_\by](c)\big| \] Since at the start of the multi-component iteration each counter \(c\in\Czero \) is equal to \(\frac{n}{\bincount} \) it follows from \(
	\epsilonbigO<1
	\) that these counters cannot become negative during any single multi-component iteration. 
	
	\textbf{Case \(c\in C_+ \):}	 From the conditioning on \(\eventbzbound\), whenever a multi-component \(\bx_j \cdot \lfloor n^{1-\epsilonone} \rfloor+\bz\) is to be iterated it holds \(|\bz|\leq n^{\epsilonbigO} \).
	
	The number of times some \(\bx_j \cdot \lfloor n^{1-\epsilonone} \rfloor+\bz\) is iterated is \(\gresets(i) \). Therefore the maximal change on the deterministic part of the computation for any counter \(c\in C_+ \) in the main-bin during the \(i\)-th loop is \begin{gather*}
		\big(\fsize(i)\cdot \sum_{\by\in\simplecompset(\A_k)} a_\by \cdot \lfloor n^{1-\epsilonone} \rfloor + \gresets(i)\cdot n^{\epsilonbigO}  \big) \cdot \max_{\by\in\simplecompset(\A_k)} \max_{c\in\countersset(\A)}\big|\E[\realeffect_\by](c)\big|
		\leq\\\leq 
		\big(\fsize(i)\cdot \sum_{\by\in\simplecompset(\A_k)} a_\by\cdot \lfloor n^{1-\epsilonone} \rfloor + \frac{\fsize(i)\cdot \lfloor n^{1-\epsilonone} \rfloor }{n^{\espilonresets}}\cdot n^{\epsilonbigO}  \big) \cdot \max_{\by\in\simplecompset(\A_k)} \max_{c\in\countersset(\A)}\big|\E[\realeffect_\by](c)\big|
	\end{gather*}
	where the inequality follows from \(\gresets(i)\leq \frac{\fsize(i)\cdot \lfloor n^{1-\epsilonone} \rfloor }{n^{\espilonresets}} \) which we get from the conditioning on  \(\eventresets\). This can be simplified as 
	\begin{gather*}
		\big(\fsize(i)\cdot \sum_{\by\in\simplecompset(\A_k)} a_\by\cdot \lfloor n^{1-\epsilonone} \rfloor + \frac{\fsize(i)\cdot \lfloor n^{1-\epsilonone} \rfloor }{n^{\espilonresets}}\cdot n^{\epsilonbigO}  \big) \cdot \max_{\by\in\simplecompset(\A_k)} \max_{c\in\countersset(\A)}\big|\E[\realeffect_\by](c)\big|
		=\\=
		\fsize(i)\cdot \lfloor n^{1-\epsilonone} \rfloor \cdot\big( \sum_{\by\in\simplecompset(\A_k)} a_\by + n^{\epsilonbigO-\espilonresets}   \big) \cdot \max_{\by\in\simplecompset(\A_k)} \max_{c\in\countersset(\A)}\big|\E[\realeffect_\by](c)\big| \leq \fsize(i)\cdot n^{\epsilonkuwait}			
	\end{gather*}
	where the last inequality follows from \begin{equation}\label{eq-eps-4}
		1-\epsilonone+\epsilonbigO-\espilonresets<\epsilonkuwait
	\end{equation}

	Therefore conditioned on \(\hat{F} \) the  decrease of any \(c\in C_+\) in the main-bin during the \(i\)-th loop is upper bounded by \(\fsize(i)\cdot (n^{1-\epsilonone-\epsilonFIXINGDIFFERENTSMALL} + n^{\epsilonkuwait}) \). 	From \eqref{eq-fsize-def} it holds
	\begin{gather}
		\fsize(i)\cdot (n^{1-\epsilonone-\epsilonFIXINGDIFFERENTSMALL} + n^{\epsilonkuwait}) \nonumber
		=\\=
		\big\lfloor\frac{2\cdot \fsize(i-1)\cdot \lfloor n^{1-\epsilontwo} \rfloor - \gresets(i-1)\cdot n^{1-\epsilonthree+\epsilonX}}{\lfloor n^{1-\epsilontwo} \rfloor}\big\rfloor\cdot (\lfloor n^{1-\epsilonone} \rfloor + n^{\epsilonkuwait})\nonumber
		\allowdisplaybreaks\\\leq 
		\big\lfloor\frac{2\cdot \fsize(i-1)\cdot \lfloor n^{1-\epsilontwo} \rfloor }{\lfloor n^{1-\epsilontwo} \rfloor}\big\rfloor\cdot (n^{1-\epsilonone-\epsilonFIXINGDIFFERENTSMALL} + n^{\epsilonkuwait})
		\nonumber\\=
		\lfloor 2\cdot \fsize(i-1)  \rfloor\cdot (n^{1-\epsilonone-\epsilonFIXINGDIFFERENTSMALL} + n^{\epsilonkuwait}) \tag{**}\label{local-eq-max-counter-change}
	\end{gather}

	But from the conditioning on \(E_{i-1} \) each counter \(c\in C_+ \) has, at the start of the \(i\)-th loop, in the main-bin size at least \(2\cdot \fsize(i-1) \cdot \lfloor n^{1-\epsilonone} \rfloor\).  Hence no counter \(c\in C_+ \) can become negative in the main-bin during the \(i\)-th loop from \begin{equation}\label{eq-eps-kuwait}
		\epsilonkuwait<1-\epsilonone
	\end{equation}

	\textbf{Lower bound for counters \(c\in C_+ \):} It remains to show that  when the \(i\)-th loop finishes the value of each counter \(c\in C_+ \) in the main-bin is at least \(2\cdot \fsize(i) \cdot \lfloor n^{1-\epsilonone} \rfloor  \). 
	
	From \eqref{local-eq-max-counter-change} we have that conditioned on \(\hat{F} \) the maximal decrease of any counter \(c\in C_+\) during the \(i\)-th loop in the main-bin is \( \lfloor 2\cdot \fsize(i-1)  \rfloor\cdot (n^{1-\epsilonone-\epsilonFIXINGDIFFERENTSMALL} + n^{\epsilonkuwait})\). Furthermore, all the multi-component iterations combined  increase each counter \(c\) by at least \[\fsize(i)\cdot \lfloor n^{1-\epsilonone} \rfloor  \cdot  \sum_{\by\in\simplecompset(\A_k)} a_\by\cdot \E[\realeffect_\by](c) \] since they iterate the full \(\bx \) a total of \(\fsize(i)\cdot \lfloor n^{1-\epsilonone} \rfloor\) times. From  \eqref{eq-multi-comp-it-big} it holds for each \(c\in C_+ \) that \[\fsize(i)\cdot \lfloor n^{1-\epsilonone} \rfloor  \cdot  \sum_{\by\in\simplecompset(\A_k)} a_\by\cdot \E[\realeffect_\by](c) \leq \fsize(i)\cdot \lfloor n^{1-\epsilonone} \rfloor  \cdot  3 \]  Hence, when the \(i\)-th loop finishes,  each counter \(c\in C_+\) is in the main-bin equal to at least \[2\cdot \fsize(i-1) \cdot \lfloor n^{1-\epsilonone} \rfloor -  \lfloor 2\cdot \fsize(i-1)  \rfloor\cdot (n^{1-\epsilonone-\epsilonFIXINGDIFFERENTSMALL} + n^{\epsilonkuwait}) +  \fsize(i)\cdot \lfloor n^{1-\epsilonone} \rfloor  \cdot  3 > \fsize(i)\cdot \lfloor n^{1-\epsilonone} \rfloor  \cdot  2 \]
	where the inequality holds from \eqref{eq-eps-kuwait}. Hence the induction step holds.
\end{proof}

\FloatBarrier
\subsubsection{Solution for the \(\epsilon[*] \) Variables}
\label{sec-eps-assign} 
In this subsection we show that, for  any \(0<\epsilon<\frac{1}{2} \), all the restrictions we put on the \(\epsilon[*] \)  variables thorough this section form a system of linear inequalities that has a solution with \(\epsilontenvar=\frac{1}{2}-\epsilon \). We list all the restrictions as well as a solution in Table~\ref{Table-eps-equations}. In Table~\ref{Table-eps-equations-substitute} we then present all the restrictions with the corresponding \(\epsilon[*] \) variables substituted with the values assigned to them in Table~\ref{Table-eps-equations}. One can easily verify that each of the inequalities is satisfied under this solution.

\begin{table*}[!htbp]
	\caption{All the restrictions upon the \(\epsilon[*]\) variables and a value assignment satisfying these}
	\centering
		\setlength{\extrarowheight}{4.5pt}
		\begin{tabular}{|l||c c| } 
			\hline
			\(\epsilon\) assignment&\multicolumn{2}{|c|}{restrictions} \\ 
			\hline\hline
			\(0<\epsilon<\frac{1}{2} \)	&\(0<\epsilon[*]< 1  \) &   \\ 
			
			
			\hline\(\epsilontwo=\frac{\epsilon}{1000} \)&
			\(\epsilonfour<\epsilonsingleiteration\)
			& \eqref{eq_eps_1}\\
			\hline	\(\epsilonthree=\frac{\epsilon}{10}	\)&
			
			\(\epsilonbigZ<2\cdot \epsilonbigY\)
			& \eqref{eq-eps-new-xtwo}\\
			\hline
			\( \epsilonone=\frac{1}{4}-\frac{\epsilon}{100}\)&
			
			\(\epsilonbigZ<1-\epsilonthree\)
			& \eqref{eq-eps-new-xthree} \\
			\hline
			\(\epsilonsingleiteration=\frac{1}{2}-\frac{\epsilon}{10}\)&
			
			\(1-\epsilontwo>1-\epsilonthree+\epsilonX\)
			& \eqref{eq-eps-new-1}\\
			\hline
			\(\espilonresets=1-\frac{\epsilon}{1000}\)&
			
			\(\espilonresets+\epsilonANYSMALL+\epsilonsingleiteration<\epsilonTECHNICALHERE-\epsilonFIXINGDIFFERENTSMALL\)
			& \eqref{eq-eps-technical}\\
			\hline
			\( \epsev=1-\frac{\epsilon}{100}\)&

			\(\epsilonfour <\frac{1}{2}-2\cdot \epsilonthree-4\cdot \epsilonANYSMALLDIFFERENT\)
			& \eqref{eq-eps-technicaldif} \\
			
			\hline
			\(\epsilonkuwait=\frac{3}{4}+\frac{\epsilon}{500} \)&			
			\(\epsilonsingleiteration+1-\epsilonone<\epsilonNEWFIXSECOND\)
			& \eqref{eq-eps-new tesitng}   \\
			
			\hline
			\(\epsilontenvar=\frac{1}{2}-\epsilon\)&
			\(\epsilonbigO>1-\epsilonthree\)
			& \eqref{eq-eps-new-3}\\
			\cline{2-3}
			&
			
			\(\epsilonsingleiteration+ \epsilonbigO<\epsilondoublebigOO\)
			& \eqref{eq-eps-newly}\\
			
			%
			
			\cline{2-3}&
			\(\frac{1}{2}-2\cdot \epsilonone-4\cdot \epsilonFIXINGDIFFERENTSMALL>0\)
			& \eqref{eq-eps-harddd}\\
			%
			
			\cline{2-3}&			
			\(1-\epsilonone+\epsilonbigO-\espilonresets<\epsilonkuwait\)
			& \eqref{eq-eps-4}\\
			
			\cline{2-3}&			
			\(\epsilonkuwait<1-\epsilonone\)
			& \eqref{eq-eps-kuwait}\\

			%
			%
			\hline
		\end{tabular}
		\par
		\label{Table-eps-equations}	
\end{table*}

\begin{table*}[!htbp]
	\caption{All the restrictions upon the \(\epsilon[*]\) variables with values substituted for the values from Table~\ref{Table-eps-equations}}
	\centering
		\setlength{\extrarowheight}{4.5pt}
		\begin{tabular}{|c c| } 
			\hline
			\multicolumn{2}{|c|}{restrictions after substitution} \\ 
			\hline\hline
			\(0<\epsilon[*]< 1  \) &   \\ 
			
			\hline
			\((\frac{1}{2}-\epsilon)<(\frac{1}{2}-\frac{\epsilon}{10})\)
			& \eqref{eq_eps_1}\\
			\hline
			\(\frac{\epsilon}{1000}<2\cdot (1-\frac{\epsilon}{100})\)
			& \eqref{eq-eps-new-xtwo}\\
			\hline
			\(\frac{\epsilon}{1000}<1-\frac{\epsilon}{10}\)
			& \eqref{eq-eps-new-xthree} \\
			\hline
			\(1-\frac{\epsilon}{1000}>1-\frac{\epsilon}{10}+\frac{\epsilon}{1000}\)
			& \eqref{eq-eps-new-1}\\
			\hline
			\((1-\frac{\epsilon}{1000})+\frac{\epsilon}{1000}+(\frac{1}{2}-\frac{\epsilon}{10})<\epsilonTECHNICALHERE-\frac{\epsilon}{1000}\)
			& \eqref{eq-eps-technical}\\
			\hline
			\((\frac{1}{2}-\epsilon) <\frac{1}{2}-2\cdot \frac{\epsilon}{10}-4\cdot \frac{\epsilon}{1000}\)
			& \eqref{eq-eps-technicaldif} \\
			
			\hline
			\((\frac{1}{2}-\frac{\epsilon}{10})+1-(\frac{1}{4}-\frac{\epsilon}{100})<\epsilonNEWFIXSECOND\)
			& \eqref{eq-eps-new tesitng}   \\
			%
			\cline{1-2}
			\((1-\frac{\epsilon}{100})>1-\frac{\epsilon}{10}\)
			& \eqref{eq-eps-new-3}\\
			\cline{1-2}
			\((\frac{1}{2}-\frac{\epsilon}{10})+ (1-\frac{\epsilon}{100})<\epsilondoublebigOO\)
			& \eqref{eq-eps-newly}\\
			
			
			\cline{1-2}
			\(\frac{1}{2}-2\cdot (\frac{1}{4}-\frac{\epsilon}{100})-4\cdot \frac{\epsilon}{1000}2>0\)
			& \eqref{eq-eps-harddd}\\
			%
			
			\cline{1-2}
			\(1-(\frac{1}{4}-\frac{\epsilon}{100})+(1-\frac{\epsilon}{100})-(1-\frac{\epsilon}{1000})<(\frac{3}{4}+\frac{\epsilon}{500})\)
			& \eqref{eq-eps-4}\\
			
			\cline{1-2}
			\((\frac{3}{4}+\frac{\epsilon}{500})<1-(\frac{1}{4}-\frac{\epsilon}{100})\)
			& \eqref{eq-eps-kuwait}\\
			
			\hline
		\end{tabular}
		\par
		\label{Table-eps-equations-substitute}	
\end{table*}

\FloatBarrier

\subsubsection{Proof of Lemma~\ref{lemma-sqrt-mustbe}}\label{sec-lemma-sqrt-mustbe-proof}
\lemmasqrtmustbe*
\begin{proof}
	Note that all the restrictions are listed in the Table~\ref{Table-eps-equations}.

	From \eqref{eq-eps-technicaldif} it holds \begin{equation*}
		\epsilonfour <\frac{1}{2}-2\cdot \epsilonthree-4\cdot \epsilonANYSMALLDIFFERENT
	\end{equation*}
	which implies 
	\begin{equation*}
		\epsilonfour <\frac{1}{2}-2\cdot \epsilonthree 
	\end{equation*}
	\begin{equation*}
		\epsilonfour <2-2\cdot \epsilonthree - 1.5
	\end{equation*}
	\begin{equation*}
		\epsilonfour <2\cdot (1-\epsilonthree) -\epsilonTECHNICALHERE
	\end{equation*}
	From  \eqref{eq-eps-technical} it has to hold \begin{equation*}
		\espilonresets+\epsilonANYSMALL+\epsilonsingleiteration<\epsilonTECHNICALHERE-\epsilonANYSMALL
	\end{equation*}
	which implies 
	\begin{equation*}
		\espilonresets+\epsilonANYSMALL+\epsilonsingleiteration<\epsilonTECHNICALHERE
	\end{equation*}
	which further gives us \begin{align*}
		\epsilonfour 
		&<2\cdot (1-\epsilonthree) -\epsilonTECHNICALHERE\\
		&<
		2\cdot (1-\epsilonthree) -(\espilonresets+\epsilonANYSMALL+\epsilonsingleiteration)
		\\&<
		2\cdot (1-\epsilonthree) -\espilonresets-\epsilonsingleiteration
	\end{align*}
	
	From \eqref{eq-eps-4} and \eqref{eq-eps-kuwait} it holds
	\begin{equation*}
		1-\epsilonone+\epsilonbigO-\espilonresets<\epsilonkuwait<1-\epsilonone
	\end{equation*}
	\begin{equation*}
		1-\epsilonone+\epsilonbigO-\espilonresets<1-\epsilonone
	\end{equation*}
	\begin{equation*}
		\epsilonbigO-\espilonresets<0
	\end{equation*}
	
	which implies \(\espilonresets>\epsilonbigO\). From \eqref{eq-eps-new-3} it holds \(\epsilonbigO>1-\epsilonthree\), hence \[ \espilonresets>1-\epsilonthree\]
	Therefore \begin{align*}
		\epsilonfour 
		&<
		2\cdot (1-\epsilonthree) -\espilonresets-\epsilonsingleiteration
		\\&<
		2\cdot (1-\epsilonthree) -(1-\epsilonthree)-\epsilonsingleiteration
		\\&=
		1-\epsilonthree -\epsilonsingleiteration
		\\&<1 -\epsilonsingleiteration
	\end{align*}

	But from \eqref{eq_eps_1} it also holds
	\(
	\epsilonfour<\epsilonsingleiteration
	\)
	hence \begin{align*}
		\epsilonfour 
		&<
		1 -\epsilonsingleiteration
		\\&<
		1 -\epsilonfour
	\end{align*}
	which can be rewritten as 
	\[	2\cdot \epsilonfour <
	1\]
	\[	\epsilonfour <\frac{1}{2}\]
\end{proof}

\subsection{Proof  of Lemma~\ref{lemma-technical-epsprime}}
\label{sec-lemma-technical-epsprime-proof}
\lemmatechnicalepsprimeee*
\begin{proof}
	Substituting for \(\epsilon' \) on the left side we get
	\begin{gather*}
		\big(1+(i-1)\cdot \epsilon\big)\cdot a[i-1]\cdot (k[i]+\epsilon')
		=\allowdisplaybreaks\\=
		\big(1+(i-1)\cdot \epsilon\big)\cdot a[i-1]\cdot (k[i] +\frac{k[i]\cdot \epsilon}{1+(i-1)\cdot \epsilon})
		\allowdisplaybreaks\\=
		\big(1+(i-1)\cdot \epsilon\big)\cdot a[i-1]\cdot k[i]\cdot (1+\frac{\epsilon}{1+(i-1)\cdot \epsilon})
		\allowdisplaybreaks\\=
		\big(1+(i-1)\cdot \epsilon\big)\cdot a[i-1]\cdot k[i]
		+
		\big(1+(i-1)\cdot \epsilon\big)\cdot a[i-1]\cdot k[i]\cdot \frac{\epsilon}{1+(i-1)\cdot \epsilon}
		\allowdisplaybreaks\\=			
		\big(1+(i-1)\cdot \epsilon\big)\cdot a[i-1]\cdot k[i]
		+
		a[i-1]\cdot k[i]\cdot \epsilon
		\allowdisplaybreaks\\=
		\big(1+i\cdot \epsilon\big)\cdot a[i-1]\cdot k[i]
	\end{gather*}
	
	Furthermore it holds \begin{gather*}
		a[i-1]\cdot k[i]
		=
		3^{(i-1)\cdot |\countersset(\A)|+|\transitions(\MEC_1)|+\ldots+|\transitions(\MEC_{i-1})|} \cdot 3^{|\countersset(\A)|+ |\transitions(|\MEC_i)|} 
		=\\=
		3^{i\cdot |\countersset(\A)|+|\transitions(\MEC_1)|+\ldots+|\transitions(\MEC_{i-1})|+ |\transitions(|\MEC_i)|}
		=
		a[i] 
	\end{gather*} 
	
	Hence \begin{gather*}
		\big(1+(i-1)\cdot \epsilon\big)\cdot a[i-1]\cdot (k[i]+\epsilon')
		=
		\big(1+i\cdot \epsilon\big)\cdot a[i-1]\cdot k[i]
		=
		\big(1+i\cdot \epsilon\big)\cdot a[i]
	\end{gather*}	
\end{proof}

\subsection{Proof of Lemma~\ref{lemma-II-solution-bad-only-once}}
\label{sec-proof-of-lemma-II-solution-bad-only-once}
\lemmaIIsolutionbadonlyonce*
\begin{proof}
	Assume that  \(\ba_1+k\cdot \ba_2,\bb_1+k\cdot \bb_2 \) does not satisfy the last objective of of \hyperref[fig-systems]{(II)} for \(p\). That means that for all \((p,\bu,q)\in\tout(p)\) it holds \[(\bb_1+k\cdot \bb_2)(q)-(\bb_1+k\cdot \bb_2)(p) + \sum_{c\in\countersset(\A)} \bu(c)\cdot (\ba_1+k\cdot \ba_2)(c)\geq 0 \]
	
	Since \(\ba_1+k\cdot \ba_2,\bb_1+k\cdot \bb_2 \) is a valid solution of \hyperref[fig-systems]{(II)} it holds from the constraint~\eqref{eq-II-prob-states} that 
	\begin{gather*}
		\sum_{t= (\stateVass,\bu,\stateVassalt) \in \tout(\stateVass)}P(t)\cdot \big((\bb_1+k\cdot \bb_2)(\stateVassalt)-(\bb_1+k\cdot \bb_2)(\stateVass)+
		\sum_{c\in \countersset(\A)}  \bu(c)\cdot (\ba_1+k\cdot \ba_2)(c)\big)\leq 0 
	\end{gather*}
	therefore for all \((p,\bu,q)\in\tout(p)\) it holds \[(\bb_1+k\cdot \bb_2)(q)-(\bb_1+k\cdot \bb_2)(p) + \sum_{c\in\countersset(\A)} \bu(c)\cdot (\ba_1+k\cdot \ba_2)(c)= 0 \]
	
	
	which can be rewritten as  \begin{gather*}
		0=(\bb_1+k\cdot \bb_2)(q)-(\bb_1+k\cdot \bb_2)(p) + \sum_{c\in\countersset(\A)} \bu(c)\cdot (\ba_1+k\cdot \ba_2)(c)
		=\\=
		\bb_1(q)-\bb_1(p) + \sum_{c\in\countersset(\A)} \bu(c)\cdot \ba_1(c) + k\cdot\big( \bb_2(q)-\bb_2(p) + \sum_{c\in\countersset(\A)} \bu(c)\cdot \ba_2(c) \big)
	\end{gather*}
	
	but this means \[k
	=
	\frac{-\bb_1(q)+\bb_1(p) - \sum_{c\in\countersset(\A)} \bu(c)\cdot \ba_1(c)}{\bb_2(q)- \bb_2(p) + \sum_{c\in\countersset(\A)} \bu(c)\cdot \ba_2(c)}
	\]
	The lemma holds.
\end{proof}

	

\end{document}